%% file: MultiScaleTopoOpt.tex
\pdfoutput=1
\documentclass[acmtog,authorversion]{acmart}
\usepackage{wrapfig}
\usepackage{algorithm}
\usepackage{amsmath}
\usepackage{amssymb}
\usepackage{soul}
\usepackage{algpseudocode}
\usepackage{wrapfig}
\usepackage{epstopdf}
\usepackage{tabularx}

\input{Commands}

\newcommand{\mydraft}{false}

\newcommand{\fourfigure}[6]{
{\hbox{\includegraphics[width=#1,draft=\mydraft]{#3}\hspace{#2}\includegraphics[width=#1,draft=\mydraft]{#4}\hspace{#2}\includegraphics[width=#1,draft=\mydraft]{#5}\hspace{#2}\includegraphics[width=#1,draft=\mydraft]{#6}}}
}

\newcommand{\sixfigure}[8]{
{\hbox{\includegraphics[width=#1,draft=\mydraft]{#3}\hspace{#2}\includegraphics[width=#1,draft=\mydraft]{#4}\hspace{#2}\includegraphics[width=#1,draft=\mydraft]{#5}\hspace{#2}\includegraphics[width=#1,draft=\mydraft]{#6}\hspace{#2}\includegraphics[width=#1,draft=\mydraft]{#7}\hspace{#2}\includegraphics[width=#1,draft=\mydraft]{#8}}}
}

\newcommand{\twofigure}[4]{
{\hbox{\includegraphics[width=#1,draft=\mydraft]{#3}\hspace{#2}\includegraphics[width=#1,draft=\mydraft]{#4}}}
}

\newcommand{\updated}[1]{{\color{black}#1}}

\newcommand{\vect}[1]{\boldsymbol{\mathbf{#1}}}	

\acmJournal{TOG}
\acmVolume{34}
\acmNumber{4}
\acmArticle{84}
\acmYear{2017}
\acmMonth{7}

\setcopyright{acmcopyright}

\setcitestyle{square}

\begin{document}

\title{Two-Scale Topology Optimization with Microstructures}
\author{Bo Zhu, M{\' e}lina Skouras, Desai Chen, Wojciech Matusik}
\affiliation{%
  \institution{MIT CSAIL}
}

\begin{teaserfigure}
    \centering
		\includegraphics[width=.99\linewidth]{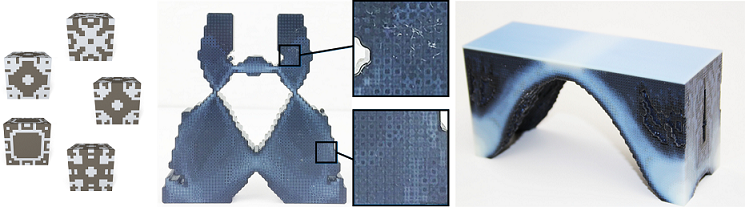}
		\caption{Our two-scale topology optimization framework allows to optimize continuous material properties mapping to printable microstructures (\emph{left}) to fabricate high-resolution functional objects (\emph{middle}) and minimum compliant structures (\emph{right}).}
    \label{fig:teaser}
\end{teaserfigure}

\renewcommand\shortauthors{B. Zhu et al.}

\input{abstract}


%
%

\maketitle

\input{Introduction}

\input{RelatedWork}

\input{Overview}

\input{MaterialModel}

\input{MicrostructureDatabase}

\input{TopoOpt}

\input{Results}

\input{Conclusion}

\section*{Acknowledgements}
This research is supported in part by the Defense Advanced Research Projects Agency (DARPA) and Space and Naval Warfare Systems Center Pacific (SSC Pacific) under Contract No 66001-15-C-4030.

\bibliographystyle{ACM-Reference-Format}
\bibliography{MultiScaleTopoOpt}
\input{Appendix}
\end{document}

%% file: Commands.tex
\newcommand{\N}{\mbox{\rm \hbox{I\kern-.15em\hbox{N}}}}
\newcommand{\R}{\mbox{\rm \hbox{I\kern-.15em\hbox{R}}}}

\def \N {\mbox{\rm \hbox{I\kern-.15em\hbox{N}}}}
\def \R {\mbox{\rm \hbox{I\kern-.15em\hbox{R}}}}

\newcommand{\bC}{\mathbf{C}}
\newcommand{\bD}{\mathbf{D}}

\newcommand{\bG}{\mathbf{G}}

\newcommand{\bK}{\mathbf{K}}

\newcommand{\be}{\mathbf{e}}

\newcommand{\bp}{\mathbf{p}}

\newcommand{\bu}{\mathbf{u}}

\newcommand{\bx}{\mathbf{x}}



%% file: abstract.tex
\begin{abstract}
In this paper we present a novel two-scale framework to optimize the structure and the material distribution of an object given its functional specifications. Our approach utilizes multi-material microstructures as low-level building blocks of the object. We start by precomputing the material property gamut -- the set of bulk material properties that can be achieved with all material microstructures of a given size. We represent the boundary of this material property gamut using a level set field. Next, we propose an efficient and general topology optimization algorithm that simultaneously computes an optimal object topology and spatially-varying material properties constrained by the precomputed gamut. Finally, we map the optimal spatially-varying material properties onto the microstructures with the corresponding properties in order to generate a high-resolution printable structure. We demonstrate the efficacy of our framework by designing, optimizing, and fabricating objects in different material property spaces on the level of a trillion voxels, i.e several orders of magnitude higher than what can be achieved with current systems.

\end{abstract}

%% file: Introduction.tex
\section{Introduction}

Many engineering problems focus on the design of complex structures that needs to meet high level objectives such as the capability to support localized stresses, optimal tradeoffs between compliance and mass, minimal deformation under thermal changes, etc. One very popular approach to design such structures is topology optimization. Topology optimization generally refers to discretizing the object of interest into small elements and optimizing the material distribution over these elements in such a way that the functional goals are satisfied \cite{bendsoe2004topology}. Traditionally, topology optimization focused on designs made of homogeneous materials and was concerned with macroscopic changes in the object geometry.
With the advent of multi-material 3D printing techniques, it is now possible to play with materials at a much higher resolution, allowing to obtain much finer designs and, thus, improved functional performances.
Unfortunately, standard techniques for topology optimization do not scale well and they cannot be run on objects with billions of voxels. This is because the number of variables to optimize increases linearly with the number of cells in the object. Since many current 3D printers have a resolution of 600DPI or more, a one billion voxel design occupies only a 1.67 inch cube.


One direction to handle this issue is to work with microstructures corresponding to blocks of voxels instead of individual voxels directly. Some recent works followed this direction and proposed to decouple macro structural design and micro material design \cite{rodrigues:2002:hierarchical,coelho:2008:hierarchical,nakshatrala:2013:nonlinear}. However, these approaches remain computationally expensive and, in most cases, limited to the well-known minimal compliance problem.
The second direction to reduce the problem complexity is to temporarily ignore the geometry of the microstructures and consider only their macroscopic physical behaviour. However, this introduces new difficulties as the space of material properties covered by all printable microstructures is much wider than the properties of the base materials. For example, microstructures made of alternating layers of soft and stiff isotropic materials exhibit an anisotropic behaviour as they are able to stretch more easily in one direction that in the others. This implies that not only the {\it ranges} but also the {\it number} of physical parameters needed to describe the physical behaviour of these microstructures increases. Therefore, in order to work with the material properties of microstructures, one needs to solve two challenging problems: (i) computing the {\it gamut} -- i.e the set -- of the material properties achievable by all microstructures, (ii) efficiently optimizing the distribution of these high-dimensional material properties inside the layout of the object.

Most previous algorithms working in the material space focused on optimizing a single material property such as density or material stiffness, for which analytical formulas describing the property bounds exist \cite{Allaire93Bounds}. On the contrary, optimizing the structure and material distribution of an object in a high dimensional material property space remains an open problem. In this work, we propose a new computational framework for topology optimization with microstructures that supports design spaces of multiple dimensions. We start by computing the gamut of the material properties of the microstructures by alternating stochastic sampling and continuous optimization. This gives us a {\it discrete} representation of the set of achievable material properties, from which we can construct a {\it continuous} gamut representation using a level set field. We then reformulate the topology optimization problem in the continuous space of material properties and propose an efficient optimization scheme that finds the optimized distributions of multiple material properties simultaneously inside the gamut.
Finally, in order to obtain fabricable designs, we map the optimal material properties back to discrete microstructures from our database.

Our general formulation can be applied to a large variety of problems. We demonstrate its efficacy by designing and optimizing objects in different material spaces using isotropic, cubic and orthotropic materials. We apply our algorithm to various design problems dealing with diverse functional objectives such as minimal compliance and target strain distribution. Furthermore, our approach utilizes the high-resolution of current 3D printers by supporting designs with trillions of voxels. We fabricate several of our designs, thus, demonstrating the practicality of our approach.

The main contributions of our work can be summarized as follows:
\begin{itemize}
\item We present a fully automatic method for computing the space of material properties achievable by microstructures made of a given set of base materials.
\item We propose a generic and efficient topology optimization algorithm capable of handling objects with a trillion voxels. The key of our approach is a reformulation of the problem to work directly on continuous variables representing the material properties of microstructures. This allows us to cast topology optimization as a reasonably sized constrained optimization problem that can be efficiently solved with state of the art solvers.
\item We validate our method on a set of test cases and demonstrate its versatility by applying it to various design problems of practical interest.
\end{itemize}

%% file: RelatedWork.tex
\section{Related Work}
\begin{figure*}[t]
    \centering
    \includegraphics[width=.95\linewidth]{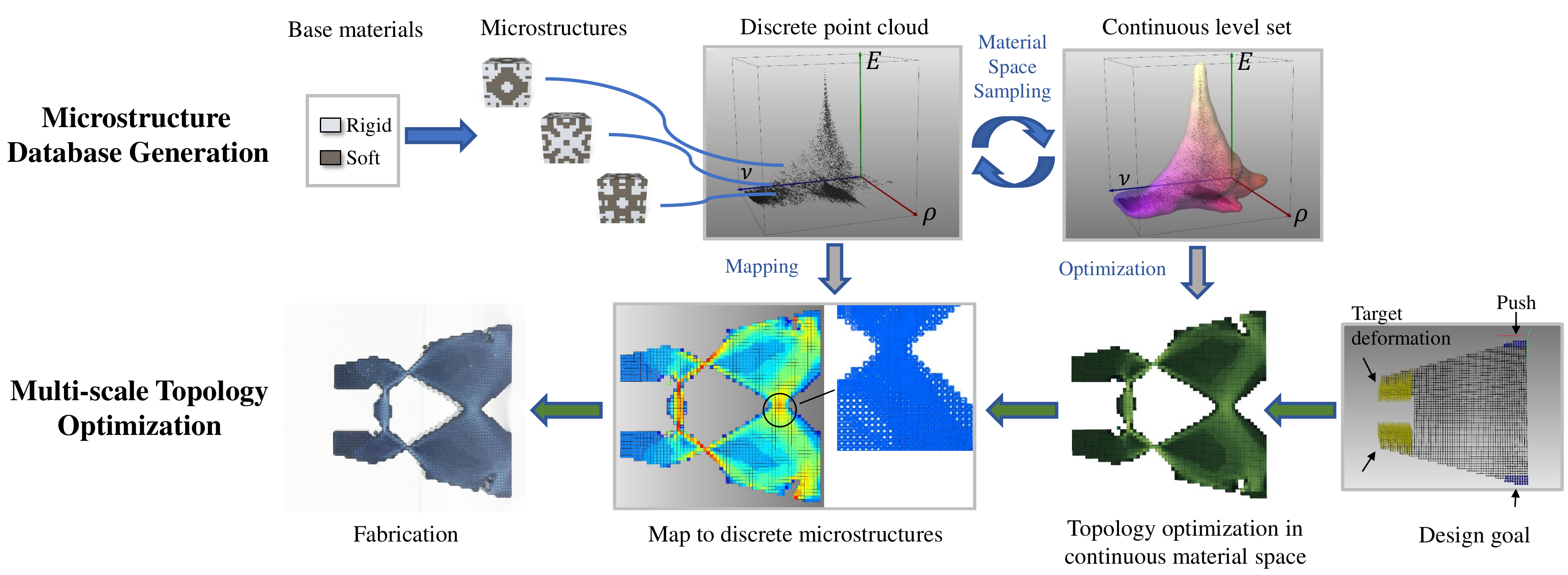}
	\caption{Algorithm overview. We start by precomputing the gamut of material properties that can be achieved with all material microstructures of a given size. Next, we run our topology optimization algorithm that optimizes the material properties of the object within this gamut such as to minimize some functional objective. Finally, we map the optimal continuous material properties back to microstructures from our database to generate a printable object.
    	}
    \label{fig:overview}
\end{figure*}

\paragraph*{Topology Optimization}
Topology optimization is concerned with the search of the optimal distribution of one or more materials within a design domain in order to minimize some input objective function while satisfying given constraints \cite{bendsoe2004topology}.
Initially applied to the structural design in engineering \cite{bendsoe:1989:optimal}, topology optimization has been extended since then to a variety of problems including micromechanism design \cite{Sigmund97Compliant},
mass transfer \cite{challis:2009:level}, metamaterial design \cite{sigmund:1996:composites,cadman:2013:design}, multifunctional structure design \cite{yan:2015:two}, coupled structure-appearance optimization \cite{Martinez:2015:SAO}.
Many algorithms have been proposed to numerically solve the optimization problem itself. We refer to the survey by Sigmund and Maute \shortcite{sigmund:2013:topology} for a complete review.
In the very popular SIMP (Solid Isotropic Materials with Penalization) method, the presence of material in a given cell is controlled by locally varying its density. A binary design is eventually achieved by penalizing intermediate values for these densities. In practice, this method works well for two-material designs (e.g., a material and a void), but generalizing this method to robustly handle higher dimensional material spaces remains challenging.
 Instead of considering only discrete structures, free material optimization \cite{Ringertz1993On,haber1994analytical} optimizes structures made of continuous material distributions constrained by analytical bounds.
Another class of methods rely on {\it homogenization}. They replace the material in each voxel of the object by a mixture of the base materials whose material properties can be analytically derived. While optimal microstructures are known for certain classes of problems (laminated composites in the case of the minimum compliance problem), this is not the case in the general setting, for which using a specific subclass of microstructures can lead to suboptimal results.
In a sense, our work is a generalization of these approaches and aims to handle a wider range of materials for which theoretical bounds on the material properties are not known a priori.

Although they are largely used in engineering, standard methods for topology optimization suffer from a major drawback : the parametrization of the problem at the voxel level makes them extremely expensive and largely impedes their use on high resolutions models such as the ones generated by modern 3D printing hardware. High-performance GPU implementations with careful memory handling can be used to push the limits of what can be done (a couple of million variables in the implementation by Wu et al. \shortcite{Wu2016}), but such approaches rely on specificities of the minimum compliance problem and are difficult to generalize. To counteract the effects of the explosion of variables in finely discretized layouts, Rodrigues et al.~\shortcite{rodrigues:2002:hierarchical} alternatively proposed an interesting formulation where microstructure designs and macroscopic layouts using the effective properties of the underlying microstructures were hierarchically coupled and treated simultaneously. This initial work has been extended in multiple ways \cite{coelho:2008:hierarchical,nakshatrala:2013:nonlinear,yan:2014:concurrent,xia:2014:concurrent}. 
Alexandersen and Lazarov \shortcite{Alexandersen2015Topology} proposed a fast simulation algorithm for optimizing complete macroscopic structures made of layered or periodic microstructures.
However, these methods still need to handle variables defined at the microstructure level and therefore they remain relatively costly. The most related work is the method proposed by Xia et al.\shortcite{xia:2015:multiscale}, which also relies on a database to speed up computations. However, their work specifically targets minimum compliance problems in the structural design which allows them to approximate the macroscale behaviour of the microstructures with a particular strain-based interpolating function.

\paragraph*{Fabrication-oriented Optimization}
The last decade has witnessed an increasing interest by the computer graphics community in the design of tools and algorithms targeting digital fabrication of physical artifacts. The range of media and applications addressed in previous literature is very diverse and we focus our discussion on systems targeting 3D printing.
The problem of optimizing the material assignment for the individual voxels of an object in order to control its large scale behaviour has been studied in different contexts. Starting with optical properties, Ha\v{s}an et al. \shortcite{Hasan:2010:PRO} and Dong et al. \shortcite{Dong:2010:FSS} provided methods for printing objects with desired subsurface scattering properties. Stava et al. \shortcite{Stava12Stress} later considered stability of 3D printed objects, Zhou et al. \shortcite{Zhou2013WorstCase} explored structural strength while Chen et al. \shortcite{Chen:2014:ANM} focused on rest shape optimization. 

Closer to our present work, frameworks for the design of objects with desired mechanical behaviours have been proposed by Bickel et al. \shortcite{Bickel:2010:DAF} and Skouras et al. \shortcite{Skouras13Computational}. Like these works, our system  allows to match given input deformations. However, while these previous systems assume a small set of available base materials and use these base materials in relatively coarse discretizations, our system combines the base materials into microstructures to expand the design possibilities. Also relevant is the tool presented by Xu et al. \shortcite{xu:2015:interactive} that allows to interactively design heterogeneous materials for elastic objects subject to prescribed displacements and forces, and the material optimization approach proposed by Panetta et al. \shortcite{Panetta:2015}. However, these methods may output materials that are not available in the real world for non-convex manifolds of material properties. By contrast, we guarantee that all the microstructures used are always realizable in such cases, which is one of the key contributions of our work. 
Lastly, in an effort to unify individual contributions when dealing with inverse modeling problems, Chen et al.~\shortcite{Chen2013} proposed an abstraction mechanism to facilitate the development of goal-based methods. The output of most of these systems is a per-voxel material composition, which cannot be efficiently represented using simple surface meshes. Vidim\v{c}e et al. \shortcite{Vidimce13} introduced a fabrication-specific language and a programming pipeline for a procedural material synthesis that lift this limitation.

\paragraph*{Microstructures and Metamaterials}
Microstructures can be defined as small scale assemblies made of one or several base materials, whose macroscale properties can be very different from those of the original materials.
Many materials found in the nature are microstructures when observed at a sufficiently small scale. Microstructures can also be engineered so as to define composites with improved capabilities or even {\it metamaterials} with exceptional properties. For example, Lakes \shortcite{lakes:1987:foam} presented in 1987 the first man-made structure with negative Poisson's ratio, i.e., a structure which transversally expands when it is axially stretched. The design of composites and metamaterials is an active research field inspiring myriads of works \cite{sigmund:1996:composites,Sigmund97Compliant,cadman:2013:design,Wang14Nonlinear,Babaee2013,Andreassen2014}. While many of these works are concerned with the inverse modeling of specific microstructures or families of microstructures, the study of the {\it space} of properties that these microstructures can achieve as a whole has been investigated much less. Theoretical bounds have been derived without experimental validation~\cite{Lipton1994Optimal,milton1995,Ting2005Poisson}. Taking into account additive manufacturing constraints, Schumacher et al. \shortcite{Schumacher:2015} and Panetta et al. \shortcite{Panetta:2015} recently investigated the design of tileable and printable microstructures. In the first part of this paper, we further explore this line of research and focus on the generation and characterization of databases of microstructures with maximal material property coverage. In particular, we present a novel approach combining a probabilistic search and a continuous optimization that allows us to {\it fully automatically} explore the gamut of material properties that can be achieved by assembling given base materials.

%% file: Overview.tex
\section{Overview}
Given as input a set of base materials, an object layout, and functional objectives, the goal of our system is to compute the material distribution inside the object in order to optimize these functional objectives. In our approach, we do not solve the problem directly, instead we work with microstructures made of the base materials and the space of physical material properties spanned by them. The complete pipeline of our system, illustrated in Figure \ref{fig:overview}, can be decomposed into three stages.

\paragraph*{Material Space Precomputation}
In the first stage, we estimate the gamut of material properties covered by all possible microstructures made by spatial arrangement of base materials. 
Since exhaustively computing the properties of all these microstructures is, in practice, intractable, we progressively increase the material space by alternating a stochastic search and a continuous optimization. The first step introduces discrete changes in the materials of the microstructures and allows emergence of new types of microstructures. The second step allows to locally push the material space boundaries by refining the microstructure shapes. After completing this stage, we obtain a discrete representation of the space of material properties and the mapping between these properties and the corresponding microstructures.

\paragraph*{Gamut-based Continuous Topology Optimization}
In the second stage, we construct a smooth continuous gamut representation of the material property space by using a level set field. We define our topology optimization problem directly in this space. Our approach minimizes the objective function over possible material parameters while asking for strict satisfaction of the physics constraints -- typically, the static equilibrium -- as well as the strict satisfaction of the physical parameter bounds. Taking advantage of our gamut representation as a level set, we formulate this last constraint as limiting the material properties to stay on the negative side of the level set. This guarantees that the material properties that we use in the optimization are always physically realizable.
\begin{figure*}[t]
\centering
\fourfigure{.25\textwidth}{.03in}{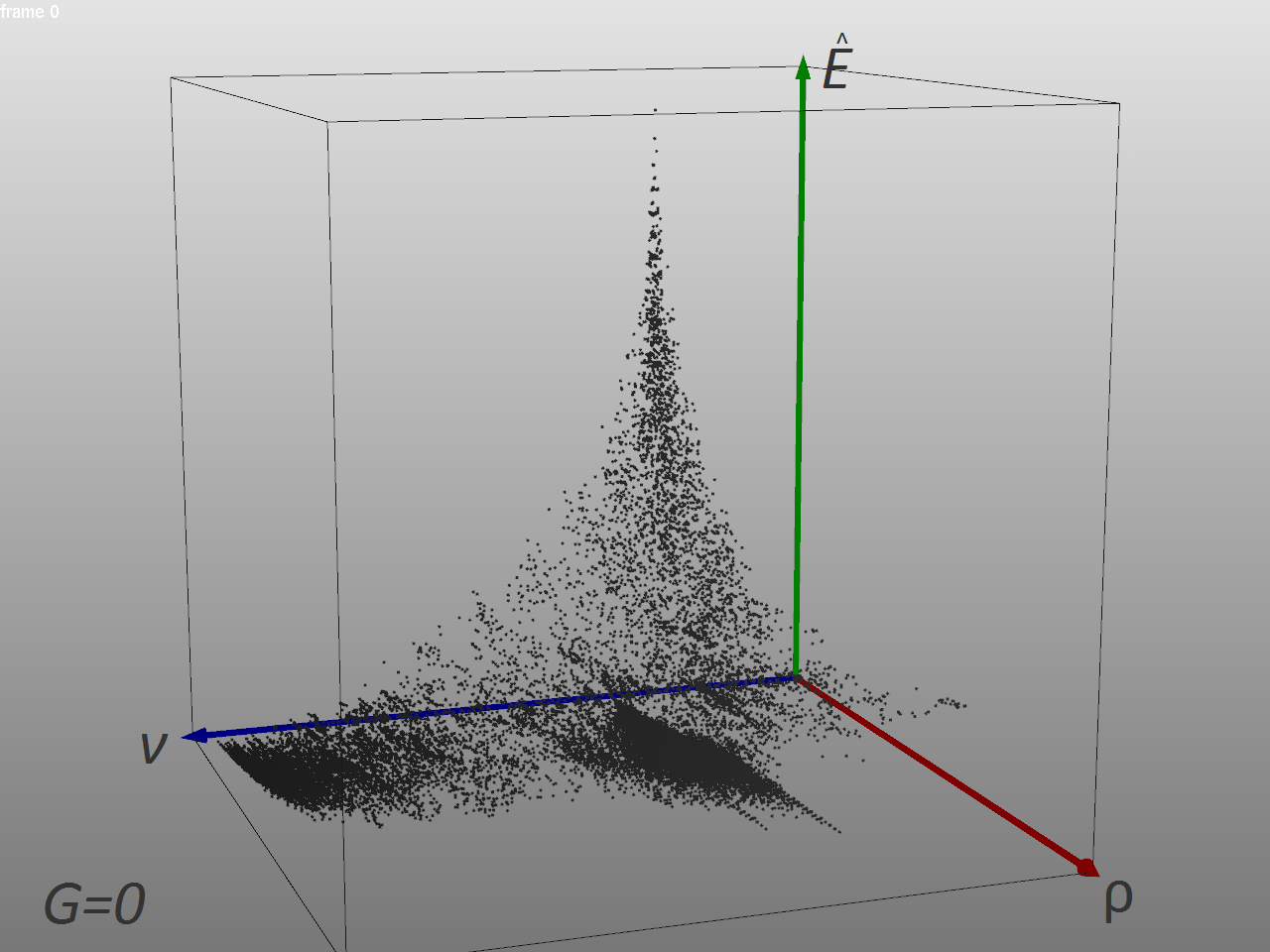}{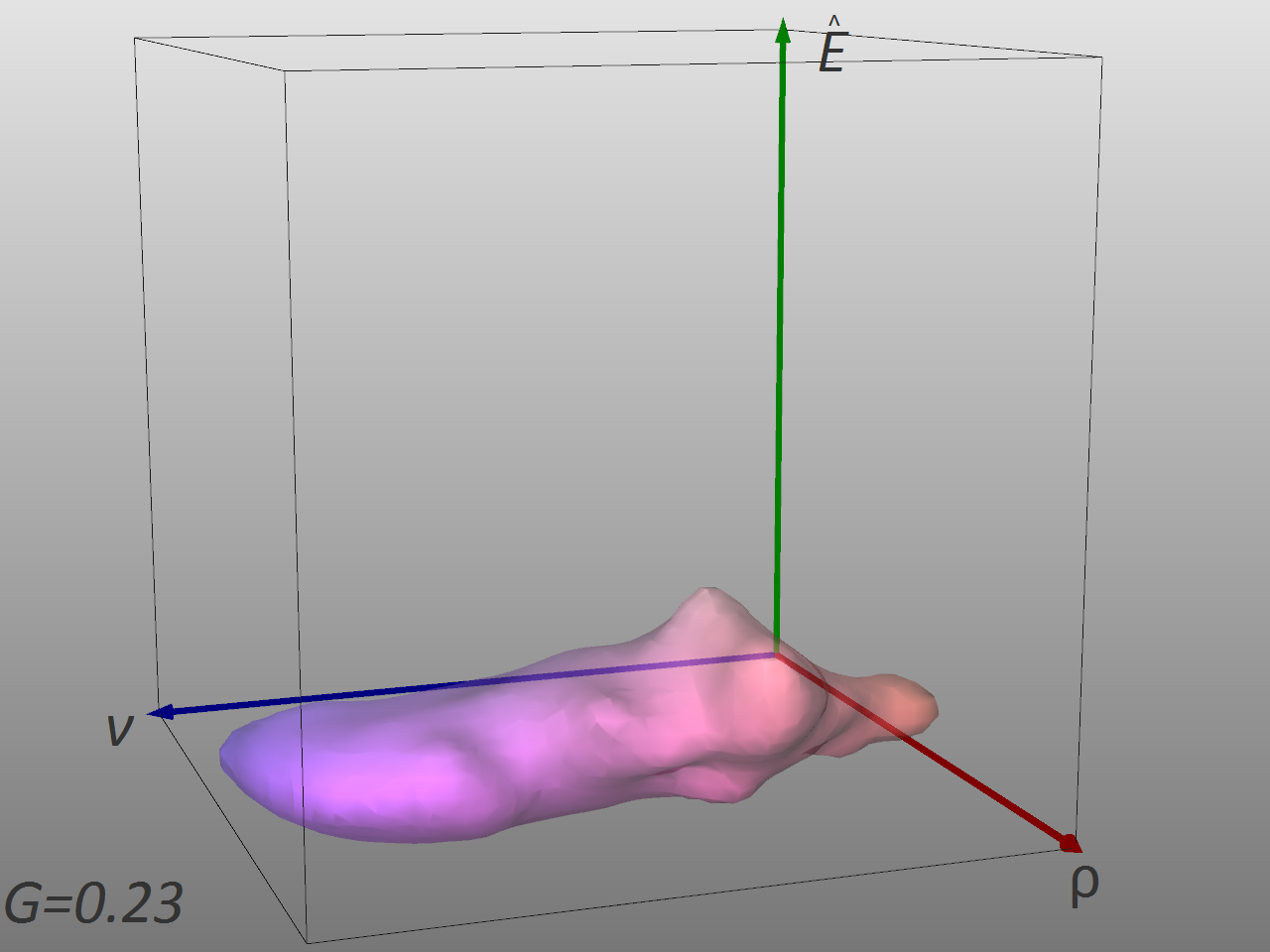}{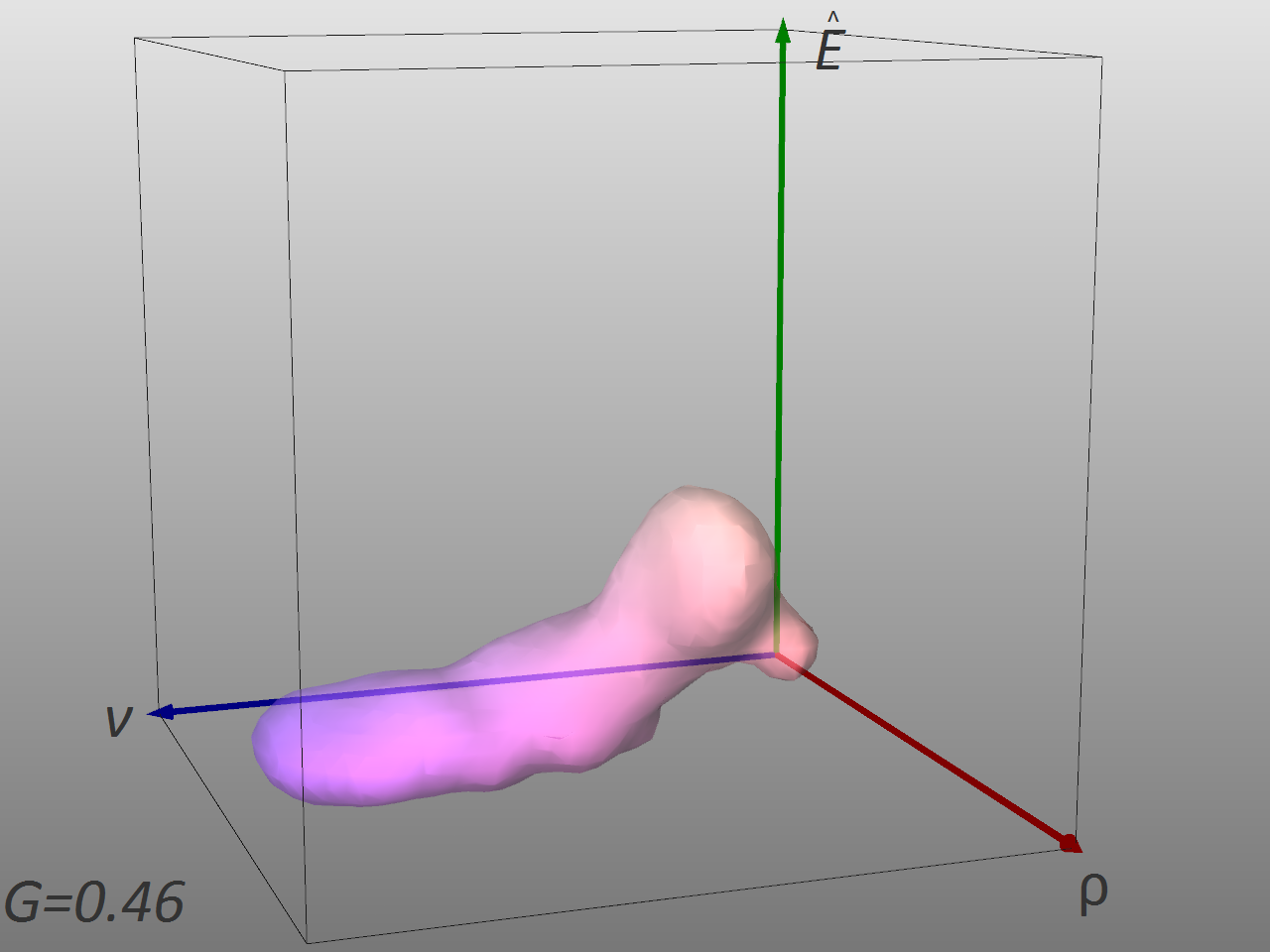}{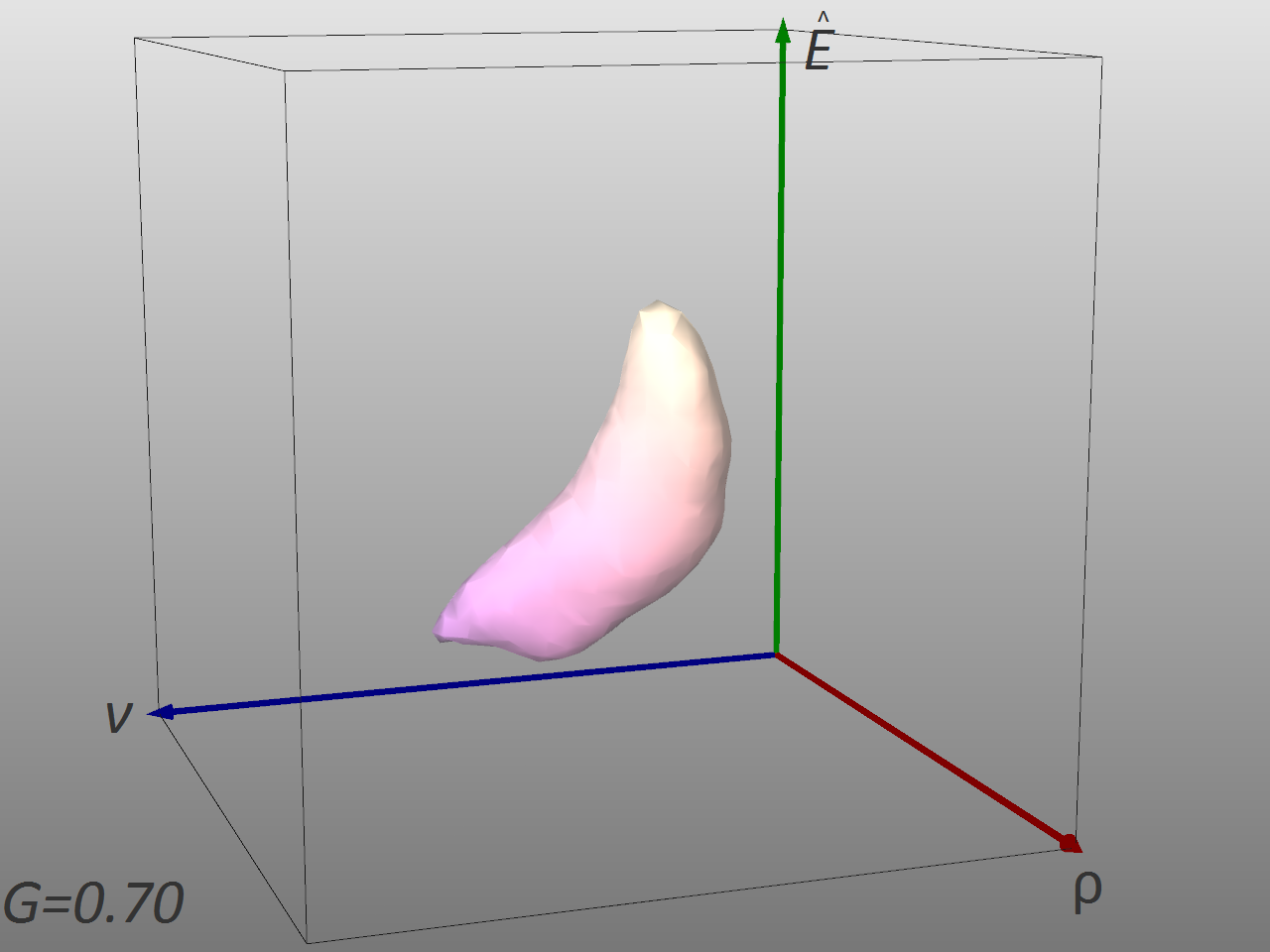}
	\fourfigure{.25\textwidth}{.03in}{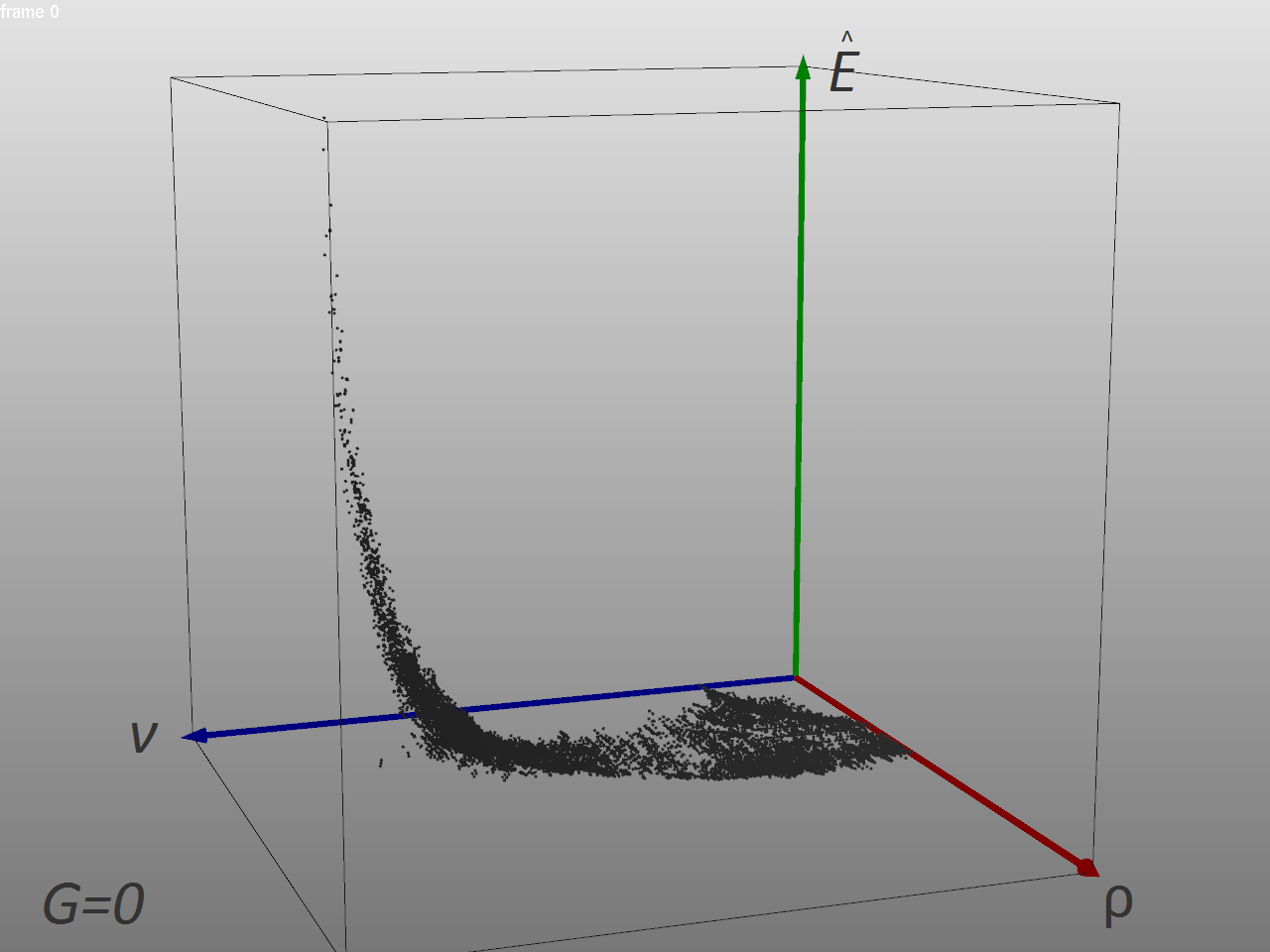}{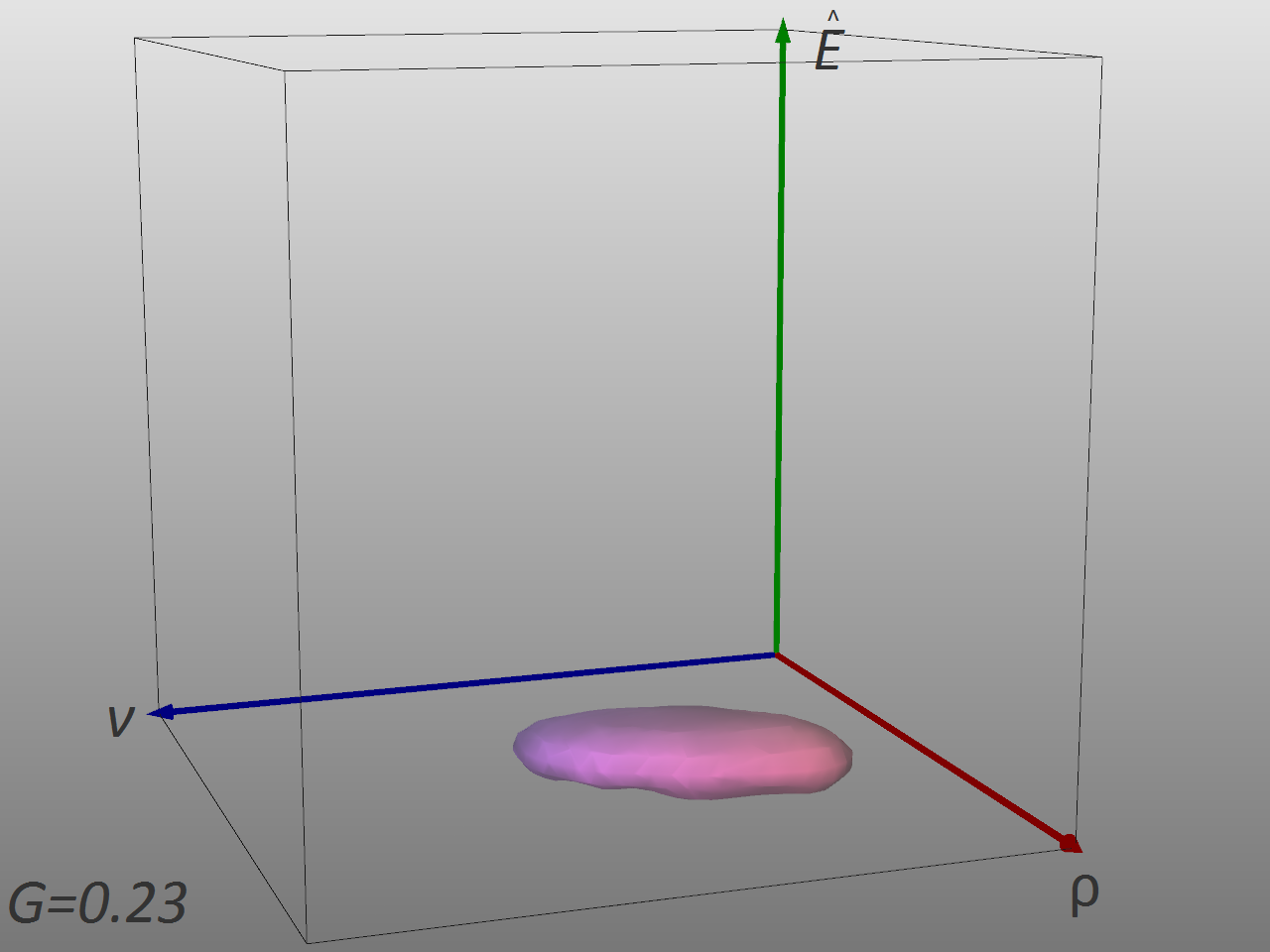}{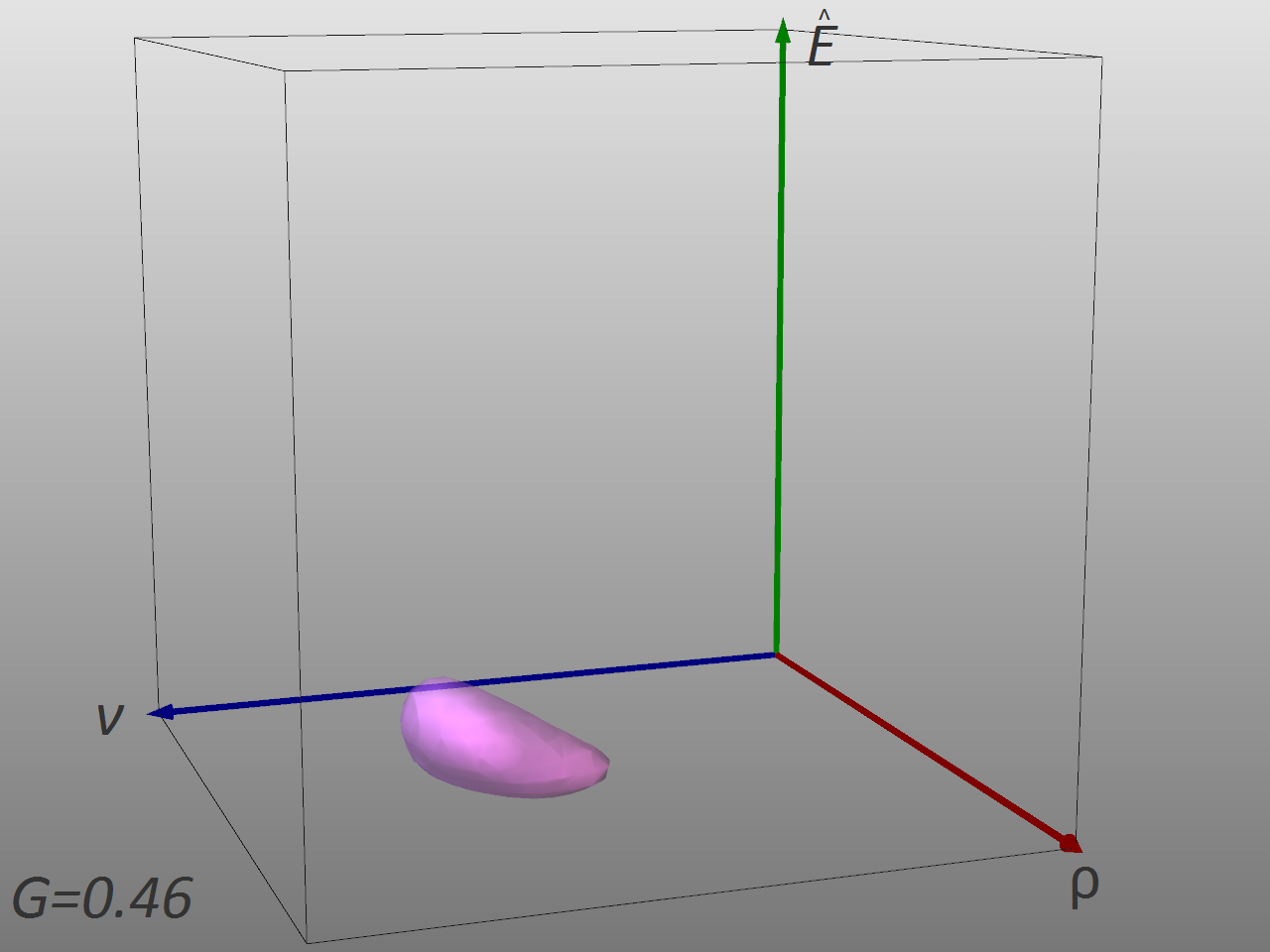}{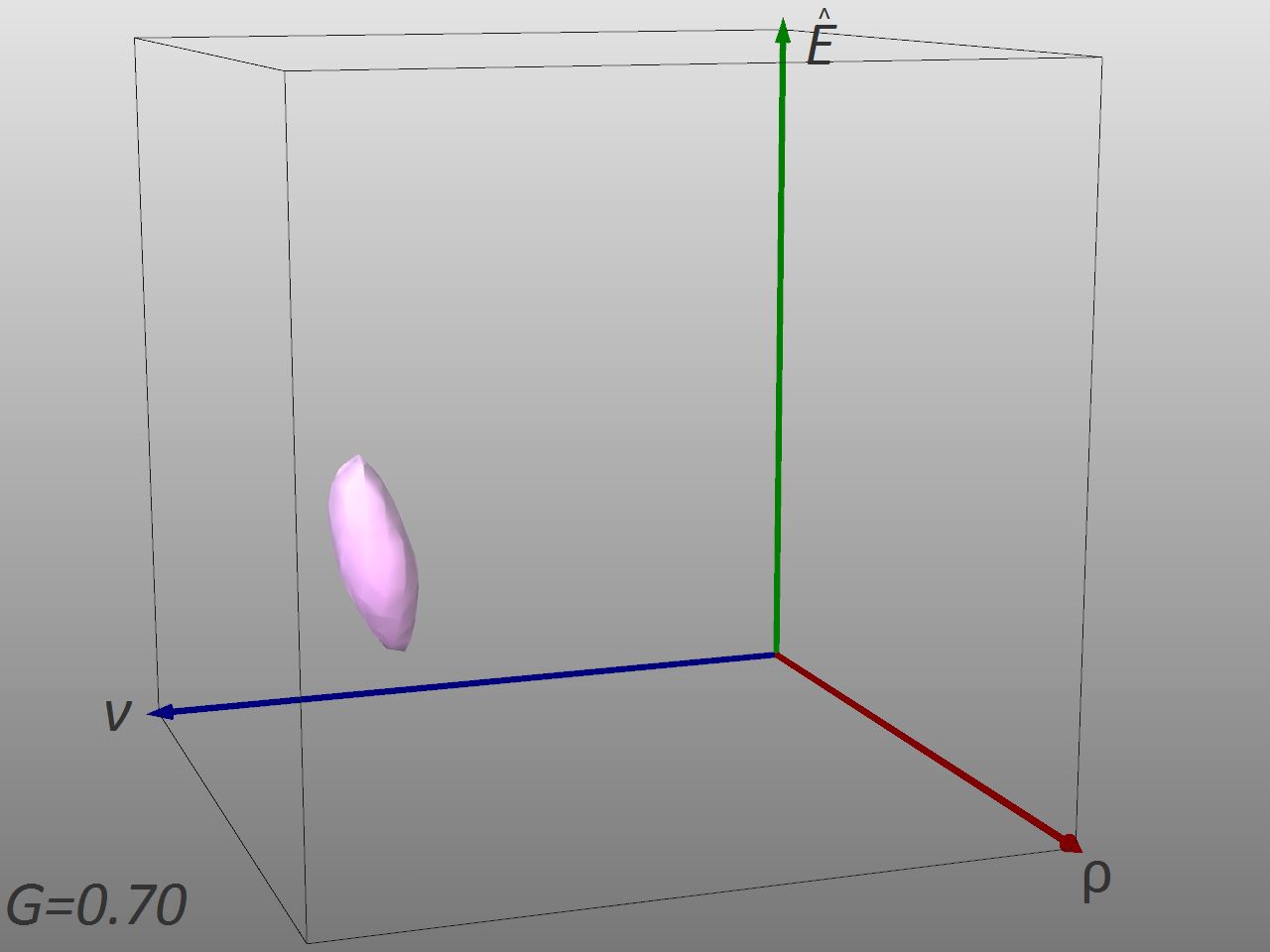}
     \caption{Level set gamuts for two dimensional cubic microstructures (\emph{top row}) and three dimensional cubic microstructures (\emph{bottom row}). The first column shows the projection of the sample points in the space parametrized by the density $\rho$, the normalized Young's modulus $\hat{E}$ and the Poisson's ratio $\nu$. The second to the fourth columns show three slices of the four dimensional level sets corresponding to different values for the shear modulus $G$.}   
\end{figure*}

\paragraph*{Fabrication-oriented Microstructure Mapping}
In the last stage, we generate a printable result by replacing each cell in the object layout with a microstructure whose material properties are the closest to the continuous material assignment resulting from the optimization. We also take into account the boundary similarity across adjacent cell interfaces to improve the connectivity between microstructures. This results in a complex, high-resolution, multi-material model with optimized functional specifications.

%% file: MaterialModel.tex
\section{Mechanics}\label{sec:mm}
In this section, we briefly introduce the background material for simulating deformable objects. We will use these concepts when computing the material properties of the microstructures (Section~\ref{sec:sampling}) and in the topology optimization algorithm (Section \ref{sec:topoOpt}).
We refer to the course by Sifakis and Barbic \shortcite{Sifakis:2012Course} for a more comprehensive exposition.

\subsection{Material Model}
Most available materials for 3D printers are elastic materials. Assuming small deformations, we use linear elasticity to compute both the mechanical behaviour of the entire object and the microstructures. In such a setting, the relation between the linear strain $\epsilon$ and Cauchy stress $\sigma$ at every material point is given by
\begin{equation}
\sigma = \bC\epsilon\ ,
\end{equation}
where $\bC$, the so-called elasticity tensor, can be described by 21 parameters \cite{Bonet97}.

Working in such a high dimensional space is prohibitive and therefore we focus on materials having a certain numbers of symmetries, such as orthotropic materials for which the elasticity tensor is defined by 12 parameters (4 parameters in 2D), cubic materials defined by 3 parameters, and isotropic materials defined by 2 parameters. For example, the tensor for a 3D cubic structure can be written (using the Voigt notation) as
\begin{align}
\bC=\left(
\begin{smallmatrix}
 (1-\nu)\hat{E}& \nu\hat{E} & \nu\hat{E} & & & \\
 \nu\hat{E} & (1-\nu)\hat{E}& \nu\hat{E}& & & \\
 \nu\hat{E} & \nu\hat{E} & (1-\nu)\hat{E}& & & \\
 & & & \mu& & \\
 & & & & \mu& \\
 & & & & & \mu
\end{smallmatrix}\right)
\label{eq:iso_E}
\end{align}
with $\hat{E}=E/((1-2\nu)(1+\nu))$ and where $E$ is the Young's modulus of the material, $\nu$ its Poisson's ratio, and $\mu$ its shear modulus.

Alternatively, one can also use Lam\'e's parameters to define the tensor $\bC$, which simplifies the derivation of the tensor with respect to the elastic parameters. In this case, the tensor has the form
\begin{align}
\bC=\left(
\begin{smallmatrix}
 2\mu+\lambda& \lambda & \lambda & & & \\
\lambda &  2\mu+\lambda& \lambda& & & \\
\lambda & \lambda &  2\mu+\lambda& & & \\
 & & & \mu& & \\
 & & & & \mu& \\
 & & & & & \mu
\end{smallmatrix}\right)\ .
\label{eq:iso_E_b}
\end{align}

 The tensor for a 2D orthotropic structure can be written as
\begin{align}
\bC=c \left(
\begin{smallmatrix}
E_x  & \nu_{yx} E_x &  \\
  \nu_{xy} E_y & E_y \\
 & & & \mu/c \\
\end{smallmatrix}\right)
\label{eq:iso_E}
\end{align}
with $c=1/(1-\nu_{xy} \nu_{yx})$, and where $E_x$ and $E_y$ are the Young's moduli along the two principal axes, $\mu$ is the shear modulus, $\nu_{xy}$ is the Poisson's ratio corresponding to a contraction in the direction y  when an extension is applied along the x axis, and $\nu_{yx}$ verifies $\nu_{yx} E_x=\nu_{xy} E_y$.

Letting $\mathcal{R}^n$ denote the space of $n$ material properties, we then write each point $\mathbf{p}\in\mathcal{R}^n$ as $\mathbf{p}=[\rho,\mathbf{e}]$, where $\rho$ is the density of the material and $\bf e$ are the other material parameters. Our gamut, i.e. the set $\mathcal{M}\subset\mathcal{R}^n$ of material properties corresponding to microstructures of a given resolution, is made of a finite number of points. However, by increasing the resolution of the microstructures this gamut gets denser and denser so that we assume that it can be approximated by the union of continuous n-dimensional manifolds and can be represented using a distance field.

\subsection{Discretization}

Following standard finite element methodology, we discretize the object in regular voxels and compute its deformed state when subject to external forces $\bf f_{ext}$ using the well-known relation
\begin{equation}
\bK\mathbf{u}=\bf f_{ext},
\label{eq:fem}
\end{equation}
where $\bf K$ is the stiffness matrix of the system, and $\bf u$ are the displacements at the nodes of the voxels.

Note that we use the same approach to simulate both the mechanical behaviour of the microstructures and the object macroscopic behaviour. However, we work at two different scales. To simulate the microstructures, we assume that each of its voxels is made of an homogeneous base material, whereas for determining the large scale behaviour of the object, we assume that each of its cells corresponds to a microstructure. The properties of the individual microstructures are determined from 6 harmonic displacements (or 2 displacements in 2D) using numerical coarsening as described by Kharevych et al. \shortcite{Kharevych:2009}.

We solve the static equilibrium Equation \ref{eq:fem} using a fast multigrid solver based on the implementation by Dick et al. \shortcite{dick2011real}.

%% file: MicrostructureDatabase.tex
\section{Material Space Exploration}
\label{sec:sampling}

\begin{figure*}[t]
    \centering
  \includegraphics[width=.95\linewidth]{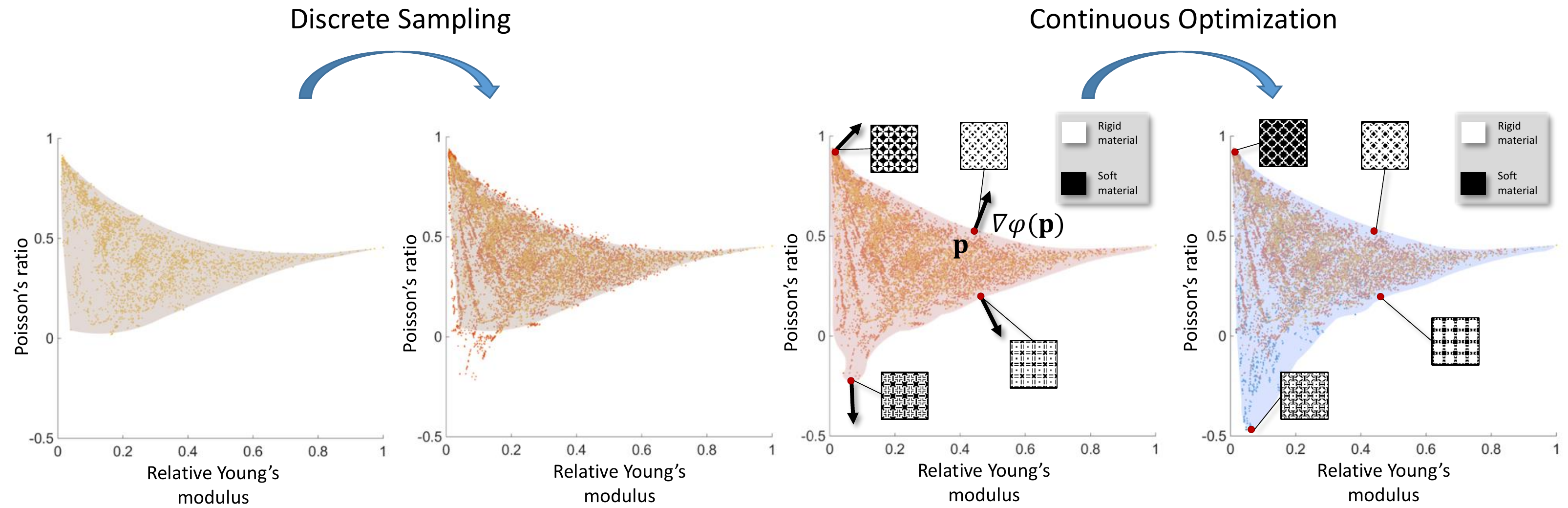}
  \caption{One cycle of computing the microstructure gamut. 
	Given a set of samples, we compute a signed distance function approximating the material gamut (\emph{left}) and randomly perturb microstructures lying near the boundary to provide new seeds to the continuous algorithm (\emph{middle left}). We then update the distance field and use the gradient of the signed distance function at the the boundary to define new target material points (\emph{middle right}). These target material points are used in a continuous optimization that generates new samples (\emph{right}).} 
    \label{fig:lssampling}
\end{figure*}

The first step in our pipeline is to determine the space of physical properties that can be achieved when combining the base materials into microstructures of a predefined size.

Computing the mechanical properties of microstructures, when arranged in periodic tilings, can be performed by probing the structure using a physical simulation. This approach, based on the homogenization theory, is a common practice and has been widely used in the past \cite{Allaire2012,Schumacher:2015,Panetta:2015}. However, while inferring the homogenized properties of individual microstructures is not particularly challenging, analyzing the space covered by {\it all} combinations of base materials is much more difficult due to the combinatorial explosion in the number of possible material arrangements. As an example, $16\times16\times16$ lattices made of only two materials  corresponds to $2^{4096}$ microstructures: exhaustively probing all microstructures is clearly an impossible task.
To address this issue, two possible avenues can be pursued:(i) we can try to sample the space of the microstructures, (ii) we can rely on the continuity between material parameters of the individual voxels and macroscopic properties of the microstructures in order to generate new microstructures with desired properties. This second option is effective in reaching locally optimal values in the material property space. However, the function that maps the material assignment to material properties is nonlinear. In particular, very different microstructures can correspond to the same point in the material property space. Additionally, since the ratio of materials in each cell is bounded between zero and one, the continuous optimization converges slowly or stops moving when material distributions in many cells are at the lower or upper bound. Being able to jump out of a local optimum and discovering different variants is important in order to provide new exploration regions. We leverage these two approaches by combining them in a scheme that alternates between a stochastic search and a continuous optimization. We provide the technical details in the rest of this section.

\subsection{Discrete Sampling of Microstructures}
We aim at sampling the space of material assignments, i.e. microstructures, in such a way that we maximize the number of samples corresponding to microstructures whose material properties lie in the vicinity of the material gamut boundaries. We do not draw all samples at once but progressively enrich the database of microstructures as we refine our estimation of the material gamut boundaries. This sampling strategy is motivated by the observation that a small change in the material assignment of a microstructure generally -- but not always -- translates to a small change of its material properties. By modifying microstructures located near the current boundaries of the material property gamut, we are likely to generate more structures in this area, some of which will lie outside of the current gamut.

Given a population of microstructures to evolve, we generate new samples from each microstructure by changing its material at random voxel locations. To rationalize computational resources, we want to avoid revisiting the same voxel twice. But we do not want to privilege any particular order either. Ritchie et al. \shortcite{Ritchie2015SOCM} recently presented a Stochastically-Ordered Sequential Monte Carlo (SOSMC) method that provides a suitable approach.
In SOSMC, a population of particles (here, our microstructures) corresponding to instances of a procedural program (here, the sequential assignment of materials to the voxels of the microstructures) are evolved so as to represent a desired distribution. During this process, the programs are executed in a random order and particles are regularly scored and reallocated in regions of high probability. In our particular settings, we use the scoring function
\begin{equation}
s(\mathbf{p_i}) = \frac{\Phi(\bf p_i)}{D(\bf p_i)} \times \frac{1}{D(\bf p_i)},
\label{eq:score_fct}
\end{equation}
where $\Phi(\bf p_i)$ is the signed distance of the material properties of particle $i$ to the gamut boundary (see Section \ref{sec:ls}) and $D(\bf p_i)$ is the local sampling density at the location $\bf p_i$. We define the sample density as
\begin{equation}
D(\bf p_i)=\sum_{k}\phi_{k}(\bf p_i)\ ,
\end{equation}
where $\phi_{k}(\bf p)=\left(1-\frac{||\bf p-\bf p_{k}||^{2}_{2}}{h^{2}}\right)^{4}$ are locally-supported kernel functions that vanish beyond their support radius $h$, set to a tenth of the size of the lattice used for the continuous representation of the material gamut (see Section \ref{sec:ls}).

The first term in Equation \ref{eq:score_fct} favors microstructures located near the gamut boundary. The normalization by $D$ allows us to be less sensitive to the local microstructures density and to hit any location corresponding to the same level-set value with a more uniform probability. The second product is used to additionally privilege under-sampled areas.

Particles are resampled using systematic resampling scheme \cite{Douc05resamplingSchmes} that is also used to initiate the population of particles. These particles are then evolved according to Algorithm~\ref{algo:proc_microstructure}. \updated{Additional details regarding the implementation of our algorithm are provided in the supplementary material.}

\algrenewcommand\algorithmicindent{1.0em}%

\begin{algorithm}[tbp]
\begin{algorithmic}
\Procedure{genMicrostructure}{input: microstructure $M_i$, output: microstructure $M_o$}
\State $M_o\gets M_i$
\While {some voxels of $M_o$ have not been visited}
\While {microstructure $M_o$ is unchanged}
    \State pick a random voxel $v$ of $M_o$ that has not been visited 
		\State assign a randomly chosen material to $v$
		\If {$M_o$ is manifold and $M_o\neq M_i$} \State{accept the change} \EndIf
\EndWhile
\EndWhile
\EndProcedure
\end{algorithmic}
\caption{Procedure for generating new microstructures}
\label{algo:proc_microstructure}
\end{algorithm}

\subsection{Continuous Optimization of Microstructures}
The goal of the continuous optimization is to refine the geometry of the microstructures located at the boundary of the gamut in order to further expand the gamut along the normal directions.
We start continuous sampling by selecting a subset of microstructures lying on the boundary of the gamut as starting points for the continuous optimization. The discrete structures are mapped to continuous values close to $0.5$. We used $0.5\pm0.3$ in our experiments. Doing so allows the topology optimization algorithms to move freely in the first steps and discover new structures.

For each starting structure, we identify target material parameters using the gradient of the level set $\Phi$ at the initial discrete sample point $\bf p$ (see Section \ref{sec:ls}) defined by ${\bf q} = {\bf p} + \nabla \Phi ({\bf p})$. We translate this target material parameters into an elasticity tensor $\bC_0$ and density $\rho_0$. Here $\rho$ is the ratio of the two base materials in the microstructure.

Note that our problem formulation does not restrict us to a particular topology optimization algorithm or material distribution parametrization. We have experimented with two objective functions that worked equally well for our purposes.
The first objective uses an energy based formulation~\cite{xia:2015:design} to compute and optimize the elasticity tensor directly. At a high level, the optimization problem is
\begin{equation}
\arg\min_{\bx} f(\bx) = (\bC(\bx)-\bC_0)^2 + w_{\rho}(\rho-\rho_0)^2, \quad \rho = \sum_{i} x_{i} ,
\end{equation}
where $\bx$ is the ratio of materials in each cell, and $w_{\rho}$ controls the weighting between the displacement term and the density term.
The authors of the method developed parameter heuristics to optimize for difficult cases such as negative Poisson's ratio structures. We naturally arrive at structures with negative Poisson's ratio without the parameter varying step in~\cite{xia:2015:design} since our discrete samples allow us to explore a wide variety of initial designs.

The second objective is formulated using harmonic displacements~\cite{Kharevych:2009,Schumacher:2015} $\bG$ instead of the elasticity tensor directly. $\bG$ is a $6\times 6$ symmetric matrix where each row corresponds to a strain in vector form. We use the target elasticity tensor $\bC_0$ to compute the target harmonic displacements matrix $\bG_0$ and minimize the objective function:
\begin{equation}
f(\bx) = (\bG(\bx)-\bG_0)^2 + w_{\rho}(\rho-\rho_0)^2.
\end{equation}
This objective matches soft structures more accurately since entries of $G$ are inversely proportional to material stiffness.

Following the work by Andreassen et al. \shortcite{Andreassen2014}, we use the method of moving asymptotes (MMA) \cite{svanberg1987method}) to optimize the objectives using an implementation provided in the NLOPT package \cite{johnson2014nlopt}. We run at most $50$ iterations since it usually converges to a solution within 20-30 steps.
MMA makes large jumps during the optimization while keeping track of the current best solution, thus causing the oscillation of the objective value.
To force continuous material ratios towards discrete values, we experimented with the SIMP model with the exponent set to $3$ and the Hashin-Shtrikman bound for isotropic materials described by Bends{\o}e and Sigmund \shortcite{bendsoe1999material}.

Either interpolation allows us to threshold the final continuous distribution and obtain a similar discrete sample. We tolerate small deviations introduced by the thresholding since our goal is to obtain a microstructure lying outside of the gamut rather than reaching a particular target. In practice, we observed that the material properties of the final discrete structures often did not change significantly after the thresholding step. 

\subsection{Continuous Representation of the Material Gamut }\label{sec:ls}
We represent the gamut of material properties using a signed distance field that is computed from the material points associated to the sampled microstructures.
First, we normalize each coordinate $p_i$ of $\bf p$ to constrain the scope of the level set to an $n$-dimensional unit cube. Then we compute the level set values on the cell centers of an $n$-dimensional Cartesian grid that encloses this unit cube. We draw inspiration from the methods for surface reconstruction used in particle fluid rendering \cite{zhu2005animating,bhatacharya2011level,ando2013highly} and extend it to $n$ dimensions. In this case, a signed distance field is generated from a set of points by evaluating an implicit distance function $\Phi$ at each point $\mathbf{p}\in\mathcal{M}$.
We initialize the signed distance field using the implicit function $\Phi(\mathbf{p})=||\mathbf{p}-\bar{\mathbf{p}}||-r$ from \cite{zhu2005animating} where $||\cdot||$ is the Euclidean distance between two points in $\mathcal{M}$, and $\bar{\mathbf{p}}$ is the average position of the neighboring points of $\mathbf{p}$ within a range of $2r$. Note that the signed distance is initially defined only near the boundary of the gamut. In order to sample the distance on the entire domain, we propagate the 0-level set surface using the fast marching algorithm and solve an explicit mean curvature flow problem defined as $\partial \Phi / \partial t = \Delta \Phi$ \cite{osher2006level} .

Having a continuous representation of the gamut of materials achievable by the microstructures, we can now reformulate the topology optimization problem directly in the material space.

%% file: TopoOpt.tex
\section{Topology Optimization}
\label{sec:topoOpt}
A classic topology optimization problem consists of optimizing the shape and structure of a given object defined by a prescribed domain in order to minimize some cost function. For example, the standard topology optimization minimizes the compliance of the object while satisfying the static equilibrium and the total weight constraint.
Since the topology of the object is unknown {\it a priori}, a method of choice is to define the shape of the object through its material distribution and to locally work with material densities. To this end, the design layout is voxelized and a density variable is assigned to every cell of the discretized domain. By penalizing intermediate values for these densities, a binary distribution corresponding to the object's final layout can be eventually obtained.

In this work, we extend the traditional topology optimization algorithm in multiple ways. First, we do not compute a binary material distribution at the cell level as commonly done. Instead, we leverage our database of microstructures and ask for each cell to be filled with one of the microstructures. By doing so we change the topology of the object at a finer scale, i.e. within each cell. This is done by working with the macro-scale material properties of the microstructures instead of their geometry directly. The second difference is that our algorithm can be used with parametrizations of the material property space that are more complex than the single density parameter per cell that is commonly used in the standard topology optimization algorithm. Indeed, in our generalized topology optimization problem, each cell $c_i$ contains an n-dimensional material parameter $\bp_i\in \mathcal{R}^n$. We use $\bp$ to denote the stacked vector of material parameters in all cells.
Given a signed distance function $\Phi(\bp_i)$ that defines the gamut,
our new topology optimization problem is then written as
\begin{equation}
\begin{aligned}
\min_{\bp} : \quad & \mathcal{S}(\bp,\mathbf{u}) & \\
s.t. : \quad & \mathcal{F}(\mathbf{\bp},\mathbf{u})=0 & \\
: \quad & \Phi(\bp_i)\leq 0,  \quad 1\leq i \leq N_c&	
\label{eq:op}
\end{aligned}
\end{equation}
where $\mathcal{S}$ is a real-valued objective function that depends on the material parameters and the displacement vector $\bu$ of the entire object at the elasticity equilibrium.
The equality constraint $\mathcal{F}=0$ requires $\bu$ to satisfy the elasticity equilibrium and the inequality constraint $\Phi\leq0$ guarantees that the material properties of each cell stay inside the precomputed gamut.

In our examples, the material parameter $\bp$ consists of the density $\rho$ and the elasticity parameters $\be$.
We split our objective function into an elasticity term $\mathcal{C}(\vect{e},\bf u)$ that controls the deformation behavior (see Section \ref{sec:obj}) and an optional density term $\mathcal{V}(\rho)$ that controls the overall mass of the object.The density term can be written as
\begin{equation}
\mathcal{V}(\rho)=(\sum_{i=1}^{N_c}\rho_iV_i-\hat{M})^2\label{eq:vol},
\end{equation}
where $V_i$ is the cell volume and $\hat{M}$ is the target overall mass.
When one of the base material is void, the use of the density term allows to modify the topology of the object at a larger scale than the one of the microstructures, and thus to change the external shape of the object. 
In fact, even for multi-material designs involving base materials with similar mass densities, we noted that we could use the density term to encourage the presence of soft material in the structure. By removing the external cells entirely made of the soft material, we could then decrease the mass of the structure without significantly changing its mechanical behaviour.
Alternatively, the density term can also be used to control other quantities related to the ratios of the different materials such as the cost of the object.

For specific problems, we can also add spatially-varying weight control terms to Equation \ref{eq:vol}.
For example, we can control the target weight of each individual cell by adding a local term $(\rho_i-\hat{\rho}_i)^2V_i$.

Assuming static equilibrium, the elasticity constraint is written as 
\begin{equation}
\mathcal{F}(\mathbf{e},\mathbf{u})=K(\mathbf{e})\mathbf{u}-\bf f_{ext}=0,
\end{equation}
where $\bf f_{ext}$ are the external loads applied to the object.

The gamut constraint for a point $\mathbf{p}_i$ in the material property space is described by an $n$-dimensional level set function $\Phi(\mathbf{p})$.
We have $\Phi(\mathbf{p}_i)<0$ for a point inside the gamut, $\Phi(\mathbf{p}_i)>0$ for a point outside the gamut, and $\Phi(\mathbf{p}_i)=0$ for a point on the boundary of the gamut.
The value of $\Phi$ represents the $n$-dimension Euclidean distance to the level set boundary. 
The gradient of $\Phi$ are evaluated by a finite difference operation on the signed distance field.

We used a standard gradient-based numerical optimizer (Ipopt \cite{Ipopt:2006} in our implementation) to solve Equation \ref{eq:op}. We enforced the elasticity equilibrium constraint using the adjoint method. The optimizer only needs to take the function values of $\mathcal{S}$ and $\Phi$ along with their gradients as input.

\begin{figure*}[t!]
    \centering
		\includegraphics[width=.245\linewidth]{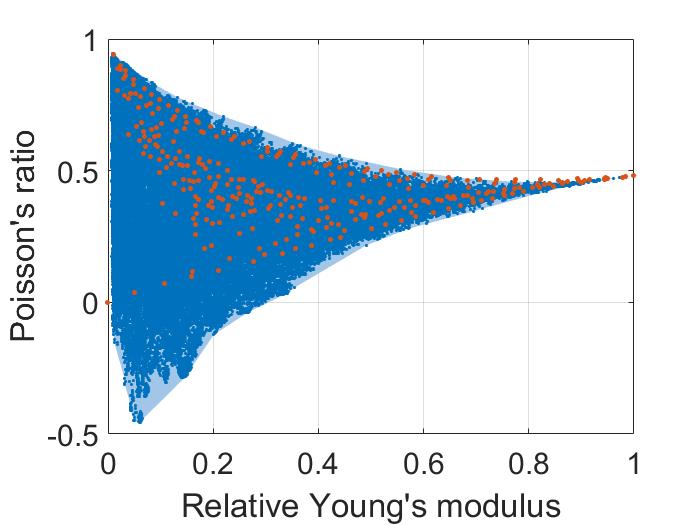}
		\includegraphics[width=.245\linewidth]{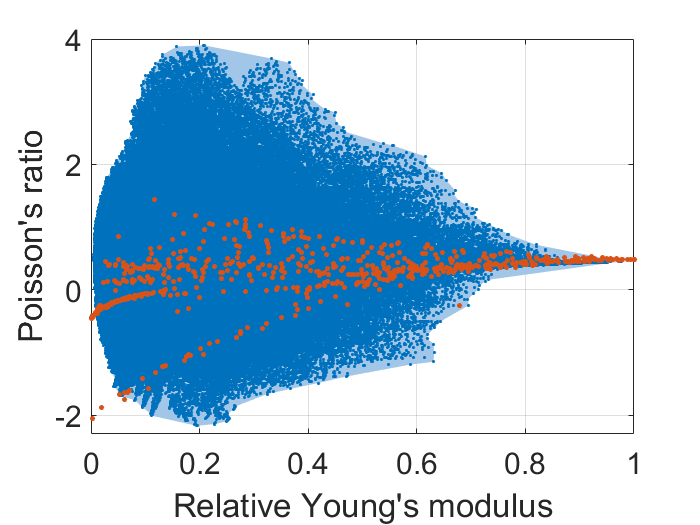} 
		\includegraphics[width=.245\linewidth]{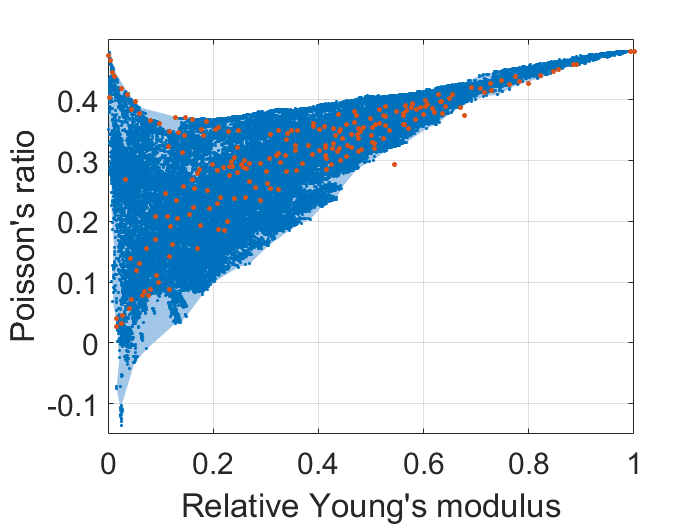}
		\includegraphics[width=.245\linewidth]{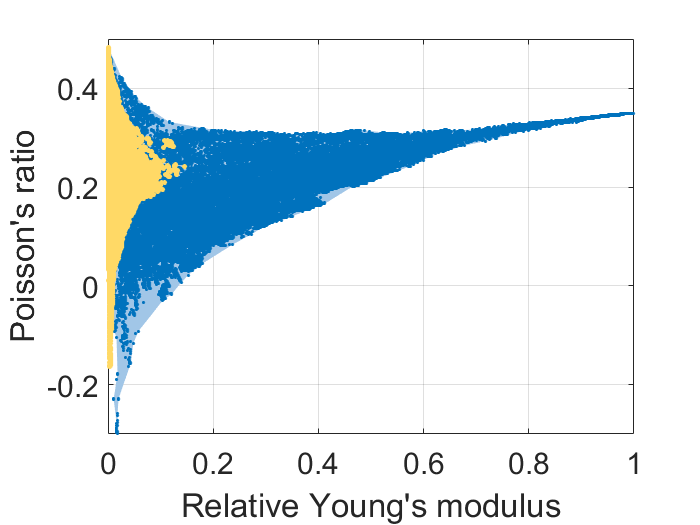}
	\caption{Gamuts computed with our discrete-continuous sampling scheme for 2D cubic structures (\emph{left}) , 2D orthotropic structures (\emph{second from left}), 3D cubic structures (\emph{second from right}) and 3D cubic structures with 0.35 as Poisson's ratio (\emph{right}). The plots show the results for the projection of the gamuts on the plane defined by the macroscale Young's modulus along the x axis (normalized by the Young's modulus of the stiffest base material) and the Poisson's ratio corresponding to a contraction along the y-direction when the material is stretched along the x-direction. The blue dots correspond to the generated samples,the orange dots correspond to the microstructures from Schumacher et al. \shortcite{Schumacher:2015} and the yellow dots correspond to the microstructures from Panetta et al. \shortcite{Panetta:2015}.}
    \label{fig:gamuts}
\end{figure*}

\subsection{Elasticity Objectives}\label{sec:obj}
We used two different types of objective functions for the elasticity term in our topology optimization algorithm. These two types of objectives allowed us to design a wide range of objects.
\paragraph*{Target Deformation}\label{sec:strain}
Our algorithm takes a vector of nodal target displacements and boundary conditions (external forces, fixed points, etc.) as input. Then, it automatically optimizes the material distribution over the object domain to achieve the desired linear deformation assuming a linear elastic behavior.

We define the deformation objective as
\begin{equation}
\mathcal{C}_d(\mathbf{e},\mathbf{u})=(\mathbf{u}-\hat{\mathbf{u}})^T \bD (\mathbf{u}-\hat{\mathbf{u}}),\label{eq:td}
\end{equation}
where $\hat{\mathbf{u}}$ is the vector of the target displacements, $\bf D$ is a diagonal matrix that determines the importance of each nodal displacement.	
We use $\bf D$ to define the subset of nodes that we are interested in.
For example, we can set most entries of $\bf D$ to zero and focus on a portion of the domain (see Figure \ref{fig:gripper}).
\paragraph*{Minimum Compliance}\label{sec:compliance}
We have experimented with the same objective as the one used in the standard topology optimization algorithm where the compliance $\mathcal{C}_c$ is defined as
\begin{equation}
\mathcal{C}_c(\mathbf{e},\mathbf{u})=\mathbf{u}^T \bK(\mathbf{e}) \mathbf{u}.
\label{eq:mc}
\end{equation}
In the commonly used SIMP algorithm, the stiffness matrix $\bK_i$ of each cell $i$ depends on the artificial density value $\rho_i$
through an analytical formula such as $\bK_i = \rho_i^3\bK_0$ where $\bK_0$ corresponds to the stiffness matrix of the base material.
In contrast, the stiffness matrix in our objective function is directly computed from the material parameters of the material space and forced to correspond to a realizable material thanks to our gamut constraints.

Like in standard algorithms, we regularized the problem to avoid checkerboard solutions by applying a smoothing kernel on the material properties that favors smooth variations of the material parameters over the object layout.
Our optimizer supports multiple objectives by linearly combining weighted objective functions.
\section{Mapping Material Properties to Microstructures}
\begin{wrapfigure}{r}{0.28\linewidth}
	\vspace{-0.4cm}
 	\makebox[\textwidth][l]{\hspace{-0.4cm} \includegraphics[width=1.0\linewidth]{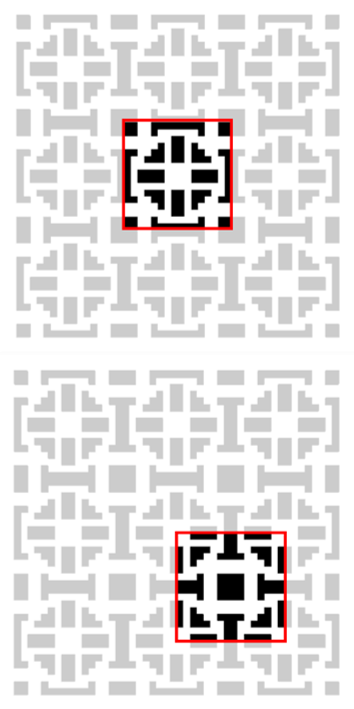}}
 	\vspace{-0.4cm}
\end{wrapfigure}
After running the topology optimization algorithm, we generate a printable result by replacing each cell in the object lattice by a microstructure whose material properties match the optimal ones.

Material properties of the microstructures are computed using the homogenization theory which is more accurate with a smooth transition between the geometries of neighboring cells.
While smoothness in the material parameters can be easily enforced, it does not imply topological similarity of nearby microstructures. 
For example, any translation of a given microstructure in a periodic tiling will result in a microstructure geometrically different but with exactly the same mechanical properties. 
Fortunately, our database is very dense and multiple microstructures generally map to similar points in the material property space, offering several variants. To further increase the number of possibilities, we also incorporate an additional exemplar for each microstructure by translating it by half its size, which preserves its cubic or orthotropic symmetry without changing its properties (see inset Figure). We then run a simple but effective algorithm that picks the microstructure exemplars that minimize the boundary material mismatch across adjacent cells. We quantify this mismatch by $\mathcal{I}=\sum_{i=1}^{N_c} \mathcal{I}_i$, where $\mathcal{I}_i$ is the contribution associated to the cell $i$ and corresponds to the number of boundary voxels filled with materials that are different from the ones of the voxels' immediate neighbours across the interfaces.

Our algorithm proceeds as follows:
\begin{itemize}
\item For each cell, we define a list of possible candidates by picking all the microstructures mapping to material points lying in the vicinity of the optimal material point and we randomly initialize the cell with one of the candidates.
\item We compute the mismatch energy $\mathcal{I}_i$ associated to each cell $i$ and sort the cells according to their energy. 
\item We pick the first cell in the sorted list, i.e. the one with the highest energy and assign to it the microstructure candidate that decreases the energy the most. If we cannot decrease the cell energy, we move to the next cell in the list.
\item We update the mismatch energies of all the impacted cells and we update the priority list.
\item We repeat the last two steps until the mismatch energy $\mathcal{I}$ cannot be decreased anymore.
\end{itemize}

\begin{figure}[t]
    \centering
		\includegraphics[width=.9\linewidth]{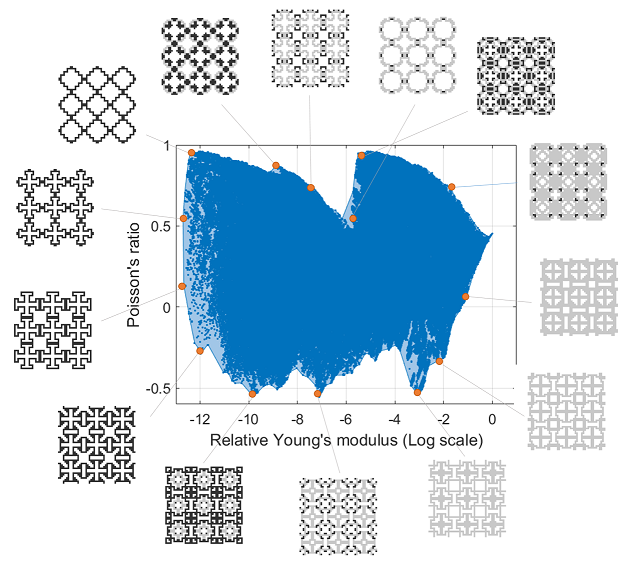}

    \caption{Gamut corresponding to 2D cubic microstructures made of two materials and void. The Young's modulus of the microstructures is plotted using a logarithmic scale. We show above some examples of microstructures lying near the estimated boundary of the gamut, i.e. with extreme material properties. The dark color corresponds to the softer material, while the light grey color is used for the stiffer material.}
    \label{fig:Cubic2D_3Materials}

\end{figure}

%% file: Results.tex
\section{Results}
We first analyzed our microstructure sampling algorithm for 2D and 3D microstructure gamuts. Then we used these precomputed gamuts and we designed and optimized a wide variety of objects with our topology optimization algorithm.

\begin{figure}[t]
    \centering
		\twofigure{.235\textwidth}{.03in}{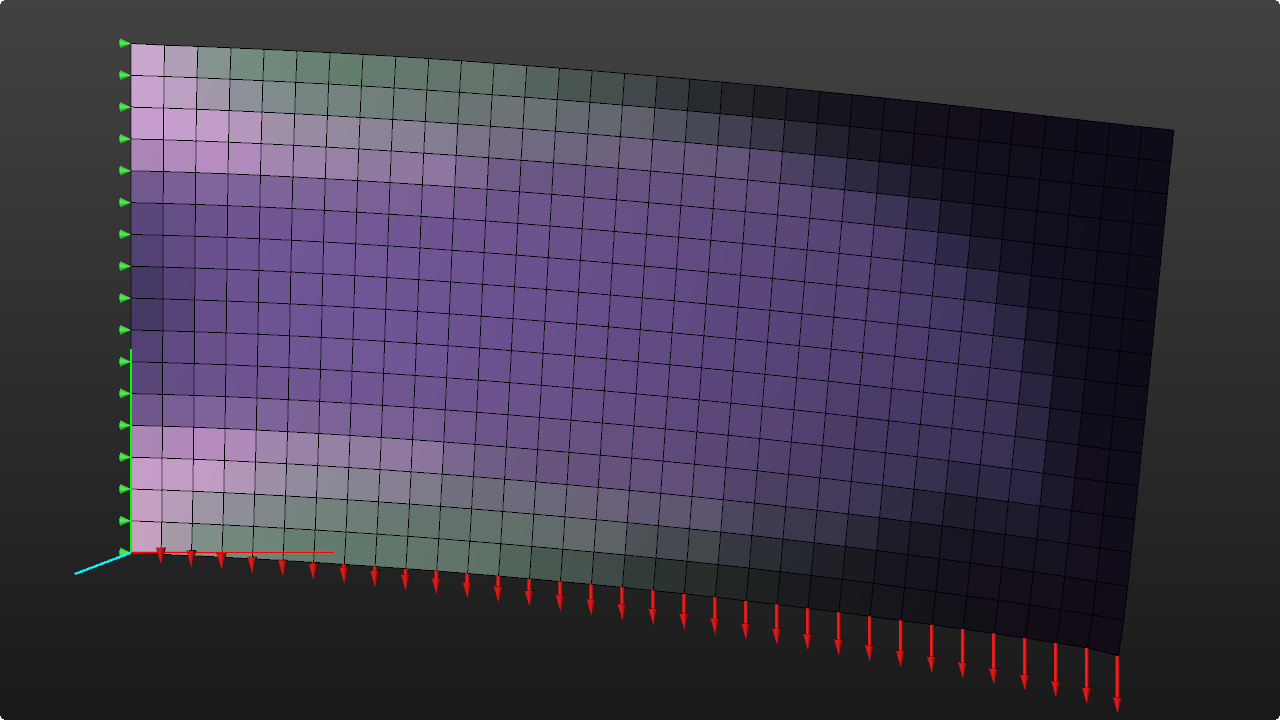}{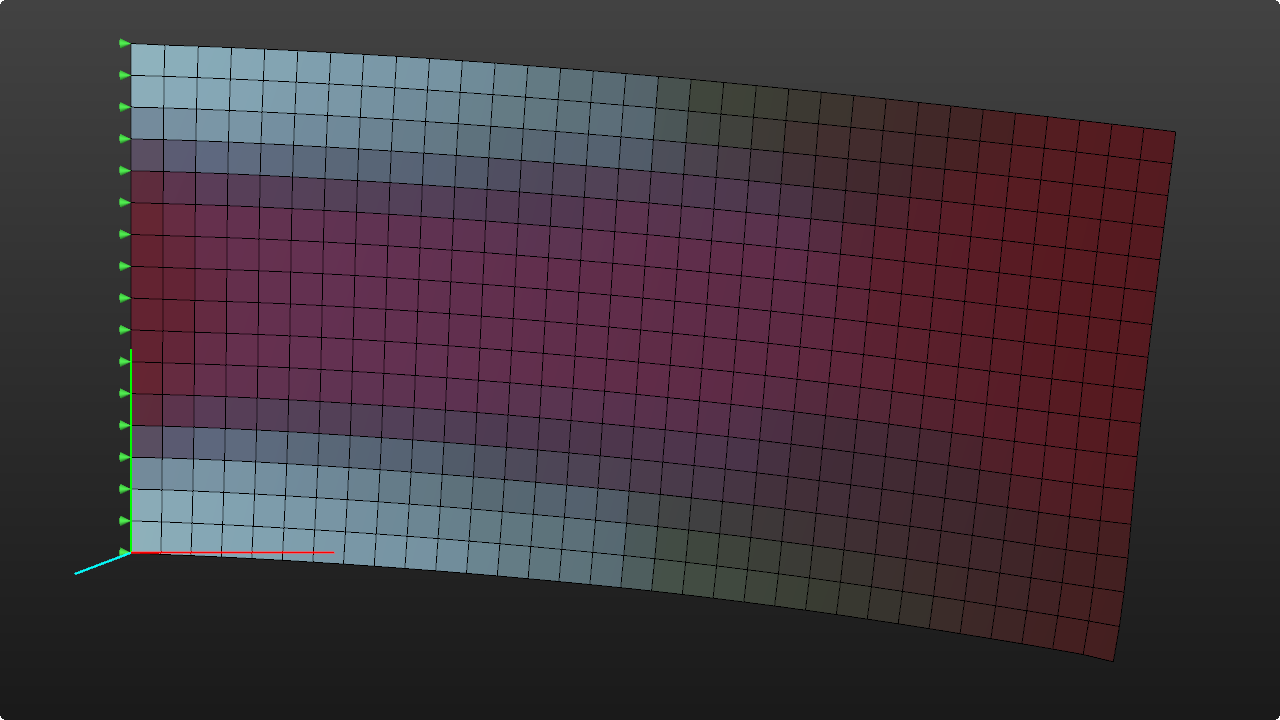}
		\twofigure{.235\textwidth}{.03in}{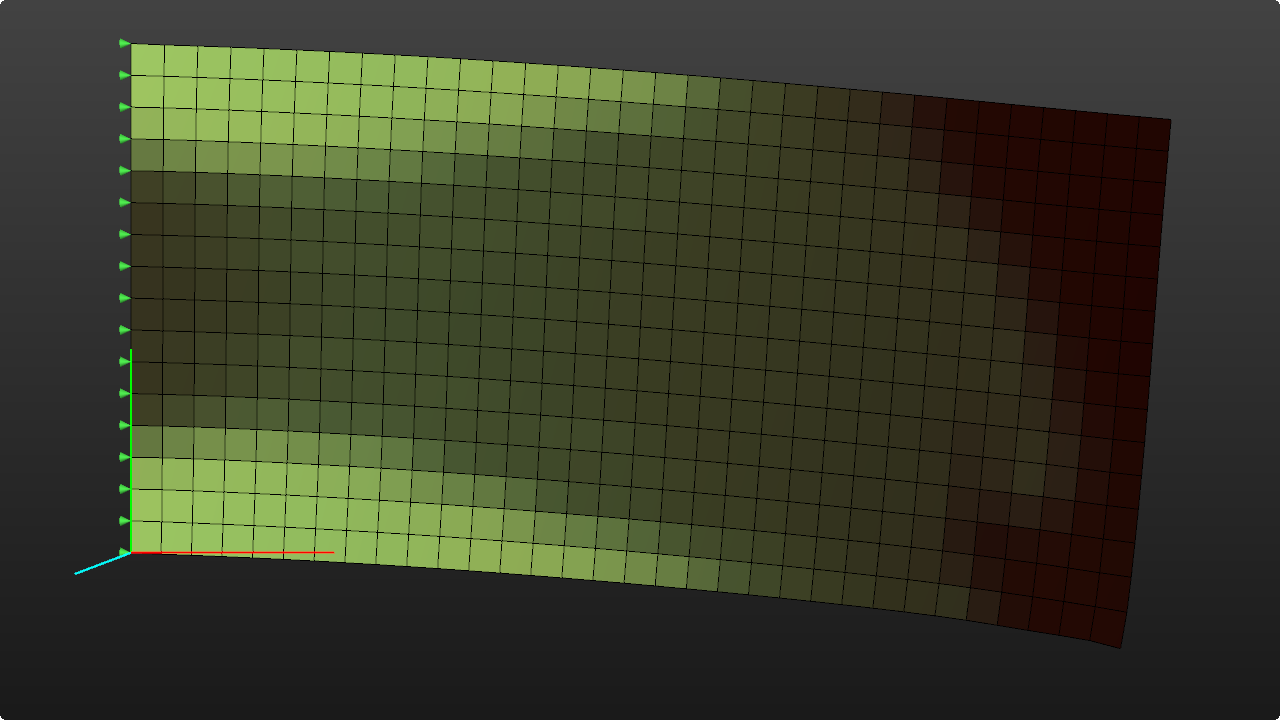}{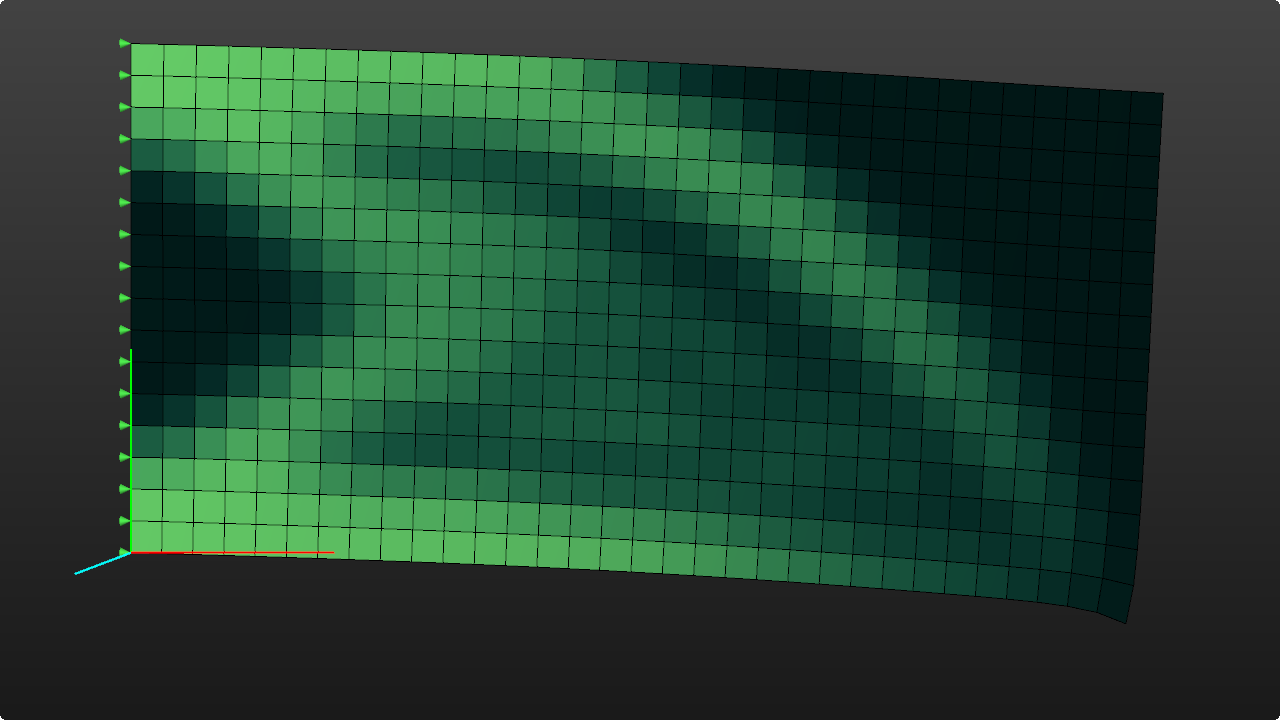}
		\twofigure{.235\textwidth}{.03in}{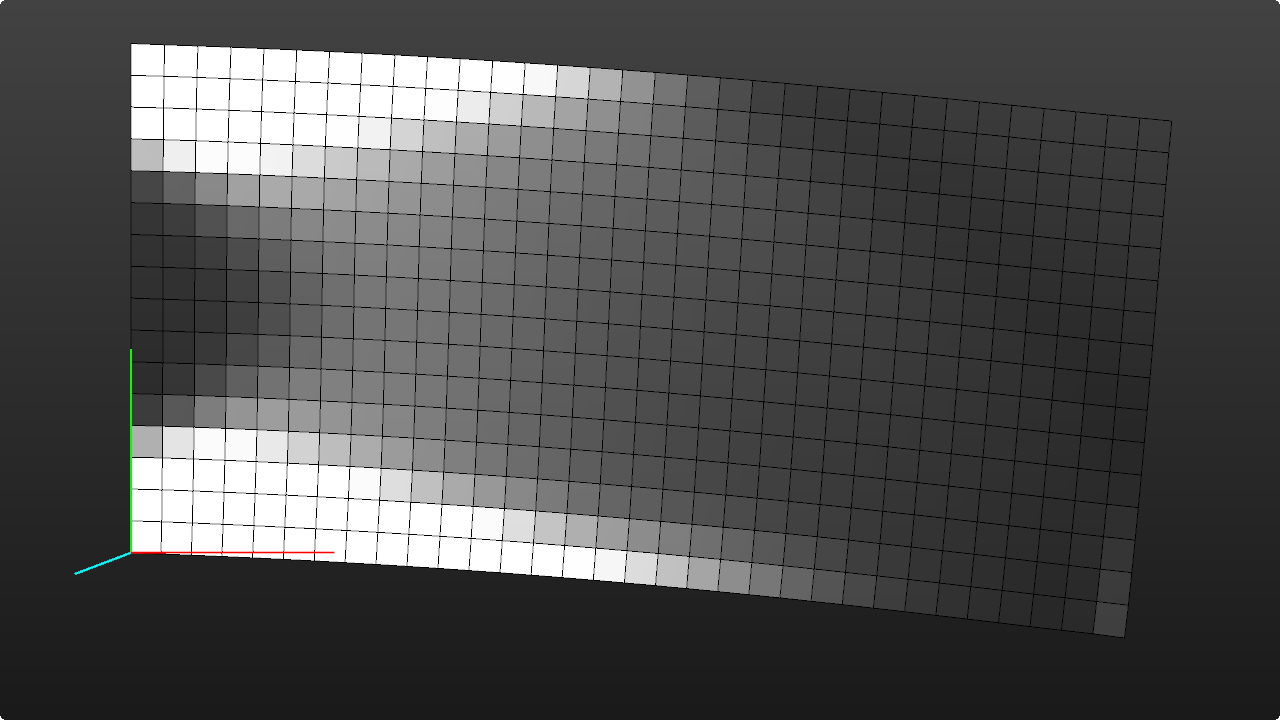}{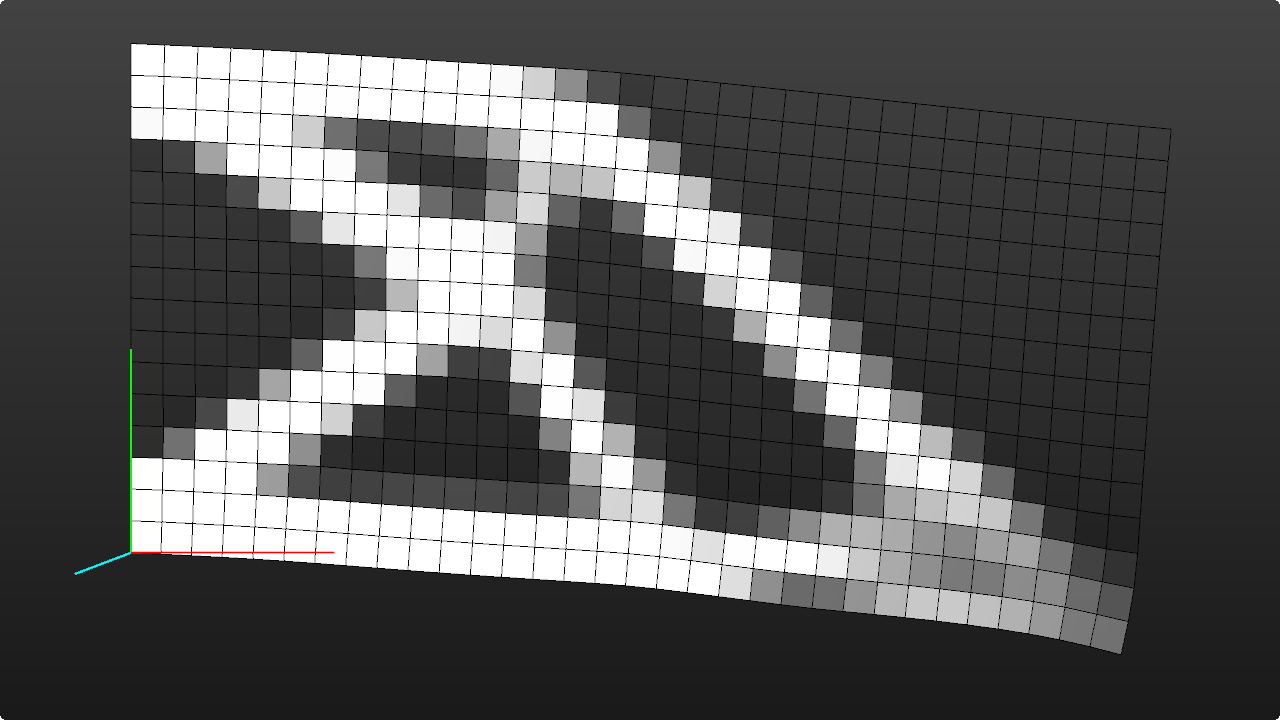}
    \caption{Resulting material property distributions when running our topology optimization algorithm in the orthotropic (\emph{top left}), cubic (\emph{top right}), isotropic (\emph{middle left}) and an analytically defined gamut $E \geq \rho^3 E_0$ (\emph{middle right}), with the material property space dimensions ranging from five to two. 
We compare our algorithm with the standard SIMP method with power index $p=1$ (\emph{bottom left}) and $p=3$ (\emph{bottom right}). For these figures, we computed the color of each cell by mapping every base vector of the normalized parameter space to a color range and linearly interpolating the colors associated to each of the parameters.
In this example, the left side of the cantilever is fixed while a discretized, linearly varying, distributed force is applied to the bottom side (see red arrows in the top left picture).}
    \label{fig:defo_cvg}
\end{figure}

\begin{figure}[h]
    \centering
		\twofigure{.235\textwidth}{.03in}{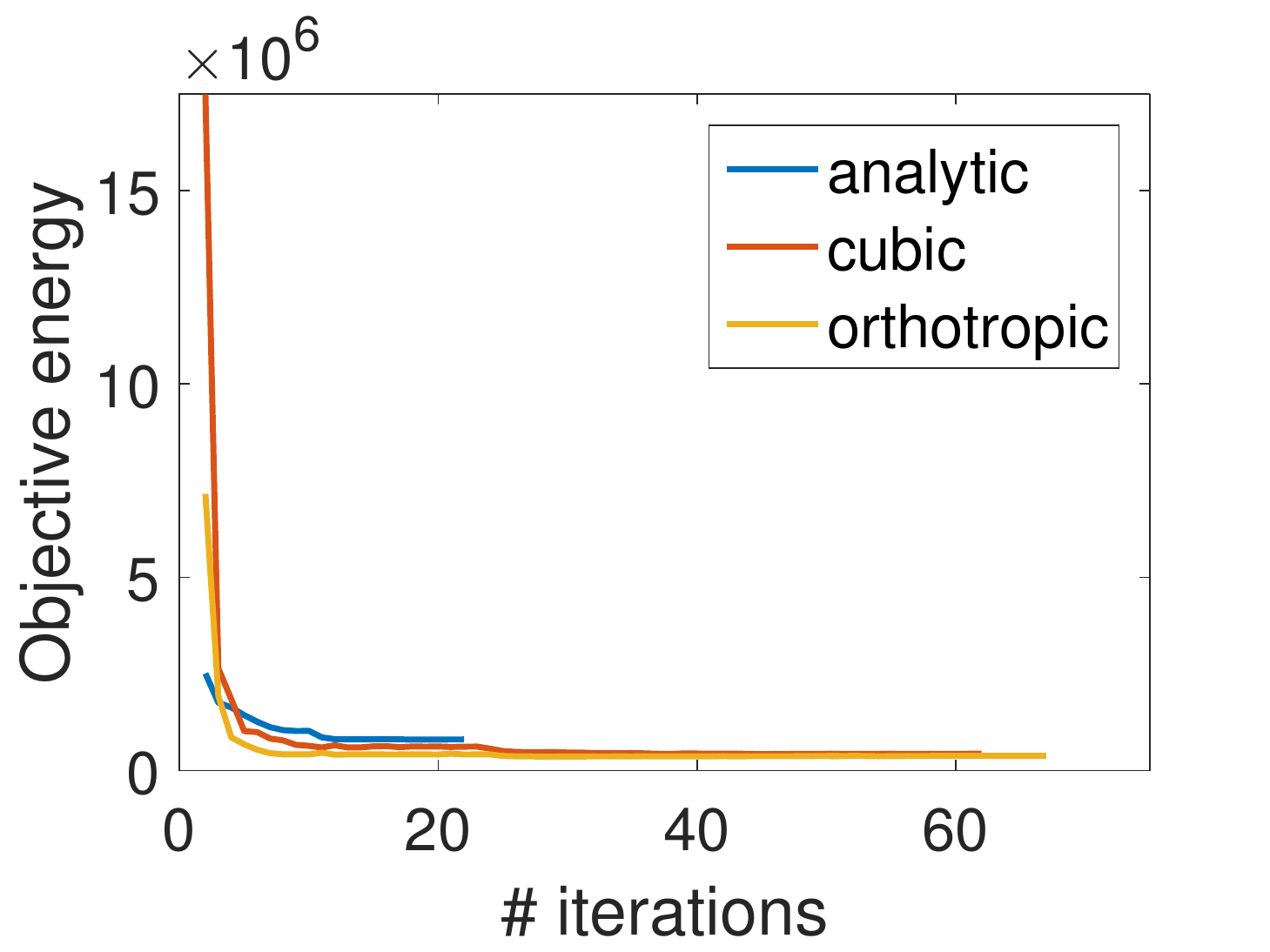}{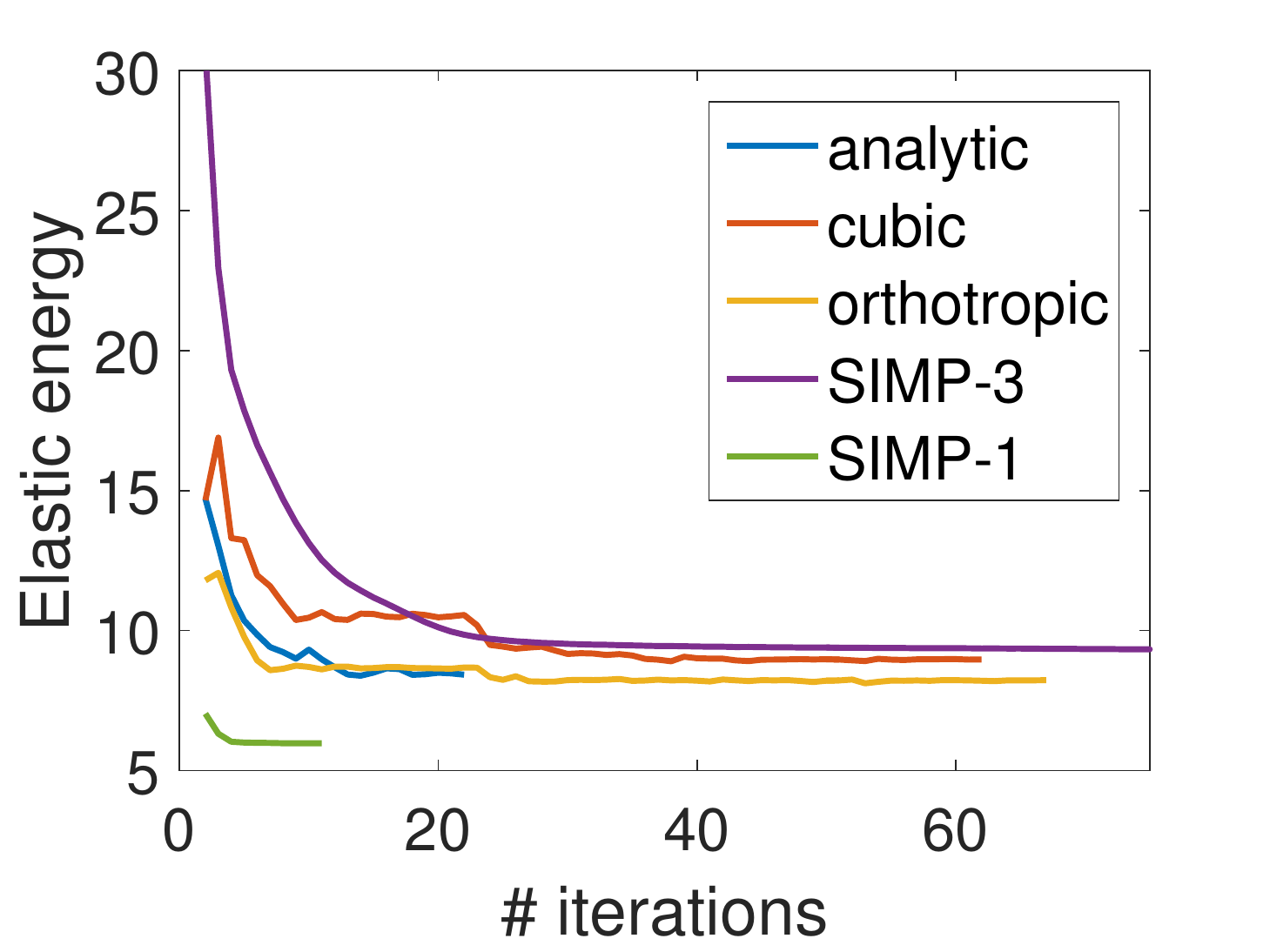}
		\twofigure{.235\textwidth}{.03in}{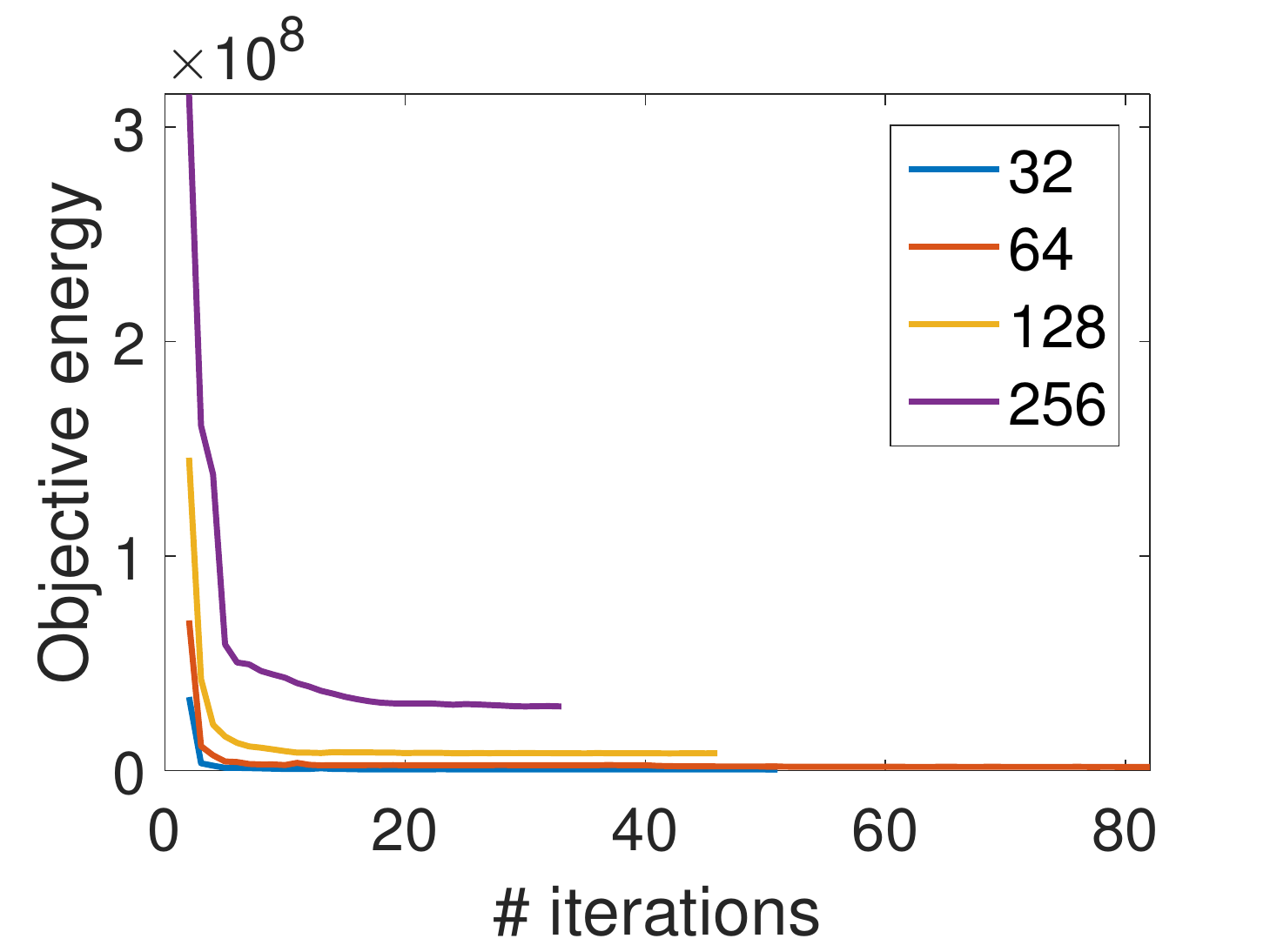}{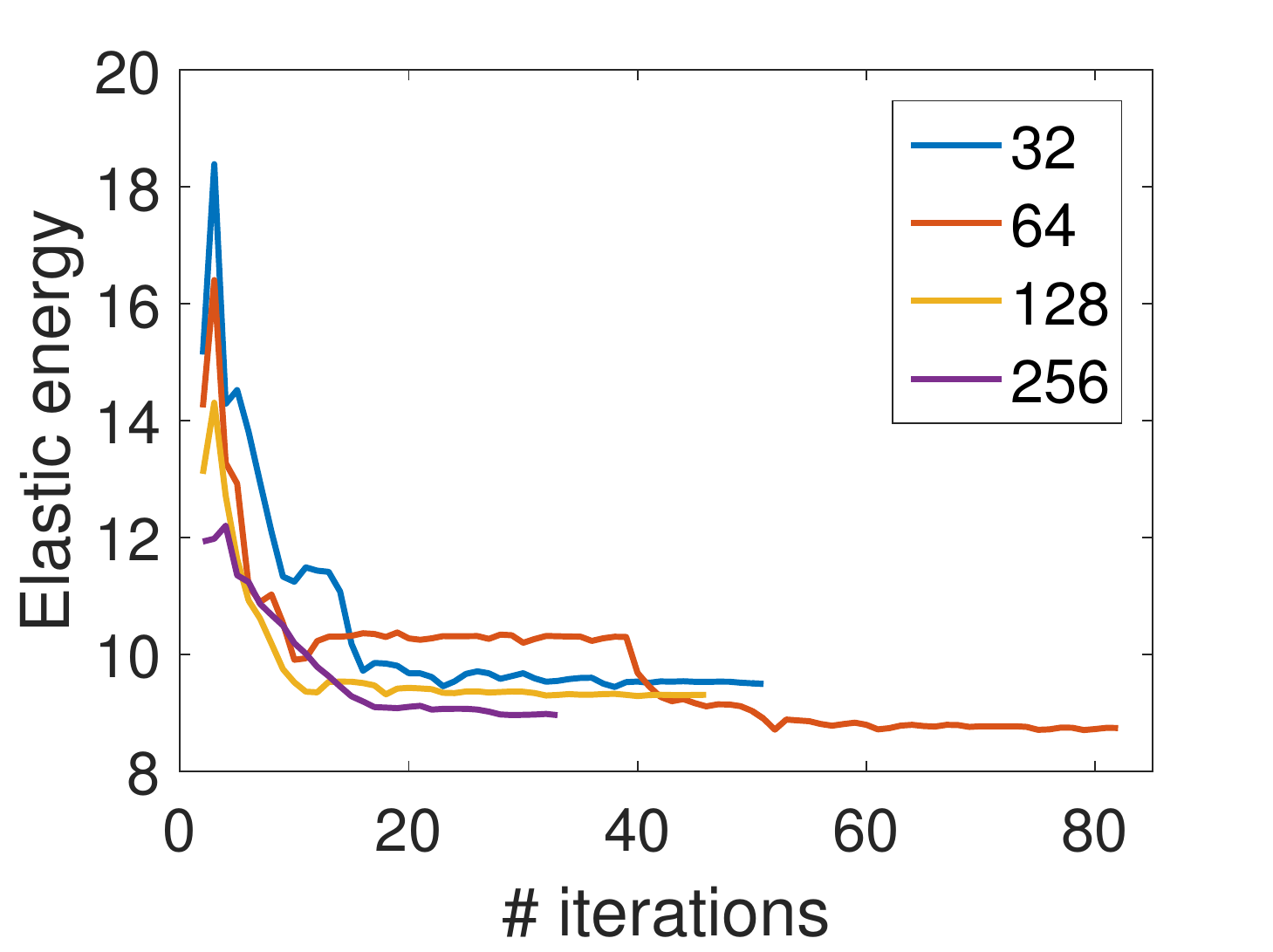}
		\twofigure{.235\textwidth}{.03in}{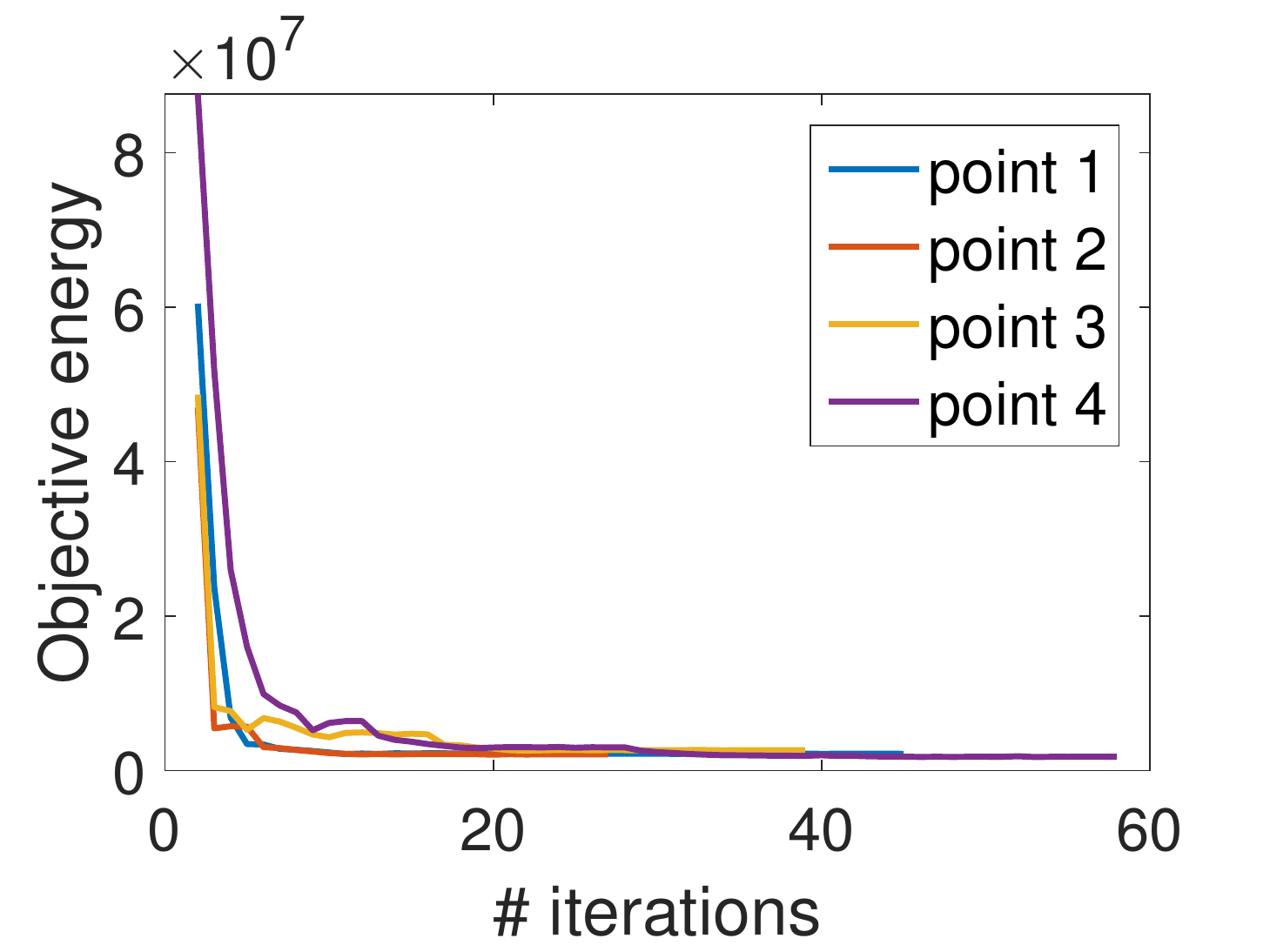}{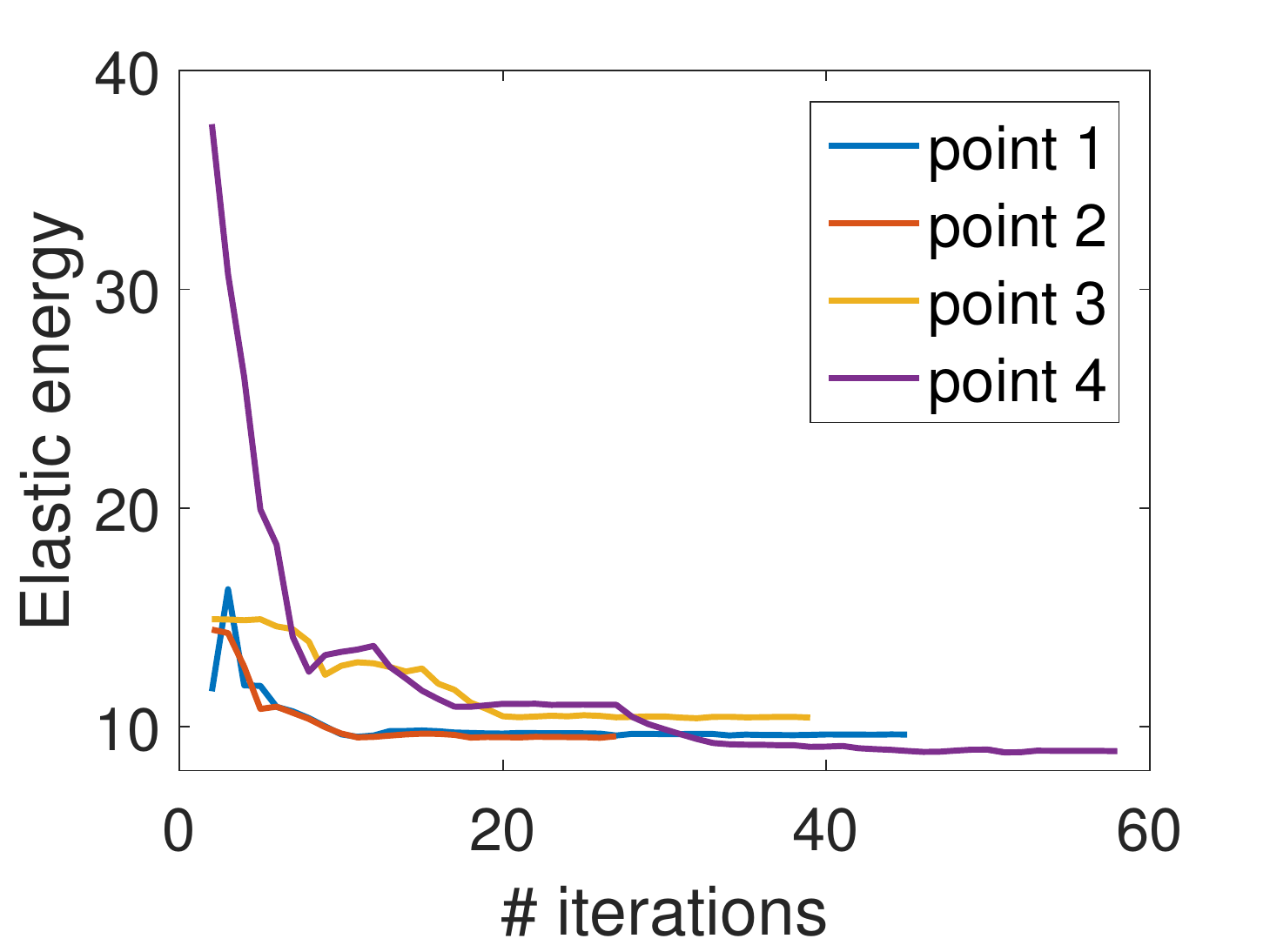}
    \caption{Convergence tests. Variation of the objective energy (\emph{left}) and the elastic energy \emph{right} of a beam being optimized for minimum compliance as the optimization progresses. The convergence plots correspond to the beam of Figure \ref{fig:defo_cvg} when optimized using different material spaces (\emph{top}), different resolutions for the beam lattice (\emph{middle}) when using cubic microstructures, and different initial material properties for the cubic microstructures (\emph{bottom}).
		}
    \label{plot:cvg}
\end{figure}

\subsection{Microstructure Sampling}
We evaluated our method on two- and three-dimensional microstructures made of one or two materials. For the 2D case we considered patterns with cubic and orthotropic mechanical behaviors that can be described with 4 parameters (3 elasticity parameters and density) and 5 parameters (4 elasticity parameters and density) respectively. In 3D we computed the gamut corresponding to cubic structures with 4 parameters. In all cases, the size of the lattice for the microstructures was set to 16 in every dimension. We used isotropic base materials whose Young's modulus differed by a factor of 1000 and having 0.48 as Poisson's ratio. We initially computed the databases for two-material microstructures, but also adapted these databases for microstructures made of a void and a stiff material. In the later case, we replaced the softer material by void, filter out all the microstructures with disconnected components and, in the 3D case, filled the enclosed voids and recomputed the homogenized properties. \updated{We provide a comparison between the initial and postprocessed databases in the supplementary material. }The resulting postprocessed gamuts are also depicted in Figure \ref{fig:gamuts}. 
Our databases contain 274k, 388k and 88k 2D cubic, 2D orthotropic and 3D cubic microstructures respectively and took from 15 hours to 93 hours to compute, which correspond to 68, 19 and 5 sampling cycles, respectively.

\begin{figure}[t]
	\centering
	\includegraphics[width=.99\linewidth]{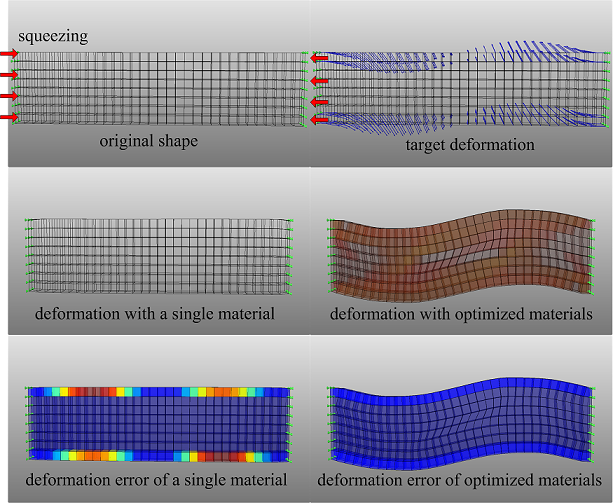}
	\caption{
Optimizing a beam to make it bend when it is squeezed. A beam with optimized material properties can take the desired ‘S’
shape (right) whereas a beam with homogeneous material properties can only axially deform under small deformation (left).  In this example, the vertices on the vertical sides are fixed in their target positions, while the other vertices are free to move. Target displacements are set on the nodes of the cells of the two horizontal boundary layers. The color plot for the bottom beams shows the deformation error of each cell as defined by Equation \ref{eq:td}. }
\label{fig:beam_S_shape}
\end{figure}

We first compared our results to the ones obtained by Schumacher et al. \shortcite{Schumacher:2015} and observed a significant increase in the coverage of the material space, even for 2D microstructures where we used a coarser discretization. This comforts us with the idea that topology optimization only, while helpful to locally improve the microstructure geometries, is suboptimal when one aims to discover the entire gamut of physical properties. The diversity of the microstructures that we obtained is also much richer, thus providing a larger set of options for the practical use of microstructures. Note that they employed some regularization to avoid thin features. For $16^3$ microstructures, we found regularization unnecessary since they are manifold and have a minimal feature size  of 1/16 of the lattice size, which is the same order of magnitude as the thinnest parts of Schumacher's microstructures. \updated{For completeness, we also compared our database of 3D microstructures to the one of Panetta et al. \shortcite{Panetta:2015} at $16^3$ and $64^3$ grid resolutions (Figure \ref{fig:gamuts}, \emph{right}). Our initial database was computed with 0.48 as Poisson's ratio and is shown in the supplementary material. For this comparison, we then recomputed the material properties of the microstructures using the same Poisson's ratio as Panetta's, i.e 0.35, which affects the extremal values of the obtained gamut. For the $64^3$ microstructures, we used morphological operations in the discrete step and sensitivity filtering with a radius of 3 voxels in the continuous step to limit the minimum feature size to 1/32 of the lattice size~\cite{sigmund:2007}. Note that this comparison is provided for reference only since our microstructures are cubic while Panetta's are isotropic (a subset of cubic). Furthermore, they target a different 3D printing technology with self-supporting constraints not imposed here.}
Finally, we also obtained a dense sampling in the interior of the space, as a result of the randomness inherent to our approach. This reduces the need of running costly optimization in these areas and occurs even if we do not explicitly enforce any sampling there.

We also experimented with three-material 2D cubic microstructures (two solid materials with Young's moduli differing by a factor of 1000 and with 0.48 as Poisson's ratio, plus a void material). The resulting database contains about 800k microstructures that can potentially be printed. The corresponding gamut and some examples of the generated microstructures are shown in Figure \ref{fig:Cubic2D_3Materials}.

\subsection{Topology Optimization}

We tested our topology optimization algorithm on a number of simple test cases and large scale examples. Detailed analysis and discussion of the results is provided below.

\paragraph*{Impact of the Material Space}
We evaluated the impact of the chosen material space on a 2D cantilever beam with optimized minimum compliance. We tested our topology optimization algorithm on isotropic, cubic and orthotropic gamuts as well as with the virtual materials used in the traditional SIMP approach and for which the stiffness of the material $E=\rho^p E_0$, $p \geq 1$ is a function of the density $\rho$ of the cell and the stiffness of the base material $E_0$. We also tested our algorithm on an analytical gamut with allowed stiffnesses $E$ defined by $E\leq \rho^3 E_0$. The results are shown in Figure \ref{fig:defo_cvg}. It can be noted that, as the dimension of the material space increases, the final energy of the system decreases. This is to be expected since higher dimensional space means larger gamuts. 
Thus, when using cubic materials, the minimum compliance objective function reaches 3\% lower energy than the standard SIMP method with power index 3.
This difference reaches 11\% when we use orthotropic materials.
It is worth noting that the lowest elastic energy is achieved when we use the traditional SIMP method with $p=1$ (as shown in Figure \ref{plot:cvg}). However, this solution does not correspond to a realizable structure since some of the optimized materials do not correspond to any microstructure.

\begin{figure}[t]
	\centering
	\includegraphics[width=.99\linewidth]{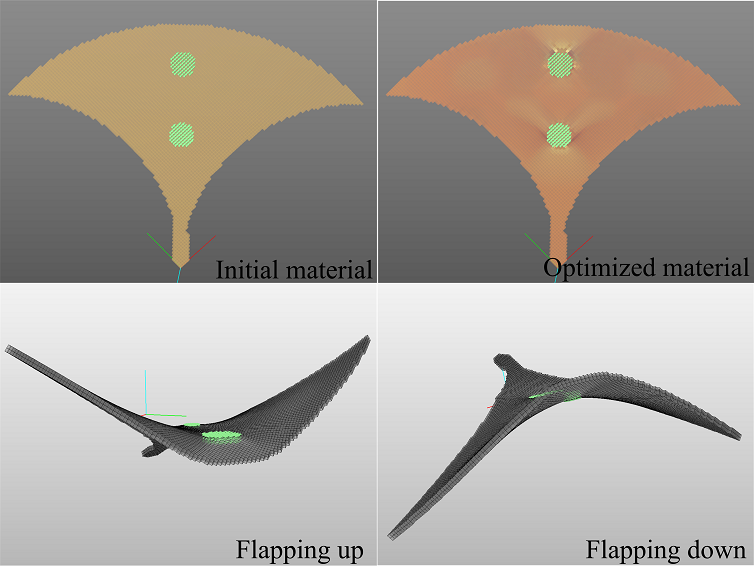}
	\caption{Designing a soft ray. The wings of the ray are asked to flap up and down when vertices on its spine contract and expand. Constrained vertices are colored in green. The deformations achieved with the optimized materials are displayed on the bottom row.}
	\label{fig:ray}
\end{figure}

\updated{
\paragraph*{Matching Quality}
We evaluated for different examples the matching quality of the target deformation optimization.

For the first test, we forced a beam to take an `S' shape when undergoing tensile forces (Figure \ref{fig:beam_S_shape}). In order to avoid overfitting, we applied target displacements on the vertices of the boundary cells only. As depicted in the figure, the use of microstructures largely improves the global shape of the beam, which closely matches the target deformed shape. This becomes even more striking when compared to the behavior of a beam made of a homogeneous material.

We also validated our algorithm by designing a soft ray whose wings can flap using a compliant mechanism (see Figure \ref{fig:ray} and accompanying video). Boundary conditions are applied on two circular areas located along the spine of the ray. Each disk has one degree of freedom for deformation, namely contracting or expanding along the disk normals. This mechanism resembles the one of many hydraulics-driven soft robots. We define two target deformation objectives corresponding to the flapping of the wings up and down, when alternatively contracting and expanding the two disks' boundaries. By running our multi-objective topology optimization framework, we can compute an optimized material design that can achieve both deformation modes when the corresponding boundary conditions are exerted.
}

\begin{figure}[t]
    \centering
		\includegraphics[width=0.99\linewidth]{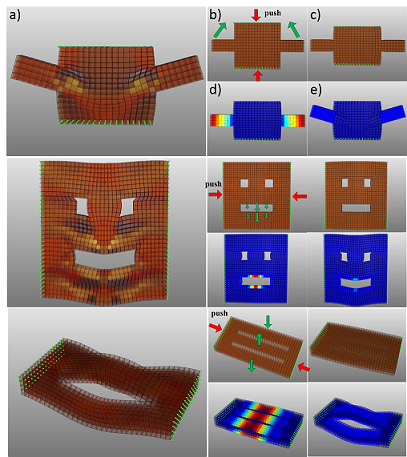}
		\includegraphics[width=0.99\linewidth]{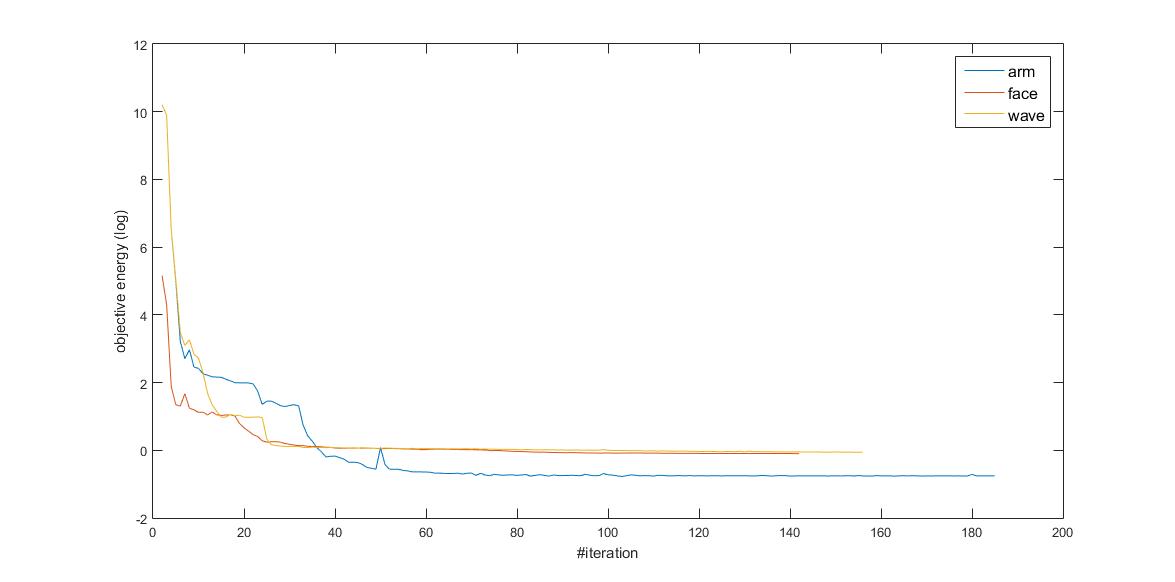}
    \caption{Target deformation examples. The pictures on the left (a) show the deformed optimized result when the boundary conditions illustrated by the pictures (b) are applied. The orange meshes (c) correspond to the simulated deformations when a homogeneous material is used. The blue-to-red meshes indicate the relative deformation error of the unoptimized (d) and optimized (e) structures. Convergence rates corresponding to the optimization of these three examples are reported on the bottom figure.}
    \label{fig:cmp_panetta}

\end{figure}

\begin{figure}[t]
	\centering
	\includegraphics[width=.99\linewidth]{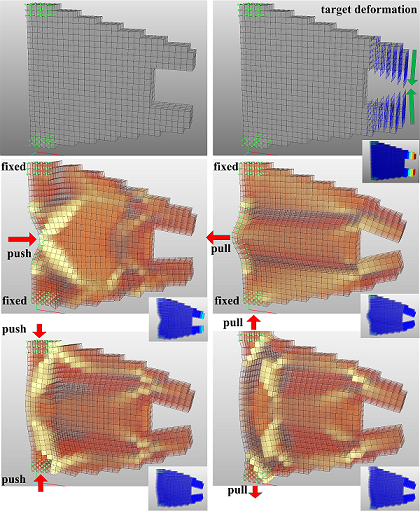}
	\twofigure{.235\textwidth}{.03in}{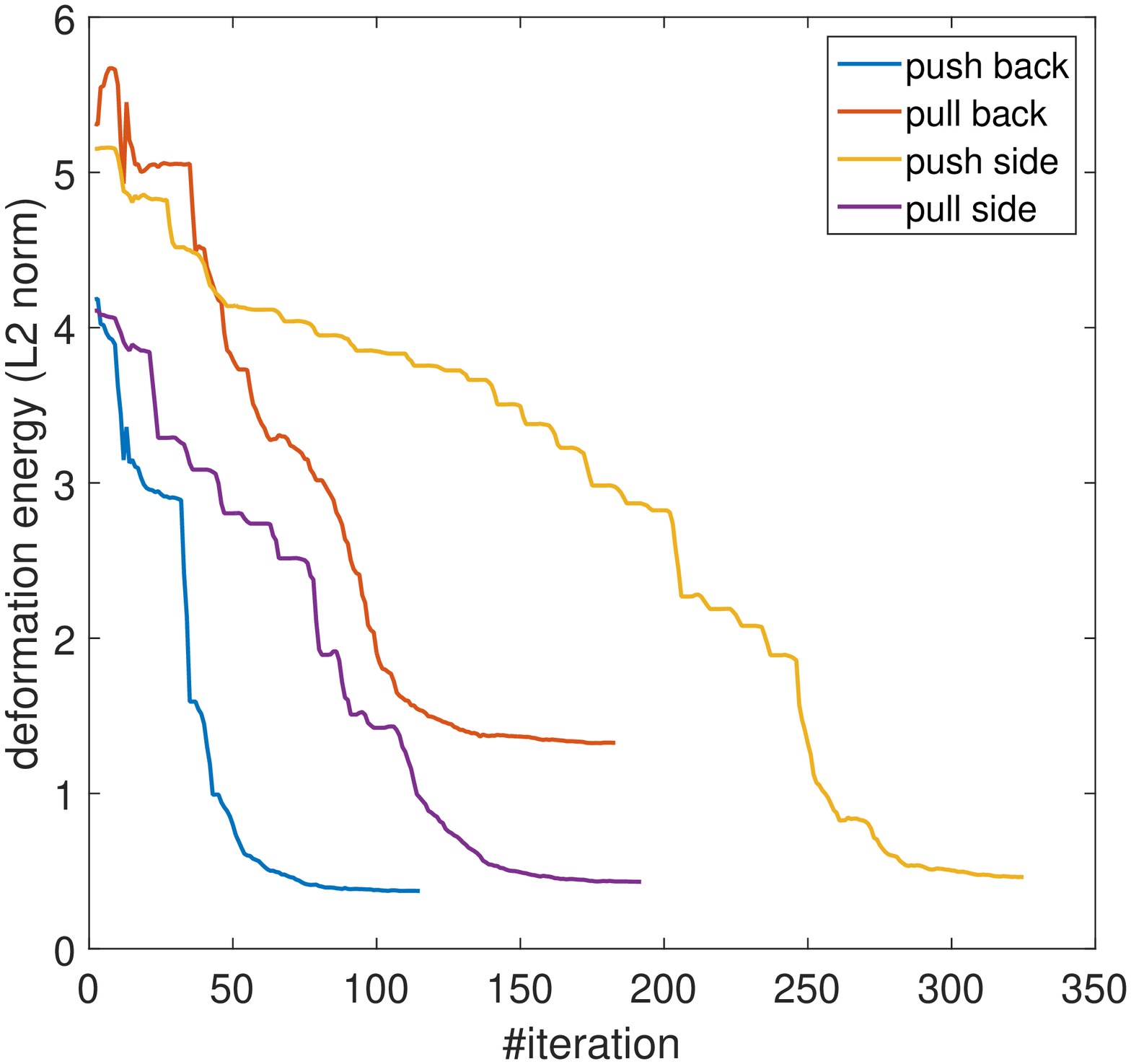}{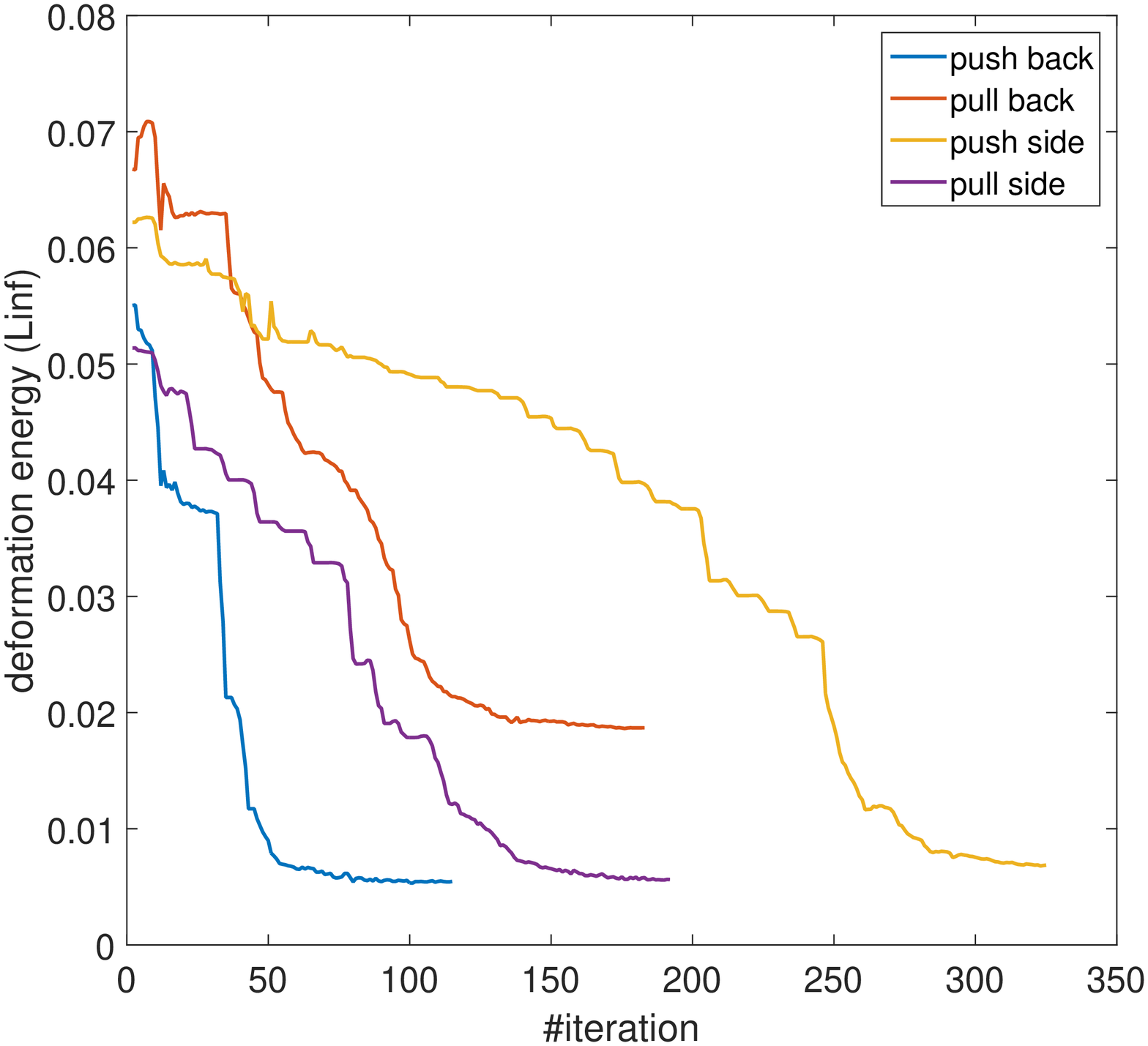}
	\caption{Designing functional grippers. The top row shows the initial shape of the gripper (\emph{left}), and the target deformation for the tip (\emph{right}). The green dots correspond to the fixed vertices while the blue arrows correspond to the target displacements. The two middle rows correspond to the optimized results obtained for the specified boundary conditions. The inset pictures color-code the initial and final deformation error for the different examples.The convergence plots in the bottom row depict the change in the sum of the deformation errors corresponding to all the cells (left) and the value of the maximal cell error contribution (right) as the optimization progresses.}  
	\label{fig:gripper}
\end{figure}

\paragraph*{Convergence and Robustness}

We evaluated the convergence rate of our topology optimization both on the minimum compliance problem and with the target deformation objective. For the minimum compliance problem, we used the same loading as the one of Figure \ref{fig:defo_cvg}. The corresponding results are shown in Figure \ref{plot:cvg} where we plot both the deformation energy of the structure as defined in Equation \ref{eq:mc} and the original objective of the problem \ref{eq:op} that also includes the volume term defined by Equation \ref{eq:vol}. For all these examples, the algorithm converged after a couple of dozen iterations, irrespectively of the lattice resolution, i.e. the number of variables and the number of non-linear constraints. This demonstrates the scalability of the our algorithm. We also tested the robustness of our algorithm by starting with different initial conditions. In this case, we initialized the material parameters of each cell with a random material point projected onto the boundary of the gamut. Similar to other topology optimization schemes, we have no guarantee that we reach the global minimum of the function, and indeed, our algorithm sometimes converges to different solutions. However we note that these different solutions have a similar final objective value and are therefore equally good.

For the evaluation of the target deformation optimization, we ran our topology optimization algorithm on scenarios similar to the ones from Panetta et al. \shortcite{Panetta:2015} where different extruded structures are asked to deform into prescribed shapes when being compressed between two plates (see Figure \ref{fig:cmp_panetta}). As shown in the figure, our optimization algorithm successfully converges to the specified deformation behavior in less than 100 iterations for all the examples. We also tested the convergence rate when optimizing for functional mechanisms. To this end, we designed several grippers that can grasp objects by moving their tips when external forces are applied to their extremities. We experimented with four sets of boundary conditions, namely, pulling and pushing the back of the gripper horizontally, and compressing and stretching the extremities of the gripper vertically. As shown in Figure \ref{fig:gripper}, these different settings lead to different material structures. The deformation errors of all the four designs converge to a low level after a couple of hundreds of iterations.

\paragraph*{Accuracy}
\begin{figure}[t]
	\centering
	\includegraphics[width=.99\linewidth]{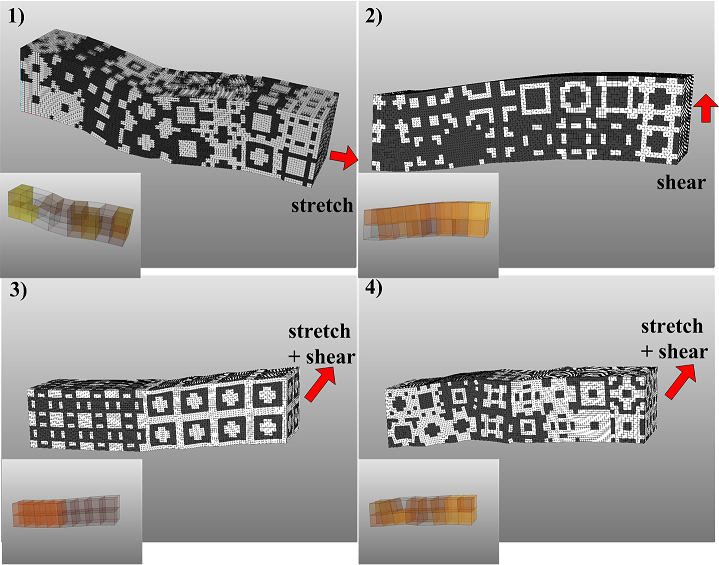}	
	\caption{Comparison of simulated beams with homogenized material properties (\emph{inset pictures}) to the ones using full microstructures (\emph{large pictures}).}
	\label{fig:hom_beam}
\end{figure}

\begin{figure}[h]
	\centering
	\includegraphics[width=.99\linewidth]{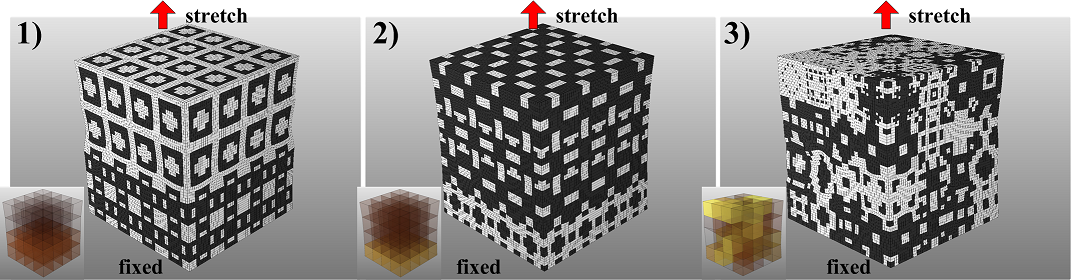}	
	\caption{Comparison of simulated cubes with different material patterns modeled by homogenized cells (\emph{inset pictures}) and full resolution microstructures (\emph{large pictures}).}
	\label{fig:hom_cube_2}
\end{figure}

We evaluated the accuracy of our algorithm on several optimized structures by comparing the deformation obtained when using the optimized homogenized material properties for each cell to the one obtained by a high resolution simulation in which every cell is replaced by its mapped microstructure.

We first evaluated the accuracy on a deformable bar, one side of which was rigidly attached while the other was subject to different sets of external conditions (see Figure \ref{fig:hom_beam}). We used a $8\times2\times2$ lattice to represent the bar with homogenized cells, which translates into a $128\times32\times32$ grid for the full resolution mesh. Similar stretching, bending and shearing behaviors were obtained for both sets of models. From a quantitative point of view, the differences amount to 5-10\% in terms of average vertex displacement and 9\%-33\% in terms of elastic deformation energy (see Table \ref{tab:hom}). We further evaluated the effects of material patterns by running a similar comparison on a cube made of periodic layers of similar microstructures and with random assignments of microstructures (Figure \ref{fig:hom_cube_2}).  As reported in Table \ref{tab:hom}, we show that the ratio between the magnitudes of the average vertex displacement differences is between 4\% and 7\%, and the elastic energy difference is between 10\% and 19\%.

\begin{figure}[t]
	\centering
	\includegraphics[width=.99\linewidth]{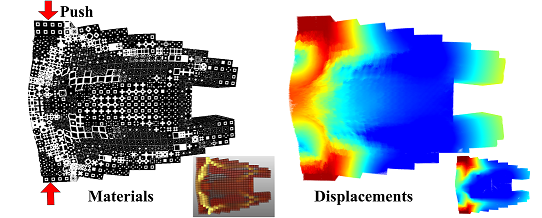}	
	\caption{Comparison of a simulated gripper with homogenized material properties (\emph{inset picture, left}) to the one using full microstructures (\emph{main picture, left}). The figures on the right show the vector field of vertex displacement of the two models. The blue-to-red colors represent the magnitudes of the displacements.}
	\label{fig:gripper_validate}
\end{figure}

Finally, we also compared the behaviors of one of the grippers (Figure \ref{fig:gripper_validate}). The original optimized gripper is made of 3k elements while the high resolution version is made of 4M voxels. Overall, the two models exhibit similar global deformation behaviors, in particular in the tip area. Some differences can be observed on the left side of the gripper for which the high-resolution model exhibits a lower effective material stiffness than its homogenized counterpart. With the same displacement boundary conditions applied, the high-resolution model deforms about 25\% more than the homogenized model.

\begin{figure}[h]
	\centering
	\includegraphics[width=.99\linewidth]{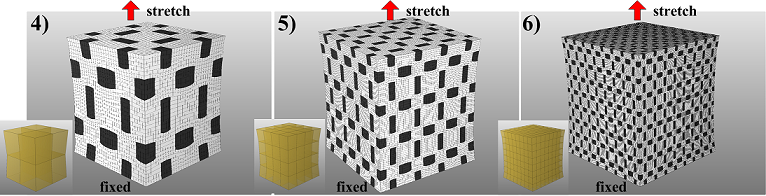}	
	\caption{Simulation of a cube made of a periodic arrangement of a single microstructure at different resolutions. Inset pictures correspond to the model with homogenized material properties, while main pictures correspond to the full resolution simulations.}
	\label{fig:hom_cube_1}
\end{figure}

The differences observed between the homogenized model behaviour and the full resolution simulation can be explained by two major factors: (i) numeral stiffness when using larger elements which tends to make the homogenized mesh slightly stiffer in particular when bending deformation arises, (ii) violation of the periodicity assumption when replacing each cell by a single microstructure. This issue can be reduced by replacing each cell by a {\it tiling} of microstructures. This was verified on a cube made of a periodic arrangement of a single microstructure (see Figure \ref{fig:hom_cube_1}). And indeed, as we increase the resolution of the simulation grid, the error between the homogenized model and the full resolution version decreases and converges to similar values. 

\begin{figure*}[t]
    \centering
		\sixfigure{.16\textwidth}{.03in}{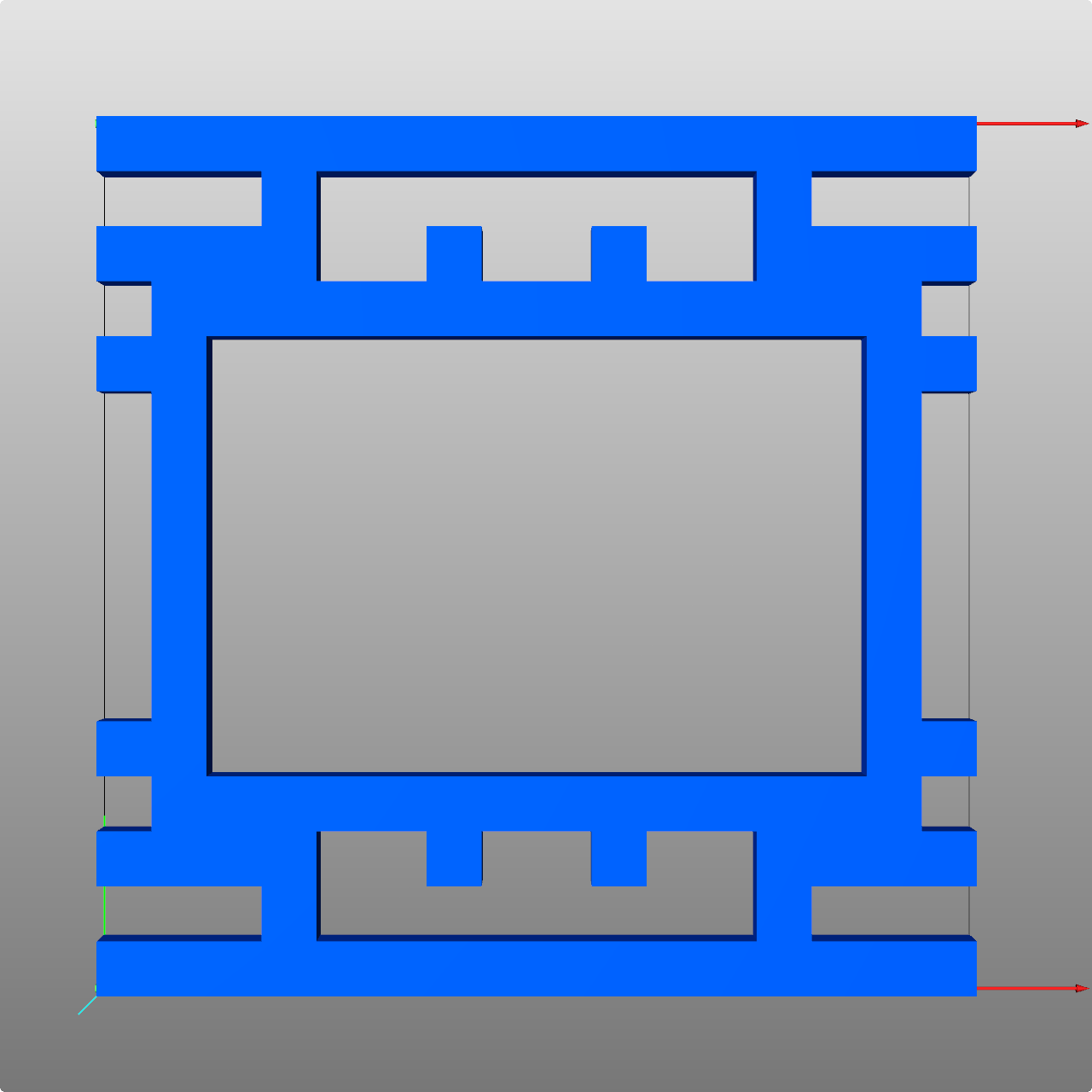}{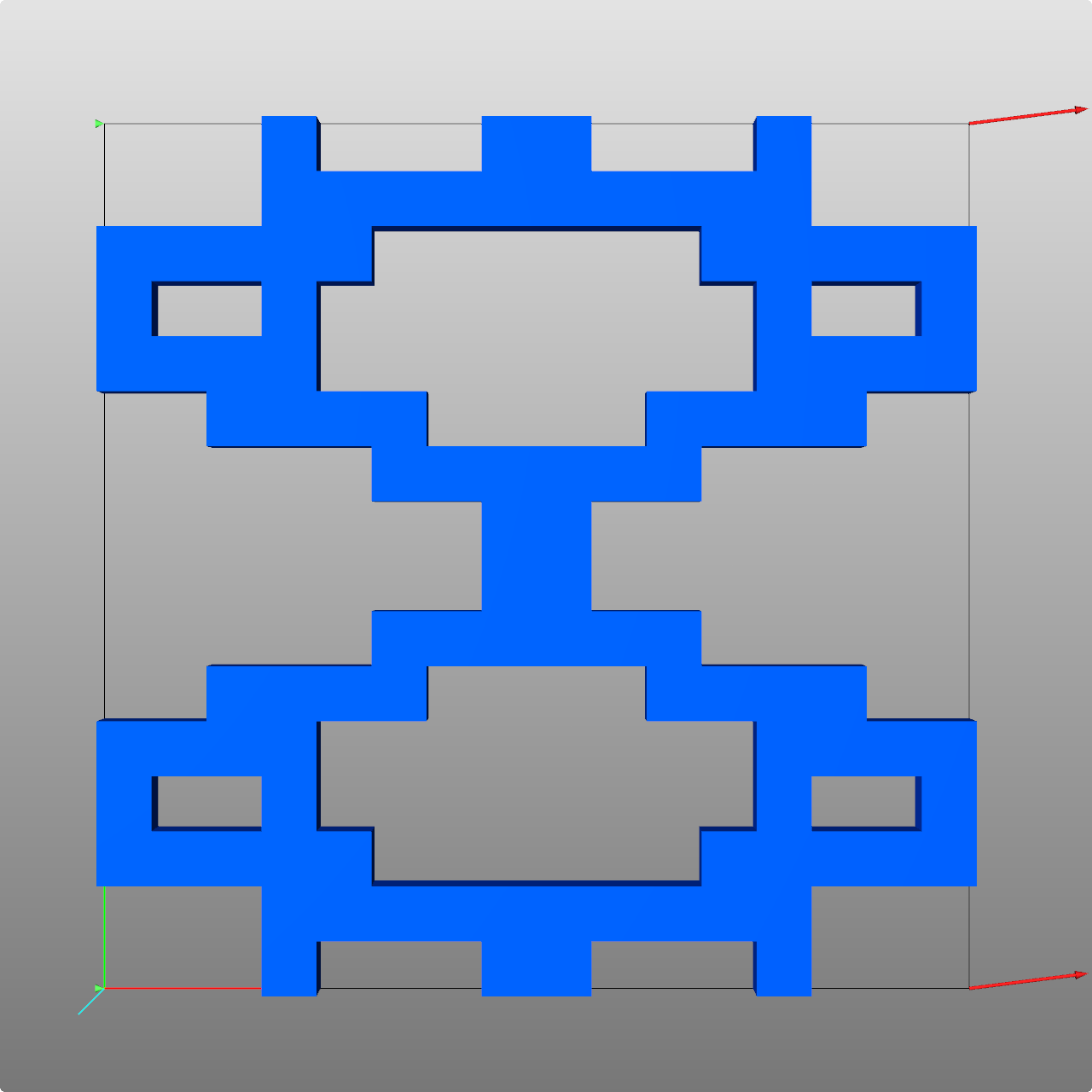}{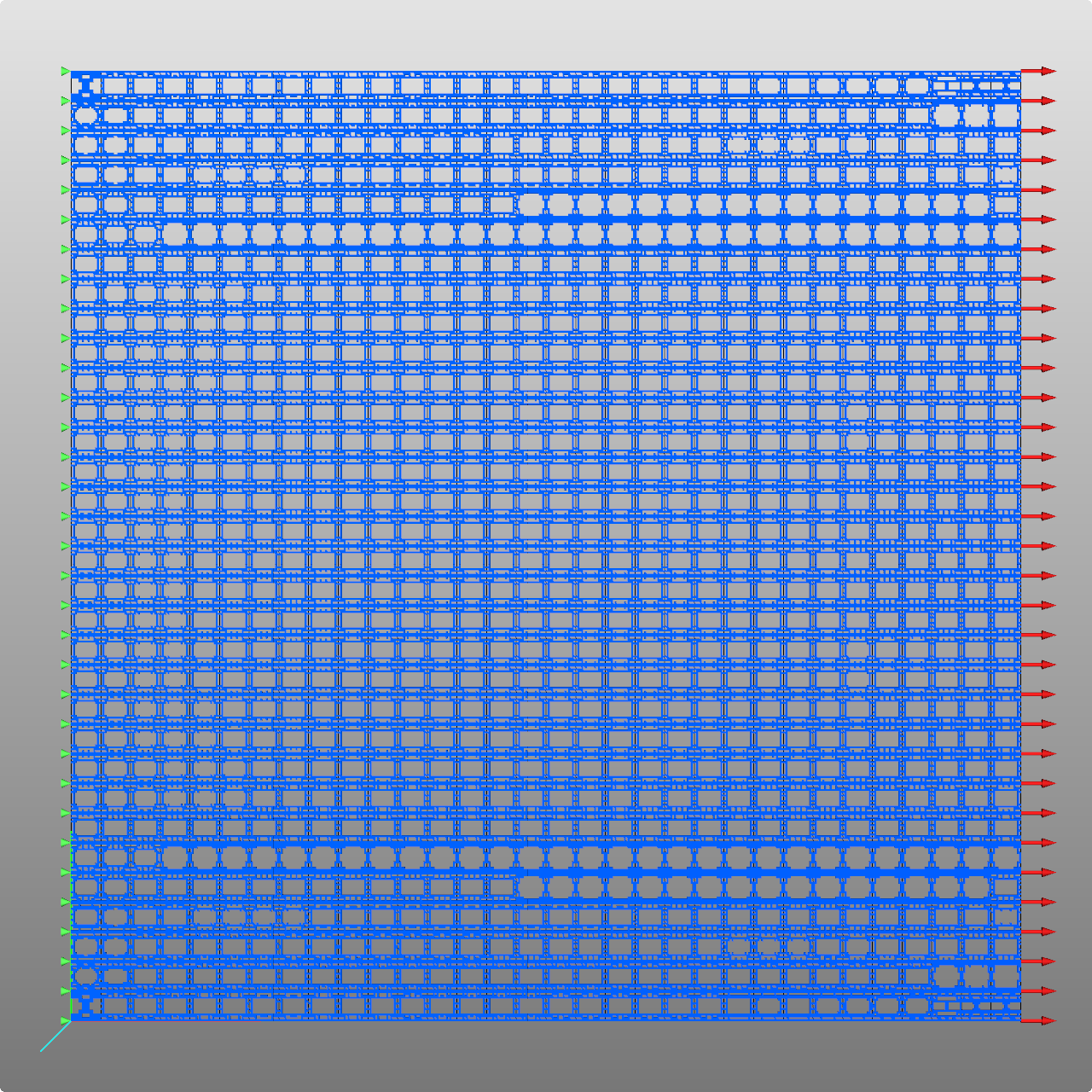}{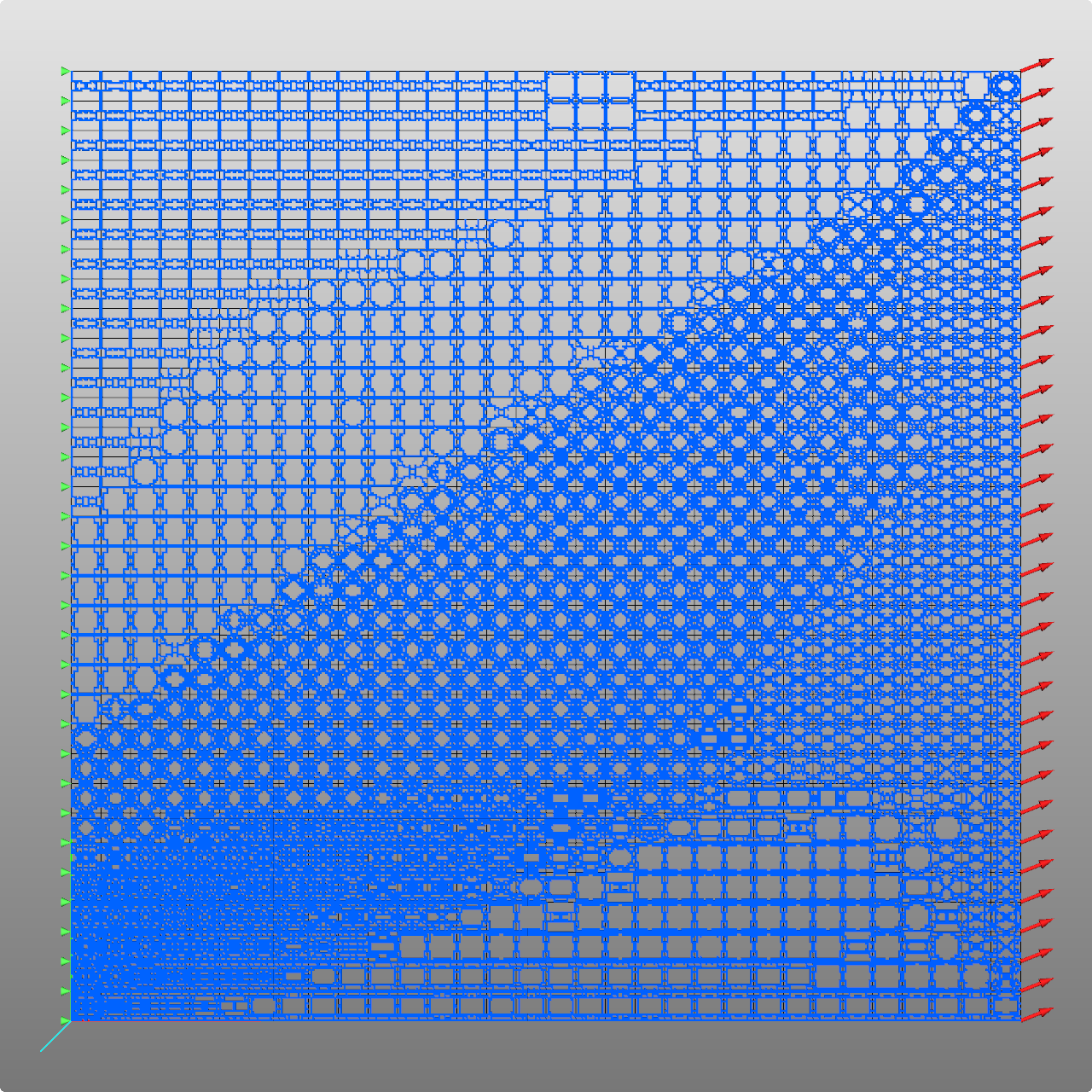}{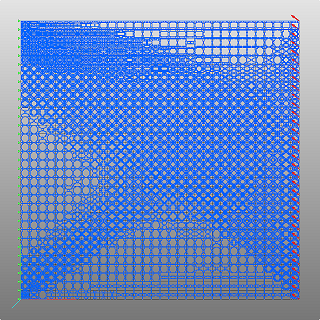}{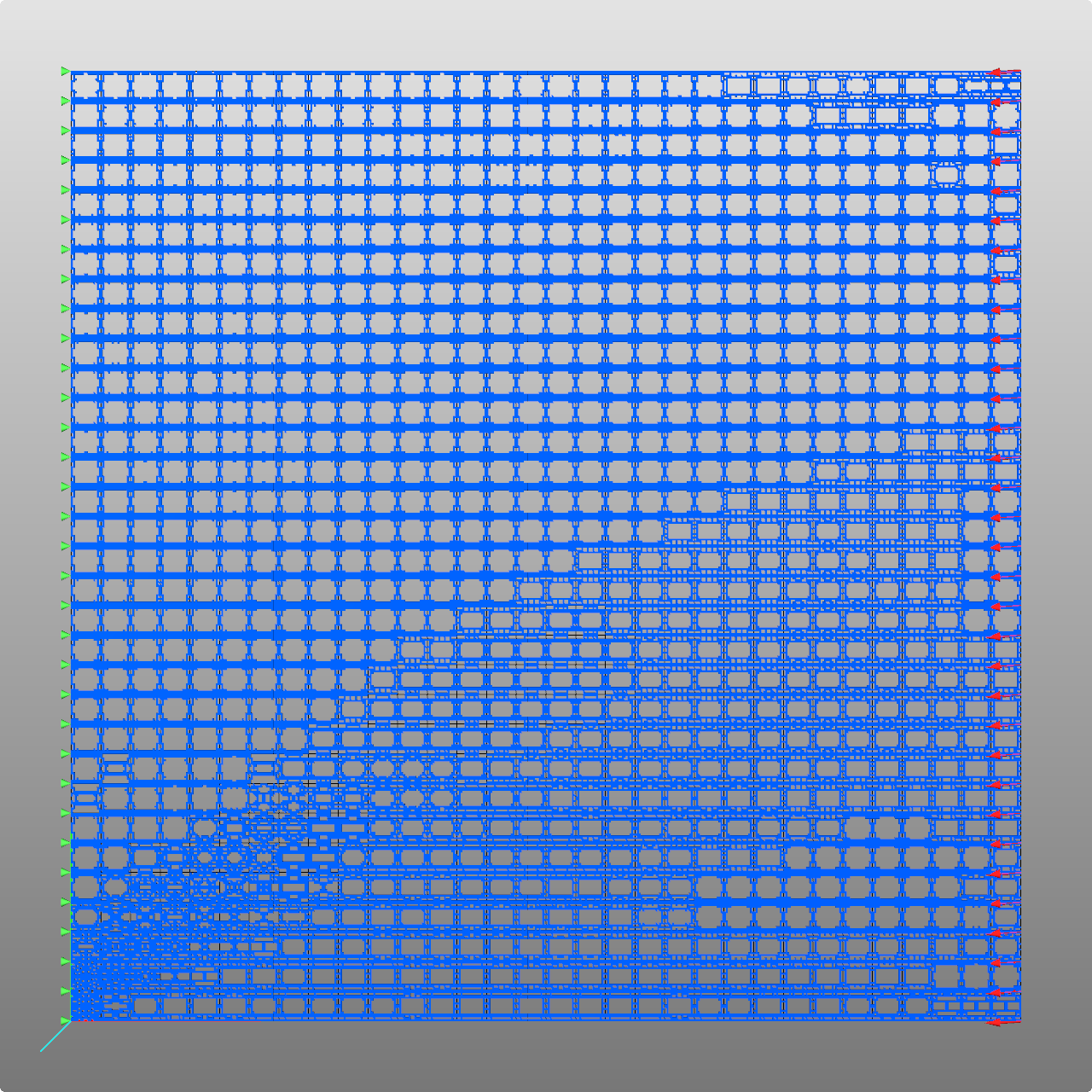}
		\sixfigure{.16\textwidth}{.03in}{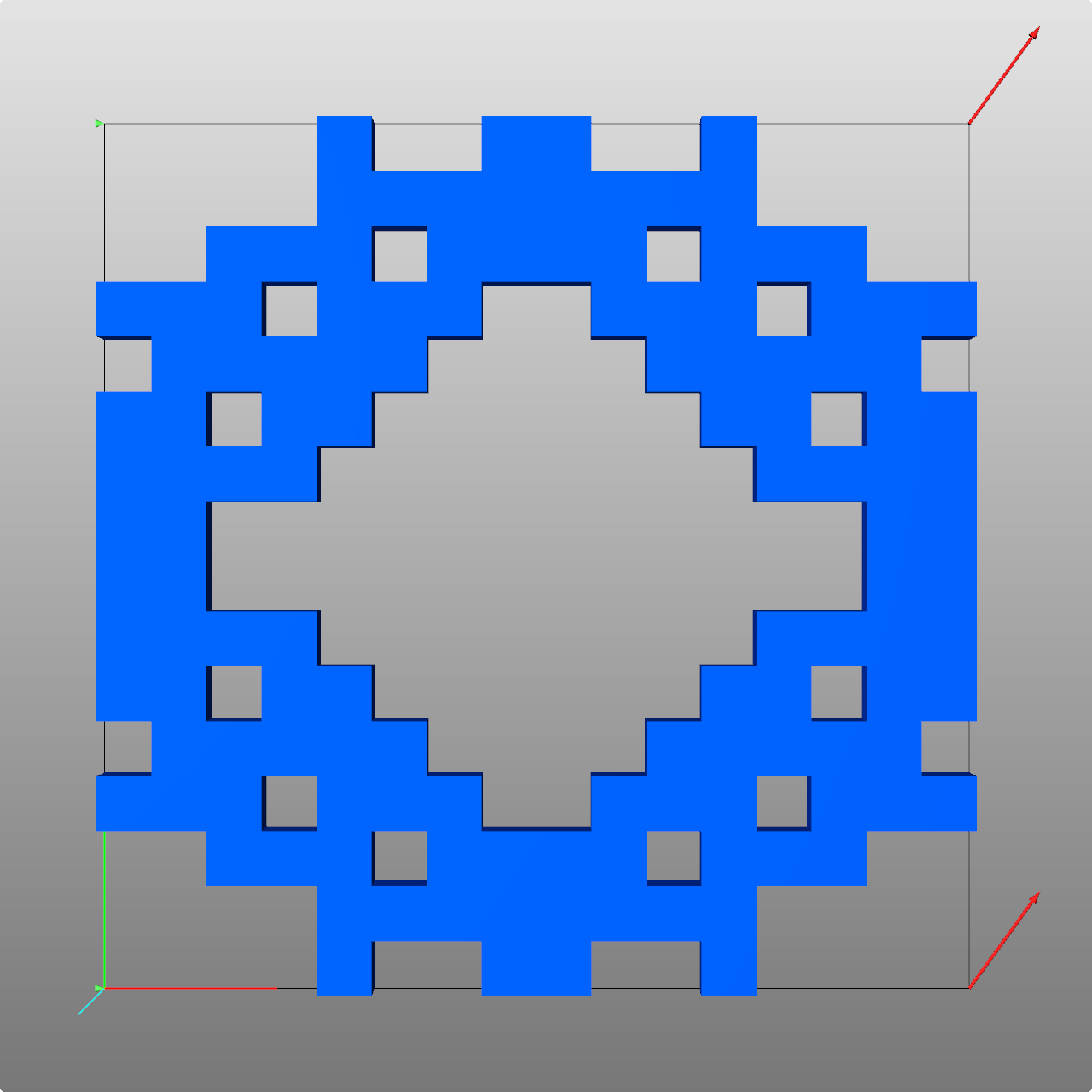}{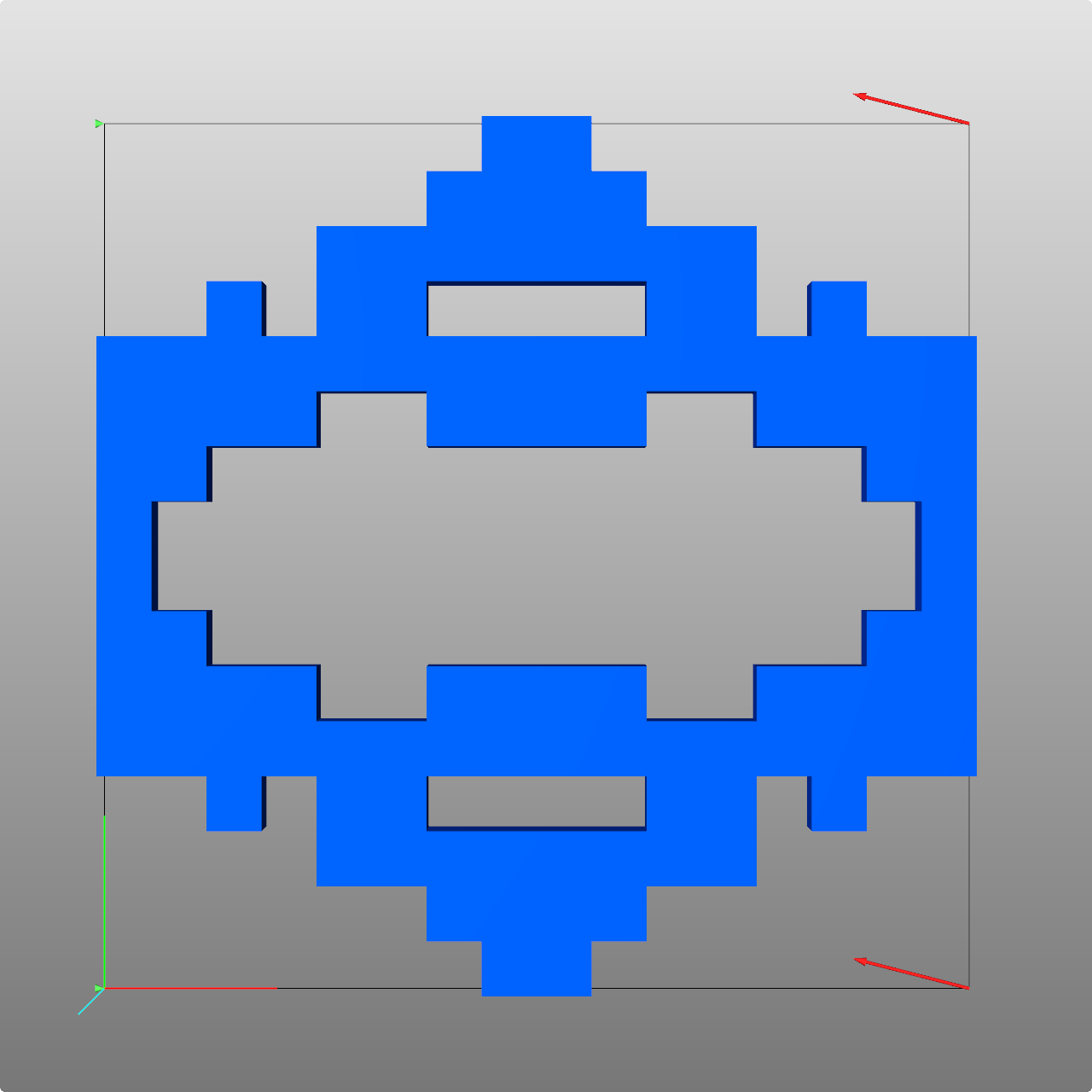}{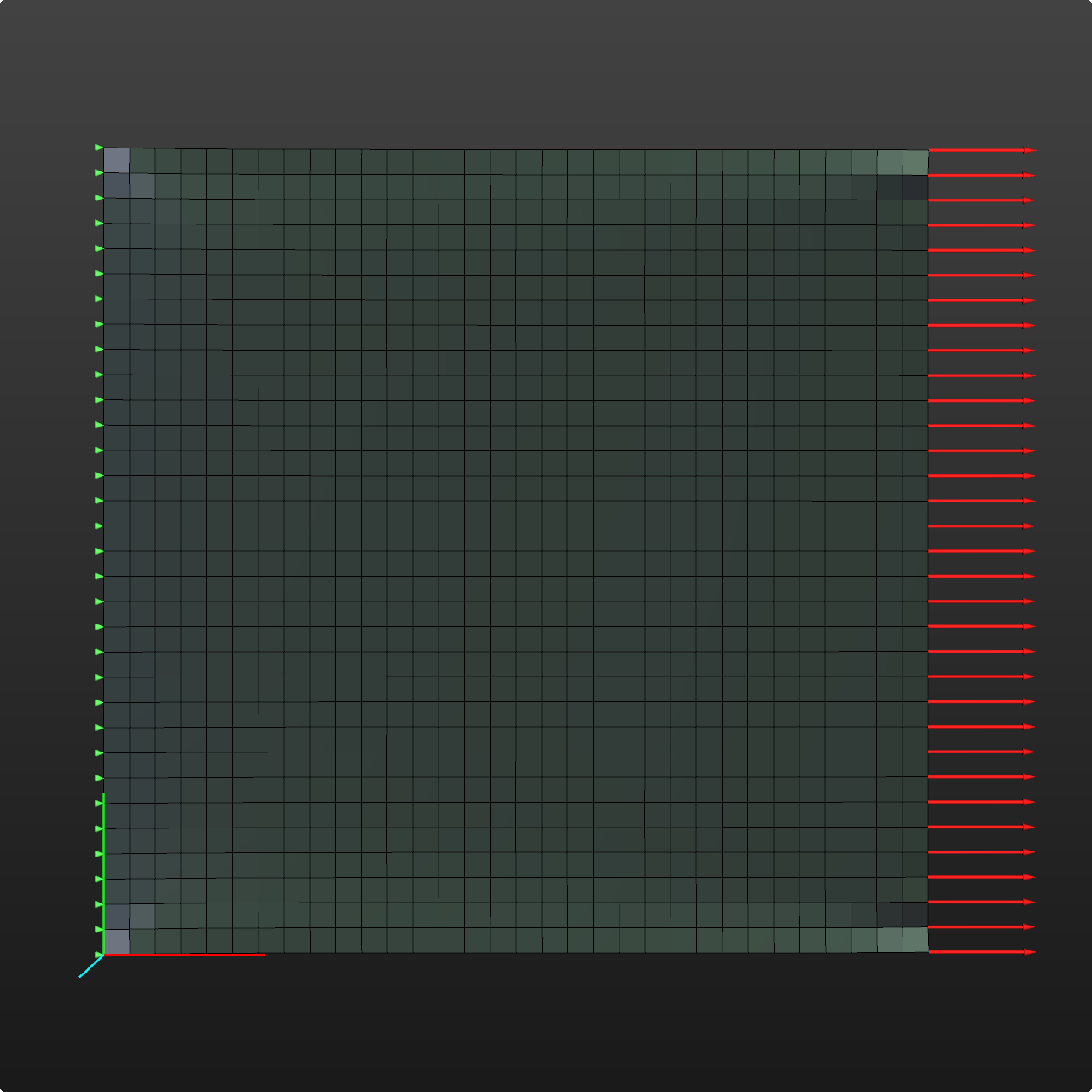}{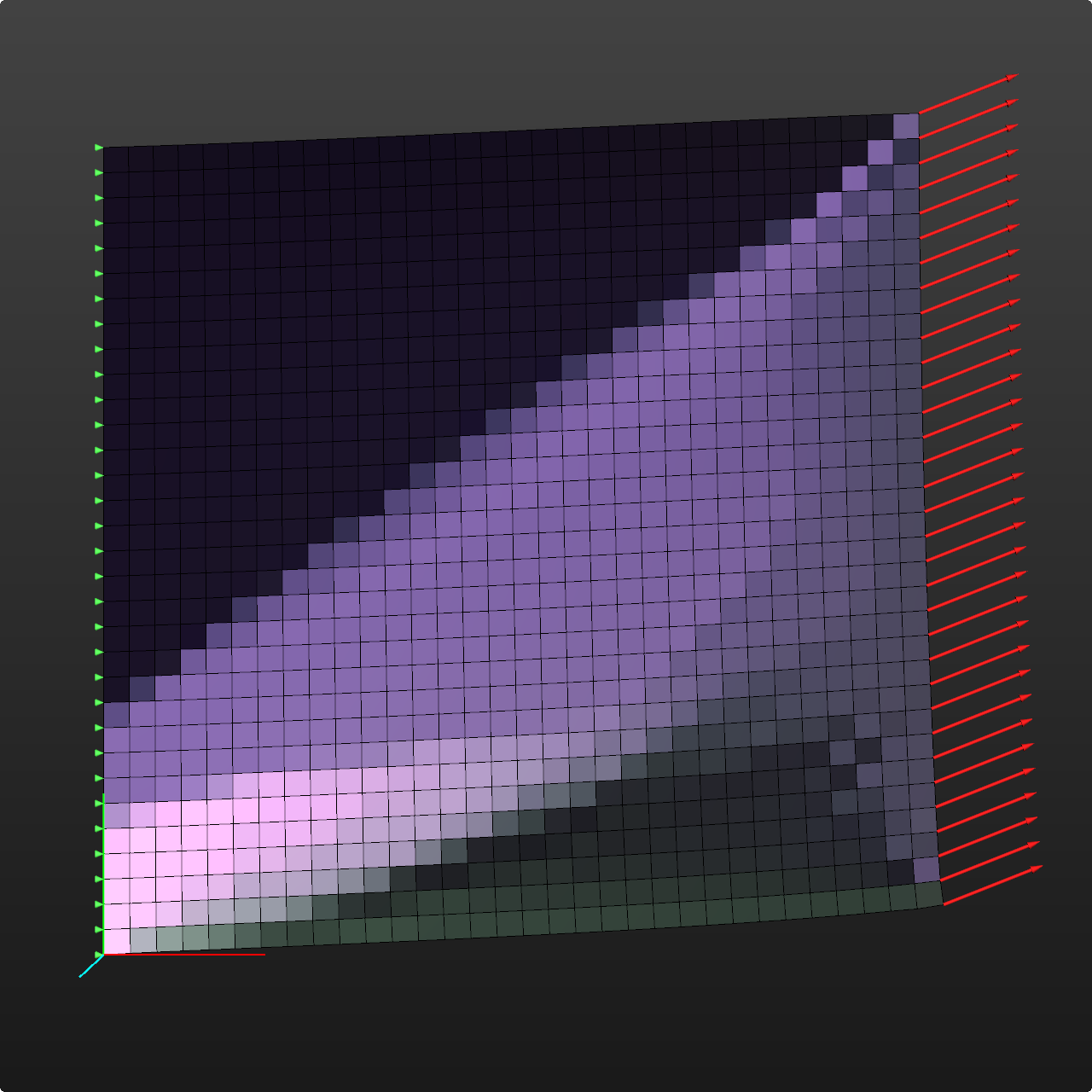}{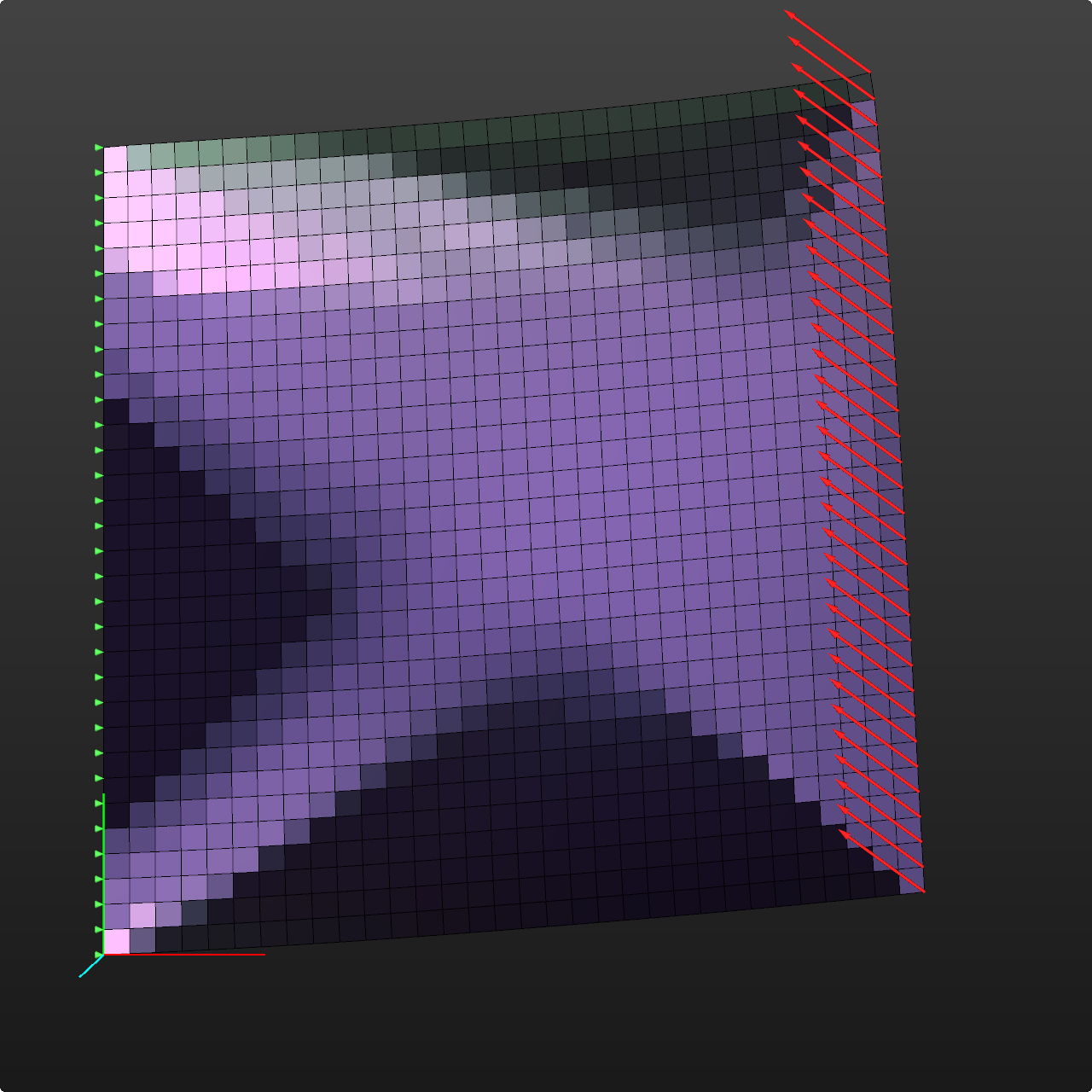}{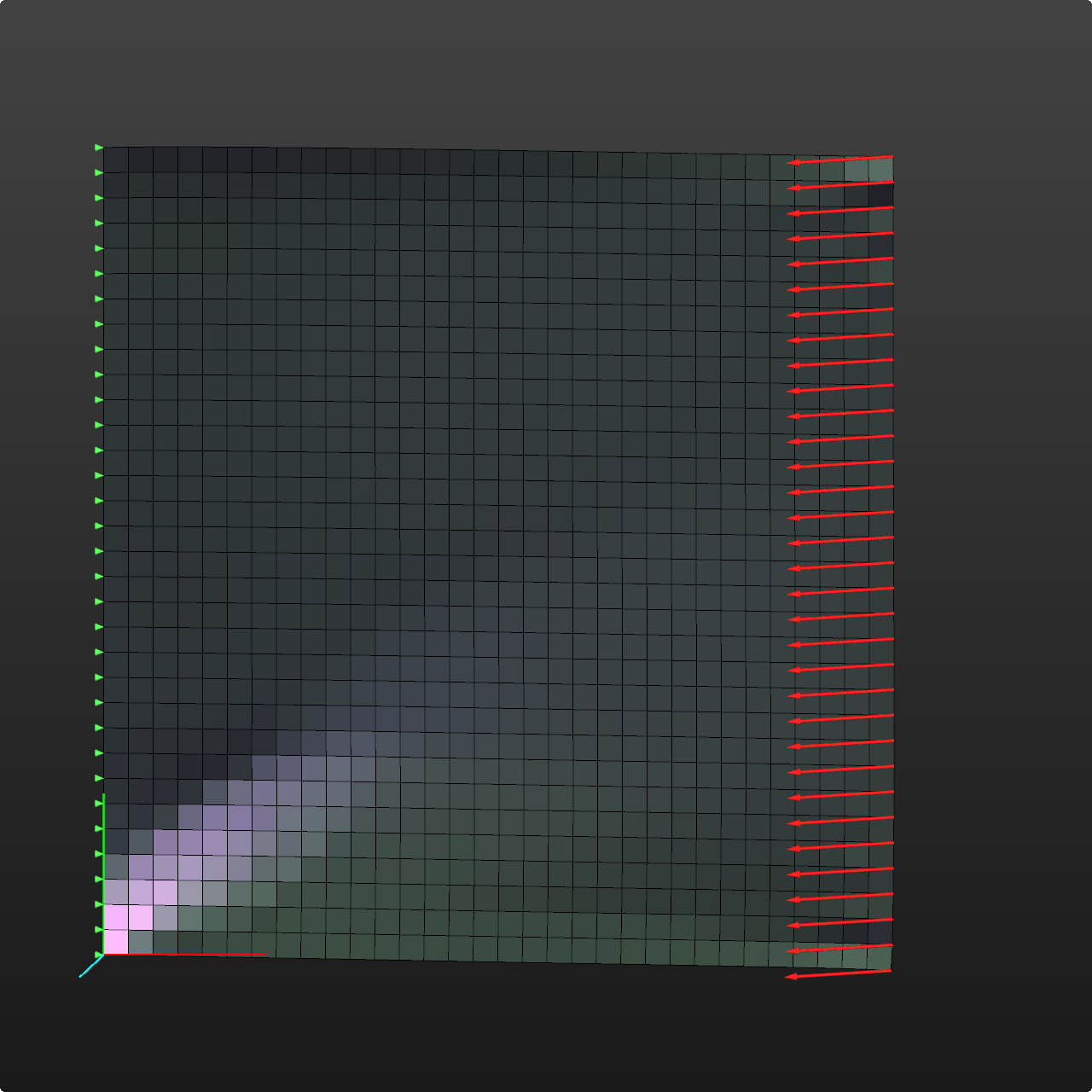}
    \caption{Optimizing the orthotropic material parameters of a single cell (\emph{left}) and a $32\times32$ lattice of cells (\emph{right}) subject to directional forces. The vertices on the left side of the layout are fixed while forces are applied on the right vertices as depicted by the red arrows. Our simple but effective tiling algorithm allows to nicely transition between microstructures of smoothly material properties (\emph{right, top}).}
    \label{plot:res}
\end{figure*}

\paragraph*{Orthotropic materials}
We tested the behavior of our algorithm in a 5-dimensional space by using the gamut of 2D orthotropic microstructures depicted in Figure \ref{fig:gamuts} (\emph{middle}). To this end, we used a regular lattice whose vertices on the left side where fixed and we applied parallel forces on the vertices of the opposite side. The goal in the test was to minimize the compliance of the structure. As can be seen in Figure \ref{plot:res} and in the accompanying video, we experimented with different force directions. Unsurprisingly, when a single cell is considered, the microstructure that we obtain has a structure that is aligned with the direction of the forces (see Figure \ref{plot:res}, \emph{left}). 
For a higher resolution lattice this is no longer true and the resulting overall structure becomes less intuitive (see Figure \ref{plot:res}, \emph{right}).
Note that the resulting material distribution varies smoothly. By considering various alternative for each material point, our tiling algorithm is able to map the material properties to microstructures which are well connected. 

\begin{table}
\centering
\footnotesize
\caption{\updated{Error statistics (SI units). The size of one microstructure is set to $1\times1\times1$.}} 
{\begin{tabularx}{\linewidth}{ |X| X | X | X | X | }
\hline
	Example & Mean displacement & Mean displacement difference & \multicolumn{2}{p{3.2cm}|}{\centering Elastic energy \newline homogenized/full resolution}\\ \hline	
	Beam 1 & 6.47$\times10^{-3}$ & 6.04$\times10^{-4}$ & 6.85$\times10^{-5}$ & 6.17$\times10^{-5}$ \\
	Beam 2 & 6.47$\times10^{-3}$ & 6.04$\times10^{-4}$ & 1.63$\times10^{-5}$ & 1.08$\times10^{-5}$ \\
	Beam 3 & 5.07$\times10^{-3}$ & 4.97$\times10^{-4}$ & 2.38$\times10^{-4}$ & 2.07$\times10^{-4}$ \\
	Beam 4 & 8.78$\times10^{-3}$ & 4.45$\times10^{-4}$ & 3.33$\times10^{-4}$ & 2.30$\times10^{-4}$ \\
	Cube 1 & 3.62$\times10^{-3}$ & 2.86$\times10^{-4}$ & 3.64$\times10^{-3}$ & 3.20$\times10^{-3}$ \\
	Cube 2 & 4.35$\times10^{-3}$ & 1.94$\times10^{-4}$ & 6.82$\times10^{-3}$ & 5.94$\times10^{-3}$ \\
	Cube 3 & 5.42$\times10^{-3}$ & 4.22$\times10^{-4}$ & 7.81$\times10^{-3}$ & 6.32$\times10^{-3}$ \\
	Gripper& 1.32$\times10^{-2}$ & 6.90$\times10^{-3}$ & 8.67$\times10^{-3}$ & 5.70$\times10^{-3}$ \\
	Cube 4 & 5.22$\times10^{-3}$ & 4.89$\times10^{-4}$ & 2.08$\times10^{-2}$ & 1.63$\times10^{-2}$ \\
  Cube 5 & 5.21$\times10^{-3}$ & 2.17$\times10^{-4}$ & 1.89$\times10^{-2}$ & 1.63$\times10^{-2}$ \\
  Cube 6 & 5.21$\times10^{-3}$ & 1.32$\times10^{-4}$ & 1.82$\times10^{-2}$ & 1.63$\times10^{-2}$ \\
\hline
\end{tabularx} }
\label{tab:hom}
\end{table}

\subsection{3D-Printed Designs}

Leveraging our two-scale approach, we used our topology optimization algorithm to generate a wide variety of high resolution models that we 3D-printed. We used a Stratasys Objet Connex 500 and the two base materials \emph{Vero Clear} and \emph{Tango Black Plus} and used the database containing the three-dimensional cubic microstructures. The sizes and computation times of the resulting models are outlined in Table \ref{tab:results}. 

Since Ipopt performs a line search at each gradient step, one single step may correspond to multiple simulations. We show the average time required for taking a step in the last column of Table \ref{tab:results}. For these large scale examples, Ipopt takes two hundred iterations in average to find a local minimum. 
Since our problem is formulated as a very general constrained continuous optimization, it is independent of the optimization package that is used and its speed could potentially be further improved by using alternative minimizers. 
We found Ipopt to be a good choice for its capability to efficiently handle a large number of inequality constraints, which is not the case of other popular minimizers used in topology optimization such as the method of moving asymptotes (MMA).

Our algorithm is mainly directed towards engineering applications and targets the design of objects undergoing small deformations. In the following examples, we sometimes intentionally exaggerated the target displacements (and scaled the external forces accordingly) for better visualization, which does not change the output of the algorithm with a linear material model.

\begin{table}
\centering
\footnotesize
\caption{Statistics on the 3D-printed models. The last row uses the database of $64^3$ microstructures.}
{
\begin{tabularx}{\linewidth}{ |X| X | X | p{1cm}| p{1cm}| }
\hline
  Example & Grid Size & \# Voxels & Time per FEM Solve [s] & Time per Step [s]\\ \hline
  Beam   & 96$\times$24$\times$4  & 38M  &  0.7 & 5\\
  Flexure & 32$\times$32$\times$16 & 67M  & 1 & 12\\
  Gripper   & 64$\times$32$\times$8  & 67M  & 1.7 & 10\\	\hline
  Bunny & 32$\times$32$\times$32 & 134M  & 0.6 & 4 \\
  Bridge & 128$\times$64$\times$32 & 1074M  & 27 & 81\\
	Bridge 2 & 320$\times$160$\times$80 & 1074G  & 1.3k & - \\
\hline
\end{tabularx} }
\label{tab:results}
\end{table}

\paragraph*{Beams with controlled deformation behaviour}

\begin{figure}[h!]
	\centering
	\includegraphics[width=.99\linewidth]{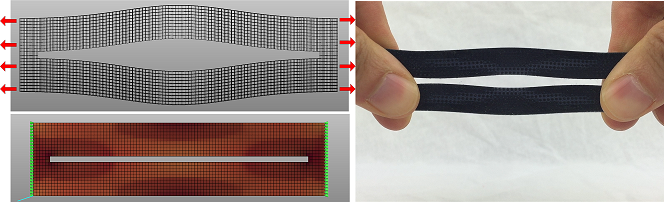}
	\caption{
	An optimized hollow beam with target deformation. The left figure shows the target deformation and optimized material distribution. The right figure shows the 3D-printed structure and the achieved deformation.}
	\label{fig:beam}
\end{figure}

We started by designing a 3D hollowed beam with a desired deformed shape. The beam was stretched by moving vertices on two opposite sides. Our topology optimization algorithm was run using a target deformation objective. The resulting optimized material properties and the 3D-printed structure are depicted in Figure \ref{fig:beam}. 

\begin{figure}
	\centering
	\includegraphics[width=.99\linewidth]{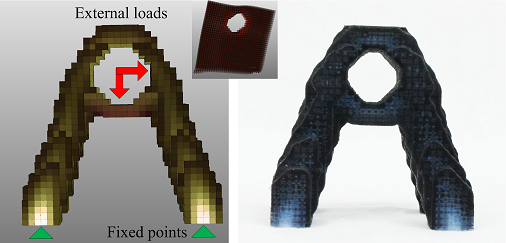}	
	\caption{Optimizing a flexure mount. The flexure is connected to an object thanks to a cylindrical connector. We leave space for this connector by keeping a cylindrical area of the design layout empty of material. The material distribution of the flexure is optimized for two sets of external forces applied to the cylindrical area. Under vertical load, the flexure should stay close to the rest shape while under horizontal load, the flexure should deform according to the inset figure.}
	\label{fig:flexure}
\end{figure}

\paragraph*{Multi-Objective Flexure Design}
We tested our algorithm on a multi-target deformation setting by optimizing the structure of a flexure mount with two different target shapes (see Figure \ref{fig:flexure}). Here, our goal is to design a flexure that resists vertical loads while remaining compliant to horizontal loads. We assume that the object mounted on the flexure is connected to the flexure using a cylindrical connector that transmits the forces to the flexure via the connecting area. In the first scenario, vertical forces are applied to the points of the cylindrical area and we ask the flexure to stay as close as possible to its rest configuration. In the second scenario, horizontal forces are applied to the points of the cylinder and we ask the flexure shape to match the shape shown in the Figure \ref{fig:flexure}.

\begin{figure}[h]
	\centering
	\includegraphics[width=.99\linewidth]{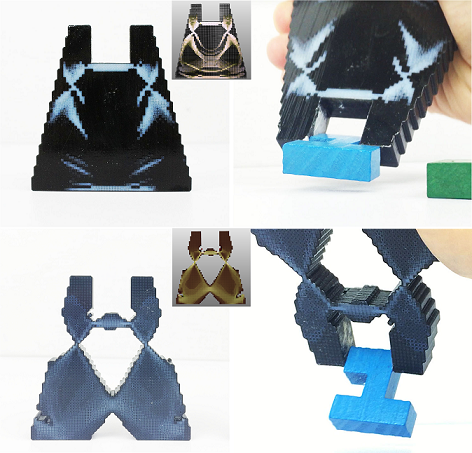}
	\caption{3D-printed functional grippers. By setting different target ratios of the rigid material, different designs can be obtained. 
	When more soft material is used the grasping behaviour of the gripper is obtained via out-of-plane bending (\emph{top}), whereas more rigid material is used, the gripper deformation remains planar (\emph{bottom}).}
	\label{fig:gripper_printed}
\end{figure}

\paragraph*{Gripper}
We verified the functionality of our grippers by fabricating two of them. For these results we ran the optimization on high resolution meshes of the version that grasps the object when the extremities of the gripper are pressed (see Figure \ref{fig:gripper_printed}). By changing the parameter controlling the ratio of the soft material, different designs based on different mechanisms can be achieved. When more soft material is used the gripper achieves its target deformation thanks to out-of-plane bending, while for stiffer designs, the grasping motion is achieved via in-plane deformation.

\begin{figure}[t!]
	\centering
	\includegraphics[width=.99\linewidth]{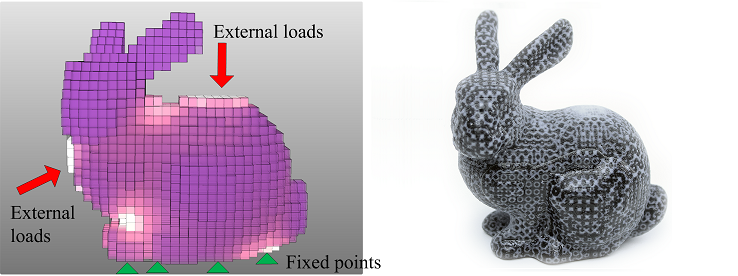}
	\caption{Stanford bunny optimized for two loading cases. Two sets of external forces are applied to the back and chest of the bunny as indicated by the arrows, and the color indicates the material distribution (\emph{left}). Material parameters are mapped to microstructures to obtain an object that can be actually printed (right).}
	\label{fig:bunny}
\end{figure}

\begin{figure}[t!]
	\centering
	
	\includegraphics[width=.99\linewidth]{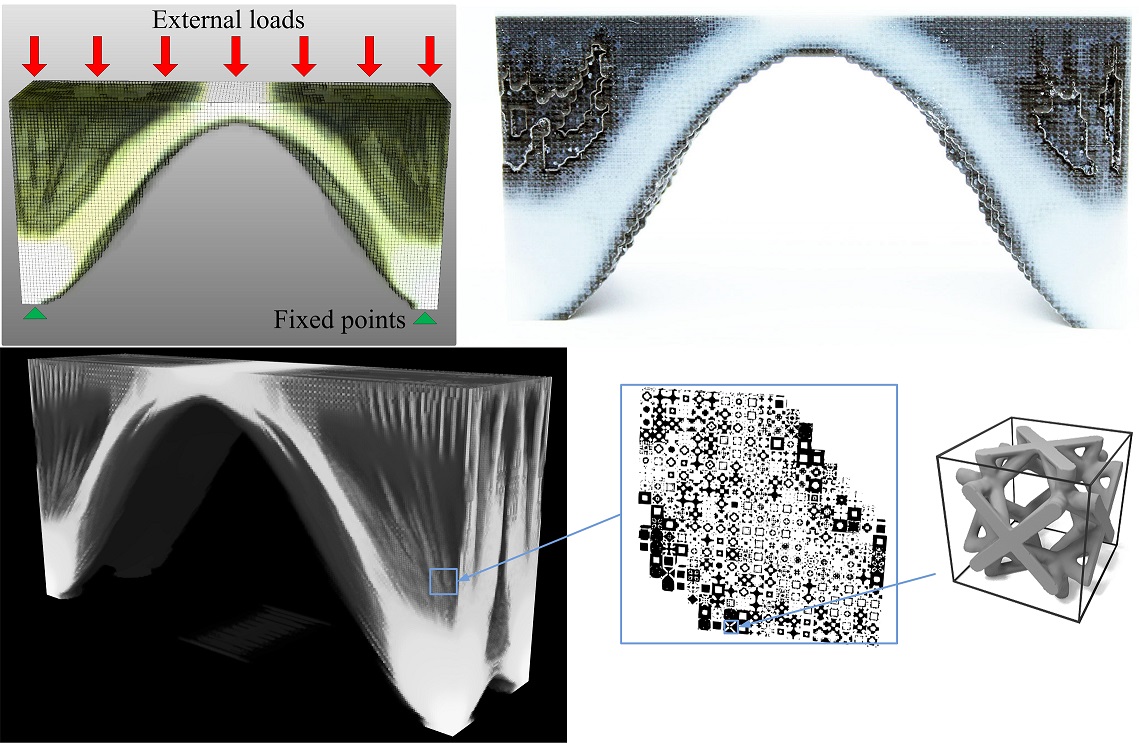}
	\caption{Optimizing a bridge. The initial layout corresponds to a 128x64x32 regular grid. We apply uniform loads on the upper plane deck. We compute the material parameters and set cells with extremely low stiffness to void (\emph{top left}). We look up the microstructures and 3D print the bridge (\emph{top right}). We scale the problem to 1 trillion voxels by using a lattice of 4 million elements where each element corresponds to a $64^3$ microstructure (bottom left). We show a $20\times20\times1$ patch on the bridge with filled microstructures and a single microstructure with $64^3$ voxels on the patch (bottom right).}
	\label{fig:bridge}
\end{figure}

\paragraph*{Minimum Compliance Examples}
To demonstrate the scalability of our algorithm, we computed a Stanford bunny (Figure \ref{fig:bunny}) made of more than 100 million voxels and subject to two load case scenarios.
We also designed two bridges of increasing resolutions. The first bridge was optimized using a lattice of half a million cells which corresponds to 1 billion voxels (Figure \ref{fig:bridge}). For the second bridge, we used the database of $64^3$ microstructures and a layout made of 4 million cells, which amounts to 1 trillion voxels. We initialized the topology optimization by running the algorithm on a lower resolution grid with 1.4 million elements and used the resulting parameters as initial material parameter values for the higher resolution optimization. 

%% file: Conclusion.tex
\section{Conclusion}
We have presented a computational framework for two-scale topology optimization. Our approach can efficiently optimize high resolution models that can be fabricated using multi-material 3D printing. Our first insight is to use a precomputation process to efficiently sample the space of microstructures and their corresponding material properties in order to define a continuous material property gamut. Our second insight is to use this gamut as a constraint in a generalized topology optimization framework to assign spatially-varying material properties throughout the optimized object. Finally, the volume with assigned material properties can be converted to a 3D model with corresponding spatially-varying  microstructures. We demonstrated the effectiveness of our approach on multiple examples and showed improvements over traditional binary topology optimization schemes.

\paragraph*{Limitations and Future Work}
First, while our sampling method outperforms current approaches in terms of the material space coverage and the approach converges to stable gamuts, we do not provide any theoretical guarantees that the gamut space cannot be further expanded. Second, we would like to investigate microstructures with additional properties, e.g., electrical or magnetic properties, their combined property gamuts, and the different applications they enable. Finally, our framework builds upon linear elasticity and optimizes the material distribution of objects subject to small deformations only. While this is enough for many engineering applications, extending our algorithm to the nonlinear regime would be an interesting direction for future work.

%% file: Appendix.tex
\appendix
\section{Appendices}
\subsection{Discrete sampling of microstructures}

Given a set of microstructures, a new population of microstructures is generated using the Stochastical\-ly-Ordered Sequential Monte Carlo (SOSMC) method introduced by Ritchie et al.~\shortcite{Ritchie2015SOCM}.

In our implementation, the samples (or particles) are our microstructures, i.e. binary assignments of the base materials, and the desired distribution is the one that maximizes the number of particles located near or outside the boundary of the gamut of material properties. We evaluate the contribution of each sample towards the desired goal thanks to the scoring function
\begin{equation}
s(\mathbf{p_i}) = \frac{\Phi(\bf p_i)}{D(\bf p_i)} \times \frac{1}{D(\bf p_i)},
\label{eq:score_fct}
\end{equation}
where $\Phi(\bf p_i)$ is the signed distance of the material properties of particle $i$ to the gamut boundary and $D(\bf p_i)$ is the local sampling density at the location $\bf p_i$.  The sample density is defined as
\begin{equation}
D(\bf p_i)=\sum_{k}\phi_{k}(\bf p_i)\ ,
\end{equation}
where $\phi_{k}(\bf p)=\left(1-\frac{||\bf p-\bf p_{k}||^{2}_{2}}{h^{2}}\right)^{4}$ are locally-supported kernel functions that vanish beyond their support radius $h$, set to a tenth of the size of the lattice used for the continuous representation of the material gamut.

As described by Algorithm \ref{algo:proc_microstructure_init}, given an initial microstructure, we generate a new microstructure by randomly swapping materials in the material assignments. However, as explained in Ritchie's paper, executing the procedure sequentially -- in our case visiting the voxels in a fixed order -- is often suboptimal since the best order is often unknown a priori. To introduce randomness in the program execution, and because of the simplicity of our procedure primitives, i.e. swapping voxel materials, we do not rely on Stochastic Future as in Ritchie's implementation, but directly modify the original program into the one described by Algorithm \ref{algo:proc_microstructure}.

\algrenewcommand\algorithmicindent{1.0em}%

\begin{algorithm}[b]
	\begin{algorithmic}
		\Procedure{genMicrostructure}{input: microstructure $M_i$, output: microstructure $M_o$}
		\State $M_o\gets M_i$
		\ForAll{voxels} 
		\State swap material of the current voxel $v$ with probability 0.5
		\EndFor
		\EndProcedure
	\end{algorithmic}
	\caption{Initial procedure for generating a new microstructure}
	\label{algo:proc_microstructure_init}
\end{algorithm}

\begin{algorithm}[h]
	\begin{algorithmic}
		\Procedure{genMicrostructure}{input: microstructure $M_i$, output: microstructure $M_o$}
		\State $M_o\gets M_i$
		\While {some voxels of $M_o$ have not been visited}
		\While {microstructure $M_o$ is unchanged}
		\State pick a random voxel $v$ of $M_o$ that has not been visited 
		\State assign a randomly chosen material to $v$
		\If {$M_o$ is manifold and $M_o\neq M_i$} \State{accept the change} \EndIf
		\EndWhile
		\State{\textcolor[rgb]{0,0.5,0}{// Synchronization point}}
		\EndWhile
		\EndProcedure
	\end{algorithmic}
	\caption{Procedure for generating a new microstructure}
	\label{algo:proc_microstructure}
\end{algorithm}

\begin{algorithm}[tbp]
	\begin{algorithmic}
		\Procedure{SOSCM}{input: set of $n$ microstructures $m$, output: set of microstructures $p\_o$}
		\State $p\_o \gets m$ 
		\For{i=1..$n$}
		\State \textcolor[rgb]{0,0.5,0}{// Evaluate scores of all the particles $m_i\in m$}
		\State $w(i)\gets s(m_i)$ 
		\EndFor
		\State \textcolor[rgb]{0,0.5,0}{// Sample $N$ particles}
		\State $p\gets universal\_sampling(m, N, w)$ 
		\For{i=1..$N$}
		\State start program genMicrostructure($p_i$, $q_i$) for the input
		\State microstructure $p_i \in p$
		\EndFor
		\While {some particles $q_i \in q$ have unvisited voxels}
		\ForAll{unterminated programs}
		\State run the program until the synchronization point is 
		\State reached
		\State $p\_o \gets p\_o \cup q$
		\For{i=1..$N$}
		\State \textcolor[rgb]{0,0.5,0}{// Evaluate scores for modified microstructures}
		\State $w(i)\gets s(q_i)$ 
		\EndFor
		\State \textcolor[rgb]{0,0.5,0}{// Sample particles}
		\State $q\gets universal\_sampling(q, N, w)$ 
		\EndFor
		\EndWhile
		\EndProcedure
	\end{algorithmic}
	\caption{SOSMC for discrete sampling of microstructures}
	\label{algo:socm}
\end{algorithm}

\begin{figure*}[tbp]
	\centering
	\includegraphics[width=.33\linewidth]{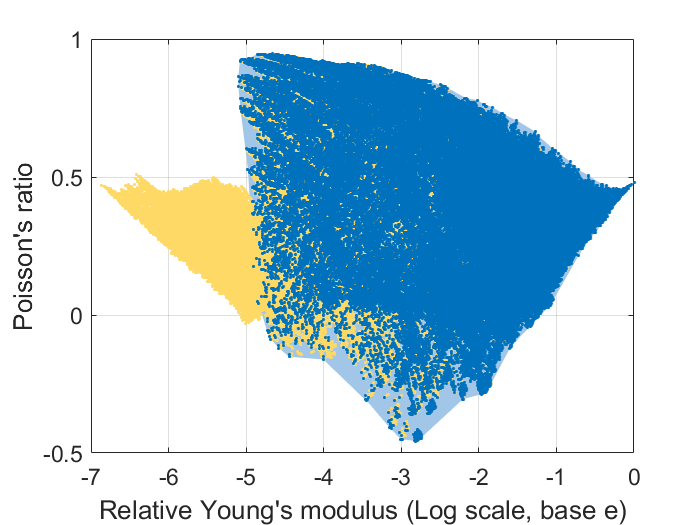}
	\includegraphics[width=.33\linewidth]{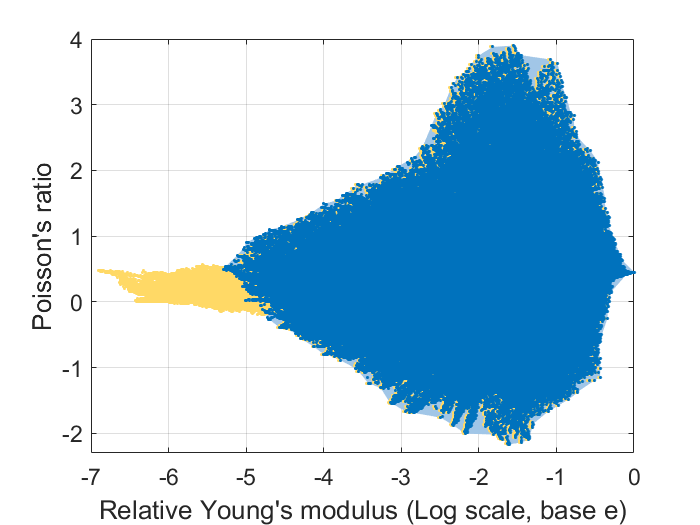}
	\includegraphics[width=.33\linewidth]{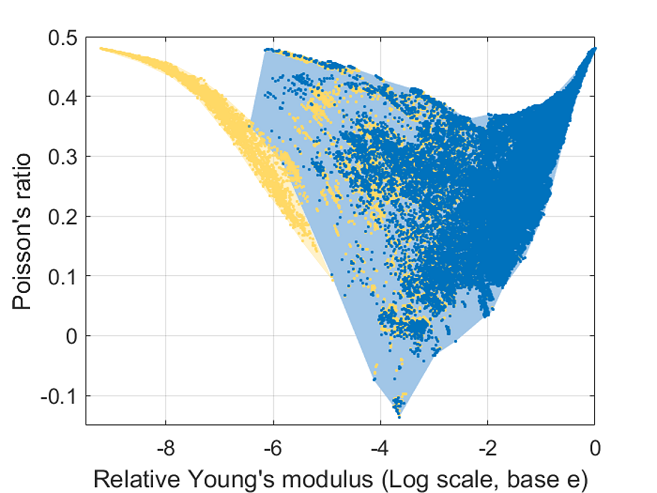}
	\caption{Gamuts computed with our discrete-continuous sampling scheme for 2D cubic structures (\emph{left}) , 2D orthotropic structures (\emph{middle}) and 3D cubic structures (\emph{right}). The plots show the results for the projection of the gamuts on the plane defined by the macroscale Young's modulus along the x axis (normalized by the Young's modulus of the stiffest base material) and the Poisson's ratio corresponding to a contraction along the y-direction when the material is stretched along the x-direction. All these plots correspond to microstructures that use 0.48 as Poisson's ratio for the base material. The blue dots correspond to the filtered one-material microstructures while the yellow dots correspond to the original two-material microstructures. }
	\label{fig:gamuts_unfiltered}
\end{figure*}

\begin{figure*}[tbp]
	\centering
	\includegraphics[width=.245\linewidth]{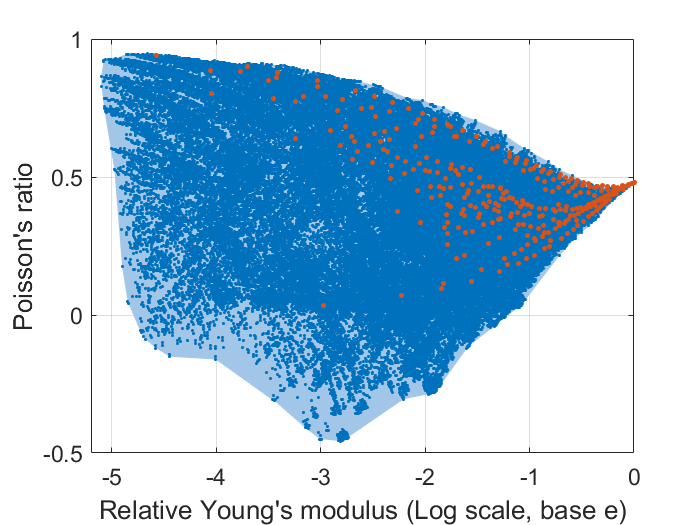}
	\includegraphics[width=.245\linewidth]{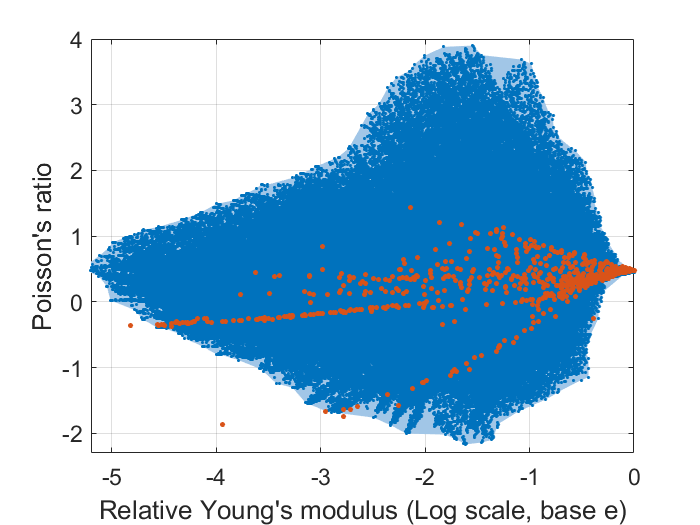} 
	\includegraphics[width=.245\linewidth]{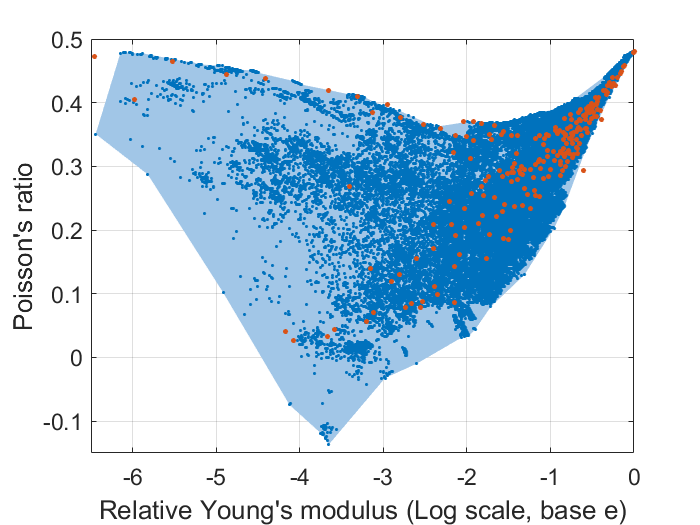}
	\includegraphics[width=.245\linewidth]{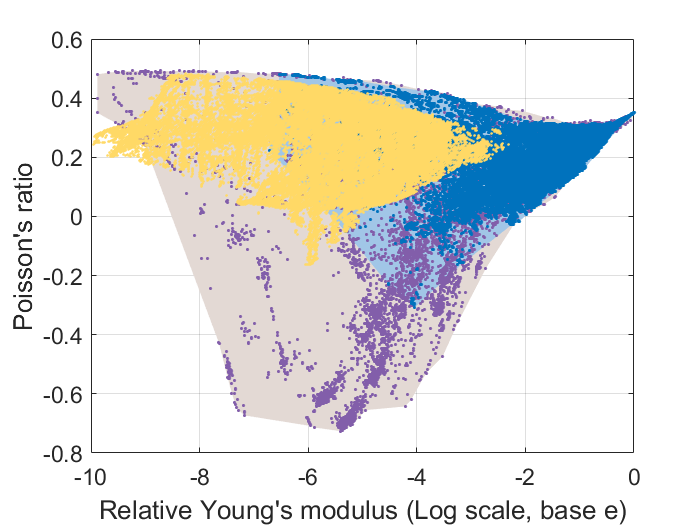}
	\caption{Gamuts computed with our discrete-continuous sampling scheme for 2D cubic structures (\emph{left}) , 2D orthotropic structures (\emph{second from left}) and 3D cubic structures (\emph{second from right}) using 0.48 as Poisson's ratio, and 3D cubic structures with 0.35 as Poisson's ratio (\emph{right}). The plots show the results for the projection of the gamuts on the plane defined by the macroscale Young's modulus along the x axis (normalized by the Young's modulus of the stiffest base material) and the Poisson's ratio corresponding to a contraction along the y-direction when the material is stretched along the x-direction. The blue dots correspond to the generated samples for $16\times16\times16$ microstructures, the purple dots correspond to generated samples for $64\times64\times64$ microstructures, the orange dots correspond to the microstructures from Schumacher et al. \shortcite{Schumacher:2015} and the yellow dots correspond to the microstructures from Panetta et al. \shortcite{Panetta:2015}.}
	\label{fig:gamuts_comparisons}
\end{figure*}

Starting with the microstructures corresponding to the entire gamut, we initialize the population of microstructures to evolve by sampling N microstructures using systematic resampling \cite{Douc05resamplingSchmes} based on their scores as computed by Equation \ref{eq:score_fct}. We then run, for each microstructure, the program described by Algorithm \ref{algo:proc_microstructure} in order to evolve the population. The program is not executed entirely but paused after the microstructure is modified, i.e. after the inner loop of the procedure has been executed, which corresponds to a so-called {\it barrier synchronization point}. When all the programs corresponding to all the microstructures of the population have reached this synchronization point, the scores of the partially modified microstructures are evaluated again and the population is resampled using systematic resampling. We again sample N microstructures, but since the scores have changed, the most interesting microstructures will appear several times, while the less promising ones will leave the evolving population. Note that the microstructures, and the associated programs, are duplicated together with their program execution history, i.e. the information regarding which voxels have been already visited, so that these voxels are not modified a second time. This ensures that interesting changes in the material assignments are preserved. After the synchronization point is reached and the microstructures have been resampled, the execution of the programs is resumed and the algorithm continues until all the voxels of all the microstructures have been visited. The entire SOSMC algorithm is summarized by Algorithm \ref{algo:socm}. In our implementation, we used $N=3000$ microstructures for the 2D and 3D cubic databases, and $N=10000$ for the 2D orthotropic database. 

\subsection{Material Gamuts}

\begin{figure}
	\includegraphics[width=.8\linewidth]{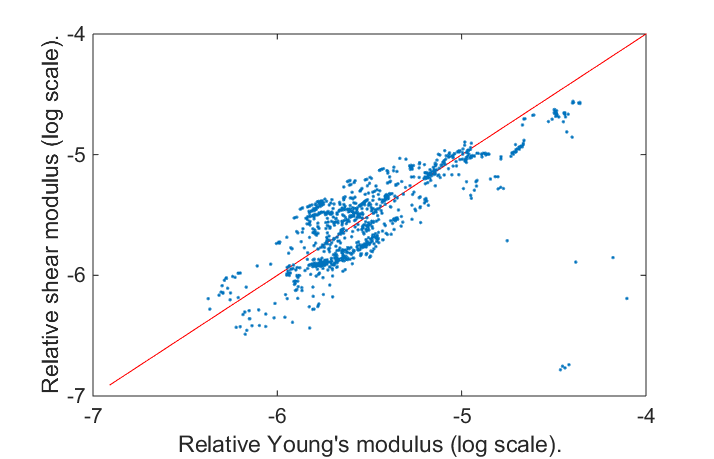}
	\centering
	\caption{The $64^3$ microstructures span a wide range of relative shear modulus even they have negative Poisson's ratios. As an example, we plotted the distribution of structures with Poisson's ratio $\nu=-0.5\pm 0.03$. Points on the diagonal line $\mu=\mu_{iso}$ are isotropic within the linear elasticity regime.}
	\label{fig:muIso}
\end{figure}

We initially targeted multi-material printers and therefore computed databases of two- and three-dimensional microstructures made of two materials. We used isotropic base materials whose Young's modulus differed by a factor of 1000 and having 0.48 as Poisson's ratio. For comparison purposes with previous research (\ref{fig:gamuts_comparisons}), we adapted these databases to one-material microstructures by replacing the softer material by void, filtering out all the microstructures with disconnected components, filling the enclosed voids in the 3D case, and recomputing their homogenized properties. The two sets of gamuts are depicted in Figure \ref{fig:gamuts_unfiltered}. We observed that the gamuts corresponding to two-material microstructures and one-material microstructures have a very similar shape, except in the area corresponding to very soft microstructures. This is to be expected since microstructures  made of soft material connecting small blocks of stiff materials are not realizable in the one-material setting. From an implementation point of view, it is worthwhile to note that, if the final intent of the user is to design one-material structures, these additional fabrication constraints can be directly accounted for during the sampling stage by preventing any change that would affect the validity of the microstructures. This avoids the need of filtering the database a posteriori \updated{and would improve the sampling density in regions that might be undersampled if one filters the database in a subsequent step. Note that we sampled the microstructures using a non-logarithmic scale for the Young's modulus and that the estimated density in the soft regions is higher than what it appears to be in Figures \ref{fig:gamuts_unfiltered} and \ref{fig:gamuts_comparisons}.
	
	Our initial databases were computed for microstructures corresponding to $16\times16\times16$ arrangements of voxels. This limits the sizes of the thinnest features of the microstructures and therefore the softness of the softer material that can be achieved. When increasing the lattice size to $64$, the gamut of the microstructure properties expands in this area and reaches what can be obtained when using other parametrization methods (see Figure \ref{fig:gamuts_comparisons}, \emph{right}). Due to high computation costs (each microstructure takes 29s to simulate in average), we limited our initial analysis of highly discretized geometries to the study of a database comprising about 10k microstructures.} Notably, our search method allows us to find a wide range of single-material structures with negative Poisson's ratio ($\nu=-0.7$). While lower Poisson's ratio is theoretically achievable, such structures contain extremely thin joints unsuitable for manufacturing.
These structures demonstrate a variety of relationships between Young's modulus and shear modulus.
To be more precise, we define $\mu_{iso}$ as the shear modulus computed from Young's modulus $E$ and Poisson's $\nu$ ratio using the relationship 
$\mu_{iso}=E/(1-2\times\nu)$. We then compare $\mu_{iso}$ with the actual shear modulus $\mu$ of each structure.
For structures with $\nu=-0.5\pm 0.03$, the ratio $\mu/\mu_{iso}$ achieves a range from $0.09$ to $1.39$ (Figure~\ref{fig:muIso}).
